\pdfoutput=1
\documentclass[11pt]{article}

\usepackage[utf8]{inputenc}
\usepackage[T1]{fontenc}
\usepackage{newpxtext}

\usepackage{tcolorbox}
\usepackage{fullpage}
\usepackage{tikz}
\usepackage{algorithmicx}
\usepackage{algorithm}
\usepackage[noend]{algpseudocode}
\usepackage[colorinlistoftodos]{todonotes}
\usepackage{soul}
\usepackage{url}
\definecolor{purple}{cmyk}{.51,.91,0,.34}
\usepackage{microtype}
\usepackage[pdftex, colorlinks,
  linkcolor=blue,
  citecolor=purple,
  filecolor=blue,urlcolor=blue]%
  {hyperref}

\usepackage{amsfonts}
\usepackage{mathrsfs}
\usepackage{mathtools}
\usepackage{amsthm}
\newtheorem{theorem}{Theorem}[section]
\newtheorem{lemma}[theorem]{Lemma}
\newtheorem{proposition}[theorem]{Proposition}
\newtheorem{corollary}[theorem]{Corollary}
\newtheorem{claim}[theorem]{Claim}
\newtheorem{open-problem}[theorem]{Open Problem}

\theoremstyle{definition}
\newtheorem{definition}[theorem]{Definition}

\newtheorem{example}[theorem]{Example}
\newtheorem{remark}[theorem]{Remark}
\newtheorem{observation}[theorem]{Observation}
\theoremstyle{remark}

\numberwithin{equation}{section}

\title{Bi-Arc Digraphs and Conservative Polymorphisms\thanks{This version of the paper is different from the previous \emph{arXiv} versions in several ways.
Apart from continuing to improve the overall presentation (in particular the algorithm
and Section 9), in this version we also discuss the complexity of the recognition problem 
for higher arity structures that admit a conservative semilattice polymorphism; we obtain 
a full dichotomy classification of this problem for general relational structures.}}

\author{Pavol Hell\thanks{Department of Computing Sciences, Simon Fraser University, Burnaby, Canada. Email: {\tt pavol@sfu.ca}. Research supported by NSERC Canada.} \and Akbar Rafiey\thanks{Department of Computing Sciences, Simon Fraser University, Burnaby, Canada. Email: {\tt arafiey@sfu.ca}. Research supported by NSERC Canada.} \and Arash Rafiey\thanks{Department of Math and Computer Science, Indiana State University, Indiana, USA. Email: {\tt arash.rafiey@indstate.edu, arashr@sfu.ca}. Research supported in part by NSF grant 1751765.}}

\date{}
\begin{document}

\maketitle


\begin{abstract}
In this paper we study the class of {\em bi-arc digraphs}, important
from two seemingly unrelated perspectives. On the one hand, they
are precisely the digraphs that admit certain polymorphisms of interest
in the study of constraint satisfaction problems; on the other hand,
they are a very broad (in a certain sense the broadest reasonable)
generalization of interval graphs.

The class of bi-arc digraphs is precisely the class of digraphs that
admit conservative semilattice polymorphisms. There is much interest
in understanding structures that admit particular types of polymorphisms,
and especially in their recognition algorithms. (Such recognition problems
are usually referred to as “metaproblems”.) It turns out that the class
of bi-arc digraphs also precisely describes the class of digraphs that admit
certain other kinds of conservative polymorphisms (cyclic, and totally
symmetric, polymorphisms of all arities). Thus solving the recognition
problem for bi-arc digraphs solves the metaproblem for digraphs for
several types of conservative polymorphisms. The complexity of the
recognition problem for digraphs with conservative semilattice
polymorphisms was an open problem, while it was known that
the problem is NP-complete for certain more complex relational
structures. We complement our result by providing a complete
dichotomy classification of which general relational structures have
polynomial or NP-complete recognition problems for the existence
of conservative semilattice polymorphisms.

The class of bi-arc digraphs also generalizes the class of interval graphs;
in fact it reduces to the class of interval graphs for digraphs that are
symmetric and reflexive. It is much broader than interval graphs and
includes other generalizations of interval graphs such as co-threshold
tolerance graphs and adjusted interval digraphs. Yet, it is still a reasonable
extension of interval graphs, in the sense that it keeps much of the appeal
of interval graphs (as we show in this paper).

Our main result is a forbidden obstruction characterization of, and a
polynomial recognition for, the class of bi-arc digraphs. This is accomplished
by a detailed analysis of possible structures in the space of ordered pairs
of vertices of a digraph.
\end{abstract}


\clearpage

{
  \hypersetup{linkcolor=black}
  \tableofcontents
}

\section{Background and Motivation} \label{section1}
\subsection{CSPs, meta-question and algebraic motivation}
The \emph{constraint satisfaction problem (CSP)} involves deciding, given a set
of variables and a set of constraints on the variables, whether or not there is an assignment to
the variables satisfying all of the constraints. This problem can be formulated in terms of homomorphims as follows. Given a pair $(\mathbb{G}, \mathbb{H})$ of \emph{relational structures}, decide whether or not there is
a homomorphism from the first structure to the second structure. 
A common way to restrict this problem is to fix the second structure $\mathbb{H}$, so that each structure $\mathbb{H}$ gives
rise to a problem CSP($\mathbb{H}$). The most effective approach to the study of the CSP($\mathbb{H})$ is the so-called
algebraic approach that associates every $\mathbb{H}$ with its \emph{polymorphisms}. This approach culminated with the recent papers by Bulatov~\cite{bulatov-dichotomy} and Zhuk~\cite{zhuk}, settling the long-standing Feder-Vardi dichotomy conjecture for finite domain CSPs~\cite{fv98}. Roughly speaking, the presence of nice enough polymorphisms leads directly to polynomial time tractability of CSP($\mathbb{H}$), while their absence leads to hardness.  
Beside decision CSPs, polymorphisms have been used extensively for approximating CSPs, robust satisfiability of CSPs, and testing solutions (in the sense of property testing)~\cite{BK, BKW,ChenVY16,DalmauKKMMO17,esa2012,kotyz,RRS}.

An interesting question arising from these studies, in particular the Dichotomy Theorem~\cite{bulatov-dichotomy,zhuk}, is known as the \emph{meta-question}. Given a relational structure $\mathbb{H}$, decide whether or not $\mathbb{H}$ admits a polymorphism from a class--for various classes of polymorphims. For many cases hardness results are known. One particular case, that is the study of this paper, is deciding whether or not $\mathbb{H}$ admits a \emph{(conservative) semilattice} polymorphism. The presence of semilattice polymorphisms lead to many positive results. As an example, it is now a classic theorem in the area that for any structure $\mathbb{H}$ having a semilattice polymorphism, the problem CSP($\mathbb{H}$) is polynomial time decidable \cite{JeavonsCG97}. In terms of approximation algorithms, Minimum Cost Homomorphism problem to $\mathbb{H}$ (when $\mathbb{H}$ is a digraph) is approximable within a constant factor if $\mathbb{H}$ admits a conservative semilattice polymorphism \cite{esa2012,RRS}. In terms of robust satisfiability, given a $(1-\varepsilon)$-satisfiable instance of CSP($\mathbb{H}$), it is easy to find a $(1-O(1/\log(1/\varepsilon)))$-satisfying assignment if $\mathbb{H}$ admits a semilattice polymorphism (in fact, the result holds for width-1 CSPs). However, on negative side, it is hard to find a $(1-o(1/\log(1/\varepsilon)))$-satisfying assignment if $\mathbb{H}$ admits a semilattice polymorphism~\cite{kotyz}\footnote{Hard instances (e.g., Horn $k$-SAT) can be found in ~\cite{GuruswamiZ11,Zwick98}. The hardness is implied by Raghavendra's result~\cite{Raghavendra08} and assuming Unique Game Conjecture.}.

Chen and Larose~\cite{benoit} proved that it is hard to decide if a relational structure admits a (conservative) semilattice polymorphism. Indeed, they proved this problem remains hard even for structures with at most binary relations. A relational structure is \emph{at most binary} if the arity of each relation is less than or equal to 2.

\begin{theorem}[\cite{benoit}]
Deciding if an at most binary relational structure (with polynomially many relations) admits any of the following is NP-complete: (1) a semilattice polymorphism, (2) a conservative semilattice polymorphism, (3) a commutative, associative polymorphism.
\end{theorem}

However, for a single binary relation, i.e., a digraph, the meta-question often turns out to be better behaved. For instance, there are forbidden induced structure characterizations for existence of conservative \emph{majority}~\cite{soda} and conservative \emph{Maltsev}~\cite{catarina,soda} polymorphisms in digraphs. The question of whether the existence of conservative semilattice polymorphism is polynomial was
explicitely raised in~\cite{bagan}. This problem is polynomial for reflexive digraphs~\cite{adjusted} and bipartite digraphs~\cite{esa2012}. In this paper, we give forbidden obstruction characterization for digraphs admitting a conservative semilattice polymorphism.
Other questions about the existence of polymorphisms of various kinds have turned out to also be
interesting \cite{nuf1,benoit,nuf2,soda,kazda,Benoit2017,maroti}. In particular, the existence of 
conservative polymorphisms is a hereditary property (if $H$ has a particular kind of conservative 
polymorphism, then so does any induced subgraph of $H$). Thus, these questions present interesting problems in graph theory. 

\subsection{Graph theoretic motivation}\label{graph-motovation}
\noindent \textbf{Digraph Generalization of Interval Graphs:} Part of our motivation also stems from a wish to generalize interval graphs. A graph $H$ is 
an {\em interval graph} if there is a family of intervals $I_v, v \in V(H),$ such that $uv \in E(H)$ 
if and only if $I_u \cap I_v \neq \emptyset$. Interval graphs constitute one of the most important 
graph classes; they admit efficient recognition algorithms, elegant obstruction characterizations, 
and frequently occur in applications \cite{booth,corneil,fg,gol,habib,lekk}. The classical digraph 
version of interval graphs \cite{sen-west} lacks many of these desirable attributes. A more
successful generalization is given in \cite{adjusted}: we say that $H$ is an {\em adjusted 
interval digraph} if there are two families of real intervals $I_v, J_v, v \in V(H)$, where for 
each $v \in V(H)$ the intervals $I_v, J_v$ have the same endpoint, such that $uv \in A(H)$ 
if and only if $I_u \cap J_v \neq \emptyset$. Adjusted interval digraphs have many of the
desirable algorithmic attributes of interval graphs, including recognition algorithms and 
forbidden structure characterizations \cite{adjusted}.

It is useful to view both interval graphs and adjusted interval digraphs as being {\em reflexive},
i.e., each vertex having a loop. (This is consistent with their definition as each $I_v$ intersects
itself, or the corresponding $J_v$.) For reflexive digraphs, the adjusted interval digraphs appear 
to be the right generalization of interval graphs. For general (not necessarily reflexive) digraphs,
the right analogue was less clear. Another special class of digraphs are {\em bipartite} digraphs, 
which are just bipartite graphs with all edges oriented from one part of the bipartition to the other 
part. It turns out there is a natural generalization of interval graphs amongst bipartite digraphs, 
namely the {\em two-directional orthogonal ray digraphs} \cite{Ueno}; which has many equivalent 
definitions \cite{jing,esa2012}, and also shares several of the desirable properties of interval graphs.

 It turns out that a digraph admits a conservative semilattice polymorphism if and only if it has a \emph{min ordering}. A reflexive graph has a min ordering if and only if it is an interval graph, a reflexive digraph has
a min ordering if and only if it is an adjusted interval digraph, and a bipartite digraph has a min 
ordering if and only if it is a two-directional orthogonal ray graph \cite{adjusted,jing,esa2012,Ueno}.
Thus it was long believed that \emph{min-orderable} digraphs are the right overall generalization of interval graphs. However, it was not known whether this class of digraphs can be recognized in polynomial 
time, whether it has an obstruction characterization, and whether it has any geometric meaning. 
Recently, two geometric representations of the class of digraphs with a min ordering have been 
given in \cite{minorder-digraphs,mfcs2018}. Min-orderable digraphs are shown there to be exactly
the same as {\em signed-interval digraphs}, which arise as a natural extension of another well
studied graphs class, the complements of so-called {\em threshold tolerance graphs}. They are
also shown to be exactly the same digraphs as {\em bi-arc digraphs}, which are defined as a digraph
analogue of the previously studied class of bi-arc graphs \cite{bi-arc}. Both these classes are defined
by intersection or inclusion of intervals or circular arcs. Thus it remained to find a forbidden structure 
characterization for, and a polynomial time recognition algorithm of, min-orderable digraphs. This is
what we accomplish in this paper, thus contributing to the argument that min-orderable digraphs are 
the right general digraph analogue of interval graphs.

\subsection{Our contributions}
In this paper we study the problem of deciding if a relational structure $\mathbb{H}$ admits a conservative semilattice (CSL) polymorphism. That is, we study for which relational structures {\sc Problem~\ref{problem1}} is polynomial time decidable and for which ones it is NP-complete.\\
\begin{tcolorbox}
{\sc Problem:\label{problem1}}\\
	\emph{Input:} A relational structure $\mathbb{H} = \langle V, R_1, \ldots , R_s  \rangle$,\\
	\emph{Goal:} Decide if $\mathbb{H}$ admits a conservative semilattice (CSL) polymorphism.
\end{tcolorbox}
  Note that any unary relation $R$ admits a CSL polymorphism. This is because if $a,b\in R$, then applying CSL polymorphism $f$ on $a,b$, would give either $a$ or $b$, and hence, $R$ is closed under $f$. So the interesting cases are when the arity of $R$ is at least two. On the positive side, we present a polynomial time algorithm that, given a relational structure with a single binary relation $\mathbb{H}=\langle V, A(V)\rangle$ i.e., digraph, decides if $\mathbb{H}$ admits a CSL polymorphism.
\begin{theorem}[Main Theorem]
 There exists a polynomial time algorithm that, given a digraph $H$, decides if $H$ admits a CSL polymorphism or not. 
\end{theorem}
    We also have a structural characterization of
    digraphs with a CSL polymorphism, in terms of a forbidden structure we
    call a \emph{strong circuit}. Recall that the class of digraphs that admit a CSL polymorphism is exactly the class of digraphs admitting a min ordering (also called bi-arc digraphs). Furthermore, this class coincides with the class of signed-interval digraphs.
\begin{corollary}
The class of min-orderable digraphs, bi-arc digraphs and signed-interval digraphs can be recognized in polynomial time.
\end{corollary}
Furthermore, we show that there is quite a bit of collapse for digraph classes in the conservative case. We will point out that the class of digraphs with a min ordering is included in the class of digraphs with a \emph{conservative set} polymorphism, which is included in the class of digraphs with a \emph{conservative and commutative} polymorphism (called CC polymorphism). We will give forbidden induced structure characterizations for all three of these digraph classes, from which it will follow that (surprisingly) the first two classes coincide. In all three cases, the characterizations yield polynomial time recognition algorithms. Although this was known for CC polymorphisms \cite{fv} (even known to be in
non-deterministic logspace \cite{benoit}), it was open for CSL and conservative set polymorphisms
\cite{bagan,benoit,monoton-proper}. Formally, we prove the following. 
\begin{theorem}
Let $\mathbb{H}$ be a digraph, then the following are equivalent: 
\begin{enumerate}
\item $\mathbb{H}$ admits a CSL polymorphism; 
\item $\mathbb{H}$ admits a conservative set polymorphism; 
\item $\mathbb{H}$ admits conservative cyclic polymorphisms of all arities.
\end{enumerate}
\end{theorem}
On the negative side, we prove that it is NP-complete to decide if a relational structure $\mathbb{H}=\langle V, R\rangle$ where $R$ is a ternary relation (arity of $R$ is three) admits a CSL polymorphism.

\begin{theorem}
Deciding if a relational structure with a single ternary relation admits a CSL polymorphism is NP-complete.
\end{theorem}
This leads us to the following dichotomy classification of the complexity of {\sc Problem \ref{problem1}}.

\begin{theorem}[Dichotomy Theorem]
Deciding if a relational structure $\mathbb{H}=\langle V, R_1,\dots, R_k \rangle$ admits a CSL polymorphism is polynomial-time solvable if all relations $R_i$ are unary, except possibly one binary relation. In all other cases, the problem is NP-complete.
\end{theorem}

\section{Preliminaries and Notation}\label{preliminaries} 
\subsection{Relational structures and polymorphisms} 
A \emph{relational structure} is a tuple
$\mathbb{H} = \langle V, R_1, \ldots,$ $R_s \rangle$ where $V$ is a non-empty finite set, called universe, and each $R_i$ is a relation of arity $r_i$ on $V$. For instance, a digraph $H$ with vertex set $V(H)$ and arc set $A(H)$ is a relational structure with universe $V(H)$ and a single binary relation $A(H)$ i.e., $H = \langle V(H), A(H)\rangle$. A \emph{polymorphism} of a structure $\mathbb{H}$ is defined as a finitary operation $f : V^k \to V$ that is a \emph{homomorphism} from $\mathbb{H}^k$ to $\mathbb{H}$. 
If $f$ is a polymorphism of $\mathbb{H}$ we also say that $\mathbb{H}$ admits $f$. 
\begin{example}[Digraphs]
Given digraphs $G$ and $H$, a {\em homomorphism} of $G$ to $H$ is a mapping $f : V(G) \to V(H)$ such that $uv \in A(G)$ implies
$f(u)f(v) \in A(H)$. A {\em product} of digraphs $G$ and $H$ has the vertex set $V(G) \times V(H)$
and arc set $A(G \times H)$ consisting of all pairs $(u,x)(v,y)$ such that $uv \in A(G)$ and
$xy \in A(H)$. The product of $k$ copies of the same digraph $H$ is denoted by $H^k$. A {\em
polymorphism} of $H$ of order $k$ is a homomorphism of $H^k$ to $H$. In other words, it is a
mapping $f$ from the set of $k$-tuples over $V(H)$ to $V(H)$ such that if $x_iy_i \in A(H)$
for $i = 1, 2, \dots, k$, then $f(x_1,x_2,\dots,x_k)f(y_1,y_2,\dots,y_k) \in A(H)$. 
\end{example}
A polymorphism $f$ is {\em conservative} if each value $f(x_1,x_2,\dots,x_k)$ is one of the arguments 
$x_1, x_2, \dots  , x_k$. A binary (arity two) polymorphism $f: V^2\to V$ that is conservative and \emph{commutative} 
$(f(x,y) = f(y,x)$ for all vertices $x,y$) is called a CC polymorphism. Notice that by definition any binary CC polymorphism is \emph{idempotent} i.e, $f(x,x)=x$.  If $f$ is additionally \emph{associative} then it is called a {\em conservative semilattice} or a 
CSL polymorphism. That is, it satisfies the
following  \emph{identities}, $f(f(x, y), z) = f(x, f(y, z)), \text{ and } f(x, y) = f(y, x) \in \{x,y\}$
for all $x,y,z \in V$.
A CSL polymorphism $f$ of digarph $H$ is naturally associated with a binary relation $\leq_f$ on the vertices of $H$ by $x
\leq_f$ y if and only if $f(x,y)=x$. By
associative property, the relation $\leq_f$ is a linear order on $V(H)$, and it can be easily checked that $\leq_f$ is in fact a \emph{min ordering} of $H$. We say that a linear order $\leq$ on $V(H)$ is a min ordering of $H$ if 
\[
uv \in A(H), u'v' \in A(H) \text{ and }  u<u', v <v' \implies  uv' \in A(H).
\] 
Conversely, with any min ordering $\leq$ of $H$, we can associate a polymorphism $f_{\leq} : H^2 \rightarrow H$ by setting $f_{\leq} (x,y) = \min(x,y)$. It is again easy to check that $f_{\leq}$ is a CSL polymorphism. We have proved the following fact.

\begin{proposition}
 A digraph H admits a min ordering if and only if it admits a CSL polymorphism. 
\end{proposition}


\subsection{Graph notation}
A digraph $H$ consists of a finite vertex set $V(H)$ and an arc set $A(H)$, each arc being an ordered pair of vertices. 
We say that $uv \in A(H)$ is an arc from $u$ to $v$. Sometimes we emphasize this by
saying that $uv$ is a {\em forward arc} of $H$, and also say $vu$ is a {\em backward arc} of $H$.
We say that $u, v$ are {\em adjacent} in $H$ if $uv$ is a forward or a backward arc of $H$ (i.e. either $uv\in A(H)$ or $vu\in A(H)$). 
A symmetric arc is an arc $uv \in  A(H)$ such that $vu \in  A(H)$; thus a symmetric arc is both a forward arc and a backward arc.

A 
{\em walk} in $H$ is a sequence $P = x_0, x_1, \dots, x_n$ of 
consecutively adjacent vertices 
of $H$; note that a walk has a designated first and last vertex. A {\em path} $P = x_0, x_1, \dots,$
$x_n$ is a walk in which all $x_i$ are distinct. A walk $P = x_0, x_1, \dots, x_n$ is {\em closed} 
if $x_0=x_n$ and a {\em cycle} if all other $x_i$ are distinct. A walk is {\em directed} if all its
arcs are forward. A directed path $P=x_1,\dots, x_n$ is a directed walk in which all $x_i$ are distinct. A vertex $u'$ is said to be {\em reachable} from a vertex $u$ in $H$ if there 
is a directed path from $u$ to $u'$ in $H$; a set $U'$ is {\em reachable} from a set $U$ if every 
vertex of $U'$ is reachable from some vertex of $U$. Note that every vertex is reachable from itself, by a directed path 
of length zero. 

For walks $P$ from $a$ to $b$, and $Q$ from $b$ to $c$, we denote by $P+Q$ the walk from
$a$ to $c$ which is the concatenation of $P$ and $Q$, and by $P^{-1}$ the walk $P$ traversed
in the opposite direction, from $b$ to $a$. We call $P^{-1}$ the {\em reverse} of $P$. For a
closed walk $C$, we denote by $C^a$ the concatenation of $C$ with itself $a$ times.

The {\em net length} of a walk is the number of forward arcs minus the number of 
backward arcs. A closed walk is {\em balanced} if it has net length zero; otherwise it is {\em 
unbalanced}. Note that in an unbalanced closed walk we may always choose a direction in which the 
net length is positive (or negative).
A digraph is {\em unbalanced} if it contains an unbalanced closed walk (or equivalently an unbalanced 
cycle ); otherwise it is {\em balanced}. It is easy to see that a digraph is balanced if and only if 
it admits a {\em labeling} of vertices by non-negative integers so that each arc goes from a vertex
with a label $i$ to a vertex with a label $i+1$. The {\em height} of $H$ is the maximum net length 
of a walk in $H$.  Note that an unbalanced digraph has infinite height and the height of a balanced
digraph is the greatest label in a non-negative labeling in which some vertex has label zero.

For a walk  $P=x_0,x_1,\dots,x_n$ and any $i \le j$, we denote by $P[x_i,x_j]$ the walk $x_i, x_{i+1}, \dots, x_j$,
and call it a {\em prefix} of $P$ if $i=0$. Suppose $P = x_0, x_1, \dots, x_n$ is a walk in $H$ of net length 
$k \geq 0$. We say that $P$ is {\em constricted from below} if the net length of any prefix $P[x_0, x_j]$ is 
non-negative, and is {\em constricted from above} if the net length of any prefix is at
most $k$. We also say that  $P$ is {\em constricted} if it is constricted both from below and 
from above. Moreover, we say that $P$ is {\em strongly} constricted from below or above, if the 
corresponding net lengths are strictly positive or smaller than $k$. For walk $P$ of net length 
$k< 0$, we say that $P$ is (strongly or not) constricted from below, or above, or both, if the above 
definitions apply to the reverse walk $P^{-1}$.

\begin{definition}[An extremal vertex]
Consider a cycle $C$ in $H$ of positive net length $k$. A vertex $v$ is {\em extremal} in $C$ 
if traversing $C$ from $v$ in the positive direction yields a walk constricted from below.
\end{definition}

We observe that a cycle $C$ of positive net length $k$ has at least $k$ extremal vertices. Namely, we can obtain such vertices $v_0, v_1, \dots , v_k$ as follows:
starting at any vertex $x$ and following $C$, the net length of the prefix $C[x,v]$ varies with $v$ from $0$ to a possibly 
negative minimum $m$, but ending with $k>0$. We  let $v_0$ be the last vertex with the 
net length of $C[x,v_0]$ equal to the minimum $m$ (possibly $v_0=x$ if $m=0$). We can let 
$v_i$ be the last vertex with the net length of $C[v_0,v_i]$ equal to $i, i=1, 2, \dots, k-1$. Note
that each walk $C[v_i,v_{i+1}]$ is constricted from below and has net length one. We also note
for future reference that any other extremal vertex of $C$ has a walk of net length zero to one of 
$v_0, v_1, \dots, v_{k-1}$.  We say vertex $x$ is an {\em extremal vertex in a digraph } $H$ if there exists a cycle $C$ in $H$ that $x$ is a extremal vertex in $C$. 
A cycle of $H$ is {\em induced} if $H$ contains no other arcs on the vertices of the cycle. In particular, an induced cycle with more than one vertex does not contain a
loop.\\
We define two walks $P = x_0, x_1, \dots, x_n$ and $Q = y_0, y_1, \dots, y_n$ in $H$ to be
{\em congruent}, if they follow the same pattern of forward and backward arcs, i.e., $x_ix_{i+1}$
is a forward (backward) arc if and only if $y_iy_{i+1}$ is a forward (backward) arc (respectively).
Suppose the walks $P, Q$ as above are congruent. We say an arc $x_iy_{i+1}$ is {\em a faithful
arc from $P$ to $Q$}, if it is a forward (backward) arc when $x_ix_{i+1}$ is a forward (backward)
arc (respectively), and we say an arc $y_ix_{i+1}$ is {\em a faithful arc from $Q$ to $P$}, if it is
a forward (backward) arc when $x_ix_{i+1}$ is a forward (backward) arc (respectively). We say
that $P$ {\em avoids} $Q$ if there is no faithful arc from $P$ to $Q$ at all. 
\subsection{Pair digraph}
Let us introduce a basic tool for this paper, the {\em pair digraph} $H^+$.
\begin{definition}[The pair digraph $H^+$]
The vertices of $H^+$ are all ordered pairs $(x,y)$ of distinct vertices of $H$. To avoid confusion with the vertices of H, we will call the vertices of $H^+$  {\em pairs}. The pair digraph $H^+$ has an arc from pair $(x,y)$ to pair $(x',y')$ just if

\begin{enumerate}
\item
$xx',yy' \in A(H)$ but $xy' \not\in A(H)$, or 
\item
$x'x, y'y \in A(H)$ but $y'x \not\in A(H)$. 
\end{enumerate}
In the former case we call the arc $(x,y)(x',y') \in A(H^+)$ a {\em positive arc}, and the second
case we call it a {\em negative arc}.
\end{definition}
 
We say a positive (negative) arc $(x,y)(x',y')$ in $H^+$, is symmetric if and only if $(y,x)(y',x')$ is a positive (negative) arc in $H^+$. In other words, when $(x,y)(x',y')$ is a symmetric arc we have    
 $xx', yy' \in A(H)$ but $xy', yx' \not\in A(H)$, or $x'x, y'y \in A(H)$ but $y'x, x'y \not\in A(H)$. 
By definition, $(x,y)(x',y')$ is a positive (negative) symmetric arc if and only if $(x',y')(x,y)$ is a negative (positive) symmetric arc in $H^+$. 
 
Symmetric arcs of $H^+$ will play an important role. A walk, strong component, or subgraph of $H^+$ is called symmetric if all its arcs are symmetric.
Note that in $H^+$ we have an arc from $(x,y)$ to $(x',y')$ if and only if there is an arc from $(y',x')$ 
to $(y,x)$. We call this the {\em skew property} of $H^+$, and call the pair $(y, x)$ the {\em dual} of the pair $(x, y)$.

A directed path $W$ in $H^+$ corresponds precisely to a pair of congruent walks $P, Q$ in 
$H$ such that $P$ avoids $Q$. The {\em net value} of the directed path $W$ is defined to be the net 
length of the walk $P$ (or $Q$). It is the difference between the number of positive and negative arcs
of $W$. We say that $W$ has {\em constricted values} if the walk $P$ (or $Q$) is constricted, i.e.,
if each prefix of $W$ has net value between zero and the net value of $W$. Walks with
values constricted below or above are defined similarly. 
Other notions for $H^+$ are also defined in the manner corresponding to the notions in $H$. 
In particular, an {\em extremal pair} of a cycle $C$ in $H^+$ is a pair $\bar{v}$ such that traversing $C$ from $\bar{v}$ in the positive direction yields a walk with values 
constricted from below. Similarly, a closed walk of $H^+$ is balanced if has net value zero, and unbalanced otherwise. 
A strong component \footnote{A strong component of digraph $G$ is a maximal set of vertices, such that $\forall \ \ u,v \in V(H)$, there is a directed path from $u$ to $v$ and a directed path from $v$ to $u$ } of $H^+$ is balanced if it does not contain an unbalanced closed walk, and 
unbalanced otherwise. A strong component $S$ of $H^+$, is balanced if every directed cycle of $S$ has net value zero. Finally, a pair is called balanced if it is in a balanced strong component otherwise it is called unbalanced. 
\begin{remark}
Note that $H^{+}$ being balanced does not necessarily mean that $H$ is balanced. For example, $H$ with $V(H)=\{a,b,c,a',b',c'\}, A(H)=\{ab,bc,ac,a'b',b'c',a'c'a'b,b'c,a'c\}$ is not balanced, but $H^+$ is balanced. 
\end{remark}

\section{Warm-up: Obstructions to CC Polymorphisms} 
As mentioned 
earlier, the existence of CC polymorphisms is well understood; it is solvable by 
methods for solving 2-SAT instances  \cite{fv,fv98} , so it is both known to be decidable in polynomial time and 
characterized by forbidden substructures \cite{papa}. In fact, it is shown in 
\cite{benoit} that it can be decided in non-determinstic logspace.
Nevertheless, we present our obstructions to the existence of CC polymorphism because
they illuminate the general obstructions to CSL polymorphisms, and underscore the
relationship between the two types of polymorphisms and their obstructions. It also 
illustrate our techniques on the easy case of CC polymorphisms.

Suppose $f$ is a CC polymorphism of $H$. If $xx', yy' \in A(H)$ but $xy' \not\in A(H)$, then
$f(x,y)=x$ implies $f(x',y')=x'$. A similar situation arises if $x'x, y'y \in A(H)$ but 
$y'x \not\in A(H)$, then again $f(x,y)=x$ implies $f(x',y')=x'$. Thus, we conclude that $(x, y)(x', y') \in A(H^+)$ means that $f(x, y) = x$ implies $f(x', y') = x'$.  
Therefore, having a directed path in $H^+$ from $(x,y)$ to $(x',y')$ means that $f(x,y)=x$
implies that $f(x',y')=x'$ in any CC polymorphism $f$ of $H$. In particular, for any strong component 
$C$ of $H^+$, and any CC polymorphism $f$ of $H$, either all pairs $(x,y) \in C$ are mapped by $f$
to the first coordinate or all are mapped to the second coordinate. Moreover, if $C_2$ is reachable
from $C_1$ in $H^+$, and $f$ maps pairs in $C_1$ to the first coordinate, then it also
maps pairs in $C_2$ to the first coordinate. 

An {\em invertible pair} of $H$ is a vertex $(x,y)$ of
$H^+$ such that $(x,y)$ and $(y,x)$ are in the same strong component of $H^+$. It is easy to see,
using the skew property of $H^+$, that if one vertex of a strong component of $H^+$ is invertible,
then so are all others, and, that if $H$ has no invertible pairs, then each component $C$ has a 
corresponding {\em dual} component \label{dual} $C'$ such that $(x,y) \in C$ if and only if $(y,x) \in C'$. 

We note that constructing a CC polymorphism $f$ for $H$ amounts to choosing one from each pair 
$C,C'$ of dual strong components of $H^+$, so that if $C_2$ is reachable from $C_1$ in $H^+$, and $C_1$ was chosen, then $C_2$ is also chosen. 
Then we can set $f$ to map each pair in the chosen strong components to its first coordinate. This can be done, for instance, by the following algorithm.

We say that a strong component $C$ of a digraph is {\em ripe} if no other strong component is reachable 
from it. The algorithm begins by selecting a ripe strong component $C$ of $H^+$, and deleting it and its 
dual $C'$ from $H^+$, continuing the same way with the remaining digraph.

This algorithm clearly selects exactly one from each pair $C, C'$ 
of dual strong components of $H^+$. It remains to show that if $C_2$ is reachable from $C_1$ in $H^+$, and 
$C_1$ was chosen, then $C_2$ is also chosen. Suppose for a contradiction, that $C_2$ was not chosen. 
Since $C_1$ was ripe when chosen, and $C_2$ is not chosen, $C_2$ must have been previously deleted when $C'_2$ was chosen.  
By the skew property of $H^+$, we see that $C_1'$ is reachable from $C_2'$ , so  $C_2'$ was chosen when it was not yet ripe, contradicting the rules of the algorithm.
\begin{theorem}\label{cc}
A digraph $H$ admits a CC polymorphism if and only if no strong component of $H^+$ contains an 
invertible pair.
\end{theorem}



\section{Obstructions to Min Ordering}
If $<$ is a min ordering of $H$, then $\min$ (with respect to $<$) is a CC polymorphism of $H$, so 
many of the observations in the previous section apply verbatim. (However, the two concepts differ: for example, the directed cycle $\vec{C_3}$ admits a CC polymorphism while it does not admit a CSL polymorphism.) In particular, if $xx', yy' \in A(H)$ but $xy' \not\in A(H)$, (or if 
$x'x, y'y \in A(H)$ but $y'x \not\in A(H)$), then $x < y$ implies $x' < y'$. (Otherwise 
$x < y, y' < x'$ would violate the min property). 

\begin{definition}[Circuit, Strong Circuit]
Let $D$ be a subset of $V(H^+)$. A {\em circuit} in $D^+$ is a set of pairs 
$(x_0,x_1), (x_1,x_2), \dots, (x_{n-1},x_n),(x_n,x_0)$ 
in $D$. A {\em strong circuit } is a circuit with 
all the pairs in the same strong component of $H^+$.
\end{definition} 
Thus, in a strong circuit there are directed path (in $H^+$) from $(x_{i-1},x_i)$ to $(x_i,x_{i+1})$ for 
all $i=1,2,\dots,n+1$, modulo $n+1$. Note that an invertible pair of $H$ is a strong circuit with $n=1$ in $H^+$. If  $H^+$ contains a strong circuit, then $H$ cannot have a min ordering, since 
$x_0 < x_1$ implies $x_0 < x_1 < x_2 < \dots < x_n < x_0$ (and similarly for $x_0 > x_1$), 
contradicting the transitivity of $<$. We have proved that if a digraph $H$ admits a min ordering then $H^+$ does not contain a strong circuit. Theorem~\ref{inv} claims that the converse also holds.

\begin{theorem}\label{inv}
A digraph $H$ admits a min ordering if and only if $H^+$ does not contain a strong circuit. Moreover, there exists an algorithm that, in time $O(|A(H)|^2)$, outputs a min ordering for $H$ if one exists. 
\end{theorem}
This is our main result, giving a polynomially testable characterization of min-orderable digraphs. It also nicely complements Theorem~\ref{cc}, highlighting the difference in the obstructions to CC polymorphisms and CSL polymorphisms. In the next section, we devise an algorithm that produces a min ordering for an input digraph $H$ if $H^+$ does not contain a strong circuit. 

\section{Our Algorithm}\label{our-algorithm}
\subsection{Informal description of the algorithm}
In this section, we introduce an algorithm to construct a min ordering $<$ of $H$, provided $H^+$ contains no strong circuit. As in the case of CC polymorphisms, we will be choosing pairs of $H^+$ to decide the ordering. Specifically, if a pair $(x,y)$ of $H^+$ is chosen, we will set $x < y$. As before, choosing a pair requires choosing all pairs reachable from it. The process of choosing is different for pairs in balanced and unbalanced strong components. However, in each case, the set of chosen pairs will be closed under reachability (see Definitions \ref{reachibility-notation} and \ref{reachability-closure}).
Then all the duals of the chosen pairs will be discarded. At any stage of the algorithm, we will have a set $V_c$ of chosen pairs, and a set $V_d$ of discarded pairs; the pairs in the set $R=V(H^+) \setminus (V_c \cup V_d)$ will be called the remaining pairs. Initially, we will have $V_c = V_d = \emptyset$, and throughout the algorithm we will maintain the following properties:
\begin{enumerate}
\item \label{property-1}
$(a,b) \in V_c$ if and only if $(b,a) \in V_d$;
\item \label{property-2}
if $(a,b) \in V_c$ and $(a,b)(a',b') \in A(H^+)$ then $(a',b') \in V_c$;
\item  \label{property-3}
$V_c$ does not contain a circuit.
\end{enumerate}
Note that we will always have $V_c \cap V_d = \emptyset$, and each strong component of $H^+$ lies entirely in one
of the three sets $V_c, V_d, R$.
Moreover, at the end of the algorithm the set $R$ will be empty; this ensures that $<$ is a total ordering. Therefore, property (\ref{property-3}) will then imply the following transitivity on the chosen pairs:
\begin{itemize}
    \item if $(a,b) \in V_c$ and $(b,c) \in V_c$ then $(a,c) \in V_c$. 
\end{itemize}
 This fact, together with property (\ref{property-2}) ensures that the chosen pairs do define a min ordering, by setting $x < y$ for all chosen pairs $(x, y)$.

\begin{definition}[Reachability Notation]\label{reachibility-notation}
We write $(u, v) \leadsto (u', v')$ in $H^+$ if $(u', v')$ is reachable from $(u, v)$ in $H^+$, and, otherwise, $(u,v) \not\leadsto (u',v')$ in $H^+$. 
\end{definition}

\begin{definition}[Closure of $S$]\label{reachability-closure} 
Suppose $S$ is a set of pairs, i.e., a subset of $V(H^+)$.  The {\em closure} of $S$, denoted by $\widehat{S}$,
is the set of all pairs in $H^+$ that are reachable from $S$ in $H^+$. Note that $\widehat{S}$ contains $S$. We say $S$ is {\it closed under reachability} if $\widehat{S}=S$. 
\end{definition} 
Algorithm~\ref{alg-main} has two phases.

\noindent \fcolorbox{black}{white}{\sc Phase One:}  In the first phase we reduce the problem to a balanced sub-digraph $H^{\#}$
of $H^+$. We accomplish this by handling all the strong components of $H^+$ that are unbalanced. At each step we consider a strong unbalanced component $S$; if  ($\widehat{S} \cup V_c$) does not contain a circuit then we discard the dual of $S$ and add $\widehat{S}$ into $V_c$ and proceed to the next unbalanced component. Otherwise, we remove $S$ from further consideration and add $\widehat{S'}$ into $V_c$ where $S'$ is the dual of $S$. This is justified in Theorem~\ref{unbalanced-correctness}.

\noindent \fcolorbox{black}{white}{\sc Phase Two:} For the balanced components we need a different strategy other than the one used for unbalanced components. Roughly speaking, the main reason is the difference between the structural properties of balanced and unbalanced components, e.g., in the balanced components we no longer have directed walks with \emph{unbounded} positive (negative) net value.


 Now consider the induced sub-digraph $H^{\#}$ of $H^+$ consisting of all pairs in the balanced strong components of $H^+$. Thus, $H^{\#}$ is itself balanced. (Recall that in $H^+$ balance refers
to the equality of the number of positive and negative arcs in each directed closed walk; this is
true, since each such walk lies in a strong component of $H^{\#}$.)
 

 We partition the vertices of $H^{\#}$ into \emph{layers} as follows. Consider an auxiliary digraph $D$ with $V(D) = V (H^{\#} )$  and $(a, b)(c, d) \in A(D)$ if and only if
$(c, d)$ is reachable from $(a, b)$ by a directed path in $H^{\#}$ with negative net value. Since all directed cycles in $H^{\#}$
are balanced, $D$ is acyclic. Layer $0$ of $H^{\#}$, say $L_0$, consists of all vertices that have out-degree zero in $D$.
Having defined layers $L_1, L_2, \dots , L_j,$ layer $L_{j + 1}$ of $H^{\#}$ consists of all vertices of out-degree zero in the
digraph obtained from $D$ by removing all the vertices in layers $L_1, L_2, \dots,L_j$. Now the vertices of $H^{\#}$ are in layers so that any vertex in $L_j$ cannot be reached by a walk of positive net value from $L_j, L_{j + 1}, L_{j + 2}, \dots$. 
Let $R$ be the set of remaining balanced pairs. Notice that $R$ is a subset of $H^{\#}$. We handle the pairs in $L_0,L_1,\dots,$ consecutively, one at a time. 
To proceed with the current layer $k \ge 0$, we look for some vertex $p \in V(H)$ that satisfies the following conditions: 
\begin{enumerate}
    \item there exists $(p,q) \in R\cap L_k $ where $(p,q) \not\leadsto (q,p)$, \label{first-condition}
    \item there exists \textbf{no} $(q',p) \in V_c \cap L_k$. \label{second-condition}
\end{enumerate}
The existence of such $p$ is justified in Lemma~\ref{p-line17-main-body}. For each choice of $p$, as long as there exists some pair $(p,r) \in R \cap L_k$ so that $(p,r) \not\leadsto (r,p)$ (at least one such vertex $r$ exists, in particular $q$) we do the following. Add $(p,r)$ into $V_c$. Define \emph{transitive/reachable closure} of $V_c$ as follows.

\begin{definition}[Transitive/Reachable closure]\label{envelope}
Let transitive/reachability closure of $V_c$, $Tr(V_c)$, be the set of all pairs that is closed under reachability and transitivity. In other words, if $(x,y) \in Tr(V_c)$, and $(x,y) \leadsto (x',y')$ then $(x',y') \in Tr(V_c)$. Moreover, if $(x,y),(y,z) \in Tr(V_c)$ then $(x,z) \in Tr(V_c)$. 
\end{definition}
Notice that $V_c \subseteq Tr(V_c)$. Now, set $V_c = Tr(V_c)$. Of course, we do update $R$ by removing all the dual pairs of $V_c$ from $R$, and all the pairs of $V_c$ from $R$. Notice that during computation of $Tr(V_c)$ we may add $(q',p) \in L_k$ into $Tr(V_c)$ and $p$ no longer have condition~\ref{second-condition}. In the next subsection, we will prove that $Tr(V_c)$ does not contain a circuit (Lemma~\ref{p-line18-main-body}).

Once we are done with $p$, we look for another vertex $p_1$ on layer $k$ satisfying conditions \ref{first-condition}, \ref{second-condition} and repeat. Once we finish processing all the pairs in $L_k \cap R$, we go on to the next layer and consider the remaining pairs in $L_{k+1}$. 

\begin{algorithm}[h]
\caption{Algorithm to find a min ordering of input digraph $H$}
\label{alg-main}
\begin{algorithmic}[1]
\Function{MinOrdering}{$H$}
       
       \Comment{{\sc Phase 1:} Handling unbalanced strong components }

       \State Construct $H^+$ and compute its (strong) components
       \If{  $H^+$ contains a strong circuit} \Return False 
       \EndIf
      \State Set $V_c = V_d = \emptyset$ and let $R = V (H^+)$

  \While{ $R$ contains an unbalanced strong component } 
   \State Pick an unbalanced strong component $C$ in $R$
   \If { $\widehat{C} \cup V_c$ has no circuit } 
    
    \State Add $\widehat{C}$ into $V_c$, and add all the dual pairs of $\widehat{C}$ into $V_d$.

     \State Remove from $R$ all the pairs that are reachable from $C$ and their dual pairs.

   \Else{ ($\widehat{C'} \cup V_c$ has no circuit) } 
    \State Add $\widehat{C'}$ into $V_c$, and add all the dual pairs of $\widehat{C'}$ into $V_d$.

     \State Remove from $R$ all the pairs that are reachable from $C'$ and their dual pairs.

   \EndIf 
  \EndWhile 
 
  \Comment{{\sc Phase 2:} Handling the remaining balanced subgraph }

     \State Let $H^{\#}$ be the set of all balanced pairs, and let $R=V(H^{\#}) \setminus V_c$
     \State Compute the layers of $H^{\#}$; $L_0,L_1,\dots,$ and set $k=0$

     \While{ $R \neq \emptyset$}
       
       \While { $R \cap L_k \neq \emptyset$} \label{line16}

     \State Find $p \in V(H)$ such that 
     $\exists (p,q) \in R \cap L_k$ so that $(p,q) \not\leadsto (q,p)$ and no 
     
     \hspace{10mm} $(q',p) \in V_c \cap L_k$ \label{line17}
      
     \While { $\exists (p,r) \in L_k \cap R$ such that $(p,r) \not\leadsto (r,p)$ }  \label{line18}

       \State Add $(p,r)$ into $V_c$ and set $V_c=Tr(V_c)$ \label{cotinue-with-p0}

       \State Remove all the dual pairs of $V_c$ from $R$, and add them into $V_d$. \label{end-with-p0}
       
       \State Set $R= R \setminus V_c$.

     \EndWhile
     
     \EndWhile 
      \State Increase $k$ by one
     \EndWhile
    \Return $V_c$
\EndFunction
\end{algorithmic}
\end{algorithm}

\subsection{Justification of correctness}
Let us define some notation for the proof of correctness of {\sc Phase One}. A subset $T$ of $H^+$ is called closure-dual-free if for any $(x,y)\in T$, $(y,x)$ is not reachable from a pair in $T$, i.e., $(y,x) \not\in \widehat{T}$. Let $C : (a_0,a_1),(a_1,a_2),\dots,(a_n,a_0)$ be a circuit in $\widehat{T}$ where $T \subseteq V(H^+)$ is  closure-dual-free. 
Let $S_0,S_1,\dots,S_n$ (not necessarily distinct) be the strong components in $T$ where $(a_i,a_{i+1})$ is in $\widehat{S_i}$, $0 \le i \le n$. We say $C$ is \emph{minimal} if there is no other circuit $(a'_0,a'_1),(a'_1,a'_2),\dots,(a'_m,a'_0)$, $m<n$, where each $(a'_i,a'_{i+1})$ is in some $\widehat{S_j}$, $0 \le j \le n$. The proof of correctness depends on several technical results typified by the
following theorem. 

\begin{theorem}\label{CLAIM-main-body}
Let $T$ be a closure-dual-free set of unbalanced components and assume $\widehat{T}$ contains a minimal circuit $C$ with $n+1$ pairs. Then $n > 1$ and the following statements hold. 
\begin{enumerate}
\item There exists some minimal circuit with extremal pairs $(b_0,b_1),(b_1,b_2),\dots, (b_{n-1},b_n),$ $(b_n,b_0)$ in $T$ such that the $i$-th, $0 \le i \le n$, pair in $C$ is in the same strong component as $(b_i,b_{i+1})$, and reachable
 from $(b_i , b_{i+1})$ by a symmetric walk of non-negative net value, and constricted from below. \label{CLAIM-main-body-1}

\item  For each $i$, $0 \le i \le n$, there exists an infinite walk $P_i$ that starts from $b_i$ and has unbounded positive net length. Furthermore, for every $i, j$, $0 \le i < j \le n$, $P_i$ and $P_j$ avoid each other. \label{CLAIM-main-body-2}

\item In statement \ref{CLAIM-1}, for a given $0 \le i \le n$, we can choose $(b_i , b_{i+1})$ to be any given extremal pair from its corresponding strong component. \label{CLAIM-main-body-3}

\item There is no directed path in $H^+$ from $(b_i,b_{i+1})$ to any of $(b_j,b_{j+1})$ 
$ i \ne j$, and to any of $(b_{j+1},b_j)$. \label{CLAIM-main-body-4}

\item There is no directed path in $H^+$ from any of $(b_{i+1},b_{i})$, $0 \le  i \le n$ to $(b_i,b_{i+1})$. \label{CLAIM-main-body-5}  
\end{enumerate}
\end{theorem}

\begin{theorem}\label{unbalanced-correctness}
Suppose $C\not\subset (V_c \cup V_d)$ is an unbalanced strong component and $V_c$ does not contain a circuit. If $\widehat{C} \cup V_c$ contains a circuit, then $\widehat{C'} \cup V_c$ does not contain a circuit.
\end{theorem}  
\begin{proof} By assumption $C' \not\subset (V_c \cup V_d)$, as otherwise, by skew property this would imply that $C \subset (V_c \cup V_d)$. Suppose for contradiction that $\widehat{C} \cup V_c$ contains a circuit $(b_0, b_1 ), (b_1 , b_2), \dots,$ $ (b_n , b_0 )$
and $\widehat{C'} \cup V_c$ contains a circuit $(d_0 , d_1 ), (d_1 , d_2),\dots ,
(d_m , d_0)$. We may assume that both are minimal
circuits. Notice that Algorithm~\ref{alg-main} selects unbalanced components one at a time and adds their closure into $V_c$. Thus, if $\widehat{C} \cup V_c$ contains a circuit then that circuit would be at $\widehat{T}$ where $T$ is a set of unbalanced components. Similar statement is true for $\widehat{C'} \cup V_c$. Observe that since $V_c$ does not contain a circuit, at least one of the $(b_i , b_{i+1})$ pairs should be in $\widehat{C}$. The same holds for $\widehat{C'}$, and at least one of the $(d_j , d_{j+1})$ pairs is in $\widehat{C'}$. Hence, without loss of generality, we assume that $(b_n,b_0) \in \widehat{C}$, and $(d_m,d_0) \in \widehat{C'}$.

We first assume that both $m,n>1$. Thus, there is no $(p,q) \in C \cup V_c$ so that $(p,q) \leadsto (q,p)$, as otherwise, we have $(p,q),(q,p)\in \widehat{C}\cup V_c$ which contradicts the minimality assumption and the assumption that $n>1$. Similarly, there is no $(p',q') 
\in C' \cup V_c$ so that $(p',q') \leadsto (q',p') \in \widehat{C'} \cup V_c$. Therefore, $C \cup V_c$, and $C' \cup V_c$ are closure-dual-free. Thus, according to the statement (\ref{CLAIM-main-body-1}) of Theorem \ref{CLAIM-main-body}, we may also assume that all the pairs on these two circuits are extremal pairs in $H^ +$. Moreover, by statement (\ref{CLAIM-main-body-3}) of Theorem \ref{CLAIM-main-body}, we assume that $(b_n , b_0) \in C$ and $(d_m , d_0 ) \in C'$, i.e., $(d_0 , d_m) \in  C$, and that $(b_n, b_0 ) = (d_0 , d_m )$.

Moreover, according to statement (\ref{CLAIM-main-body-4}) of Theorem \ref{CLAIM-main-body}, we may assume that $(b_n, b_0 )$ is the only pair of the first circuit in $C$ and $(d_m, d_0 )$ is the only pair of the second circuit in $C'$. Now, consider the following circuit (where $(b_{n-1},b_n)=(b_{n-1},d_0)$, $(d_{m-1} , d_m )= (d_{m-1},b_0)$)
\begin{align*}
(b_0 , b_1), (b_1 , b_2 ), \dots , (b_{n-1} , d_0) ,(d_0,d_1), 
(d_1, d_2 ), \dots , (d_{m-1} , b_0 )
\end{align*}
all pairs of which are in $V_c$ . This contradicts the assumption that $V_c$ has no circuit. In what follows we consider separately the cases when $n$ or $m$ is $1$. 
\begin{observation}\label{Observation_n=1}
If $\widehat{C} \cup V_c$ contains a circuit $(b_0,b_1),(b_1,b_0)$ (i.e., $n=1$) then by definition we have $C \cup V_c \leadsto (b_0,b_1)$, and $C \cup V_c \leadsto (b_1,b_0)$. Now by skew property, we have $(b_1,b_0)\leadsto C' \cup V_d,(b_0,b_1) \leadsto C' \cup V_d$. Therefore, $C \leadsto C'$, and hence, there is also a circuit $(p,q),(q,p)$ where $(p,q) \in  C$, and $(p,q) \leadsto (q,p)$ (it is not possible that, $(p, q)$ or $(q,p)$ in $ V_c$ because $C \cup C' \not\subset V_c \cup V_d$). 
\end{observation}
If both circuits have $n = m = 1$ then by the above observation we have $C \leadsto C'$ and also $C' \leadsto C$, implying a strong circuit in $H^+$, a contradiction.  
Finally, if $n = 1$, but $m > 1$, then the first circuit is $(b_0, b_1), (b_1, b_0)$ and by Observation~\ref{Observation_n=1} and skew property we have $(b_0, b_1) \leadsto C'$ and $C' \leadsto (b_0,b_1)$. Now again since $m>1$, by statement (\ref{CLAIM-main-body-3}) of Theorem~\ref{CLAIM-main-body} we may assume that $C'$ contains $(b_1, b_0)$. This means $(b_0,b_1) \leadsto (b_1,b_0)$ which is in contradiction to statement (\ref{CLAIM-main-body-5}) of Theorem~\ref{CLAIM-main-body} (i.e., reverse of a pair on the circuit does not reach that pair).
\end{proof}
Theorem~\ref{unbalanced-correctness} concludes that {\sc Phase One} is correct, and the following two lemmas conclude that {\sc Phase Two} is correct. Lemma~\ref{p-line17-main-body} justifies Line~\ref{line17}, and Lemma~\ref{p-line18-main-body} justifies Line~\ref{line18}.


\begin{lemma}\label{p-line17-main-body}
Suppose $V_c$ does not contain a circuit, and furthermore, $R \cap L_{k}\neq \emptyset$. Then there exists a vertex $p\in V(H)$ (Line~\ref{line17}) such that: 
\begin{itemize}
    \item there exists $(p,q) \in R\cap L_k $ where $(p,q) \not\leadsto (q,p)$, 
    \item there exists \textbf{no} $(q',p) \in V_c \cap L_k$.
\end{itemize}
\end{lemma}

\begin{lemma}\label{p-line18-main-body}
By adding pair $(p,r)$ on line \ref{line18} of Algorithm~\ref{alg-main}, and computing $Tr(V_c)$, there will not be a circuit in $Tr(V_c)$. 
\end{lemma}

\begin{theorem}
Algorithm~\ref{alg-main} correctly decides if a digraph $H$ admits a min ordering or not. Furthermore, it correctly outputs a min ordering for $H$ if one exists.
\end{theorem}
\begin{proof}
The proof follows from Theorem~\ref{unbalanced-correctness}, Lemma~\ref{p-line17-main-body}, and Lemma~\ref{p-line18-main-body}. 
\end{proof}

\section{Summary of the Rest of the Paper}
The rest of the paper is organised as follows. In Sections~\ref{walks-structure} and \ref{circuit-structure-property} we will discuss the structural properties of walks and circuits in $H^+$. Apart from that the results in Sections~\ref{walks-structure} and \ref{circuit-structure-property} are interesting on their own, they are used to prove the correctness of Algorithm~\ref{alg-main}. The correctness of Algorithm~\ref{alg-main} is discussed in Section~\ref{tools} and its time complexity is discussed in Section~\ref{sec-time-complexity}. In Section~\ref{sec-k-min}, we consider the extension of a min ordering which is closely related to when $H$ is homomorphic to a directed cycle of length $k>1$. In Section~\ref{connection-other-polymorphism}, using the structural properties of a minimal circuit we show that how the class of digraphs with a CSL polymorphism coincides with other classes of digraphs admitting other important polymorphiosms. Finally, in Section~\ref{np-complete}, NP-completeness for general relational structures is discussed and a full complexity classification for {\sc Problem~\ref{problem1}} is given.

\section{Structure of Walks in $H^+$}\label{walks-structure}
\subsection{Implication of four congruent walks}
The following lemma is well known. (For a proof, see \cite{don,zhu}
or Lemma 2.36 in \cite{homobook}).

\begin{lemma}\label{typer}
Let $P_1$ and $P_2$ be two constricted walks of net length $r$.
There exists a constricted path $P$ of net length $r$ that admits
a homomorphism $f_1$ to $P_1$ and a homomorphism $f_2$ to
$P_2$, such that each $f_i, i=1, 2$ takes the starting vertex of $P$ 
to the starting vertex of $P_i$ and the ending vertex of $P$ to the
ending vertex of $P_i$.
\end{lemma}

We call $P$ a {\em common pre-image} of $P_1$ and $P_2$. 
Note that $f_1(P)$ is a walk on the vertices of $P_1$ and $f_2(P)$ is a walk on the vertices of $P_2$, and the walks $f_1(P)$ and $f_2(P)$ are congruent. 
We use the term {\em pre-image} of a walk $P'$ to be a
path $P$ that admits a homomorphism to $P'$ taking the first vertex of $P$
to the first vertex of $P'$ and the last vertex of $P$ to the last vertex of $P'$. 
We use the term {\em embedded pre-image} of a walk $P'$ to be a walk on $V(P')$ starting with the starting 
vertex of $P'$ and ending with the ending vertex of $P'$.

We note for future reference that if two congruent walks $P, Q$ avoid each other, 
then the same is true for any congruent embedded pre-images $P',Q'$. 
(However, note that if $P$ avoids $Q$, it is not necessarily true that $P'$ avoids $Q'$ because of the back steps involved in the pre-images.)

We now begin to provide the structural information that justifies the algorithms.  We first focus on
walks in $H^+$ corresponding to $(p,q) \leadsto (a,b)$, and $(r,s) \leadsto (b,d)$. 
In other words, we have in $H$ four
walks $A, B, C, D$ that start in four distinct vertices, $p, q, r, s$ respectively, such that the end vertices
of walks $B$ and $C$ coincide, and such that $A$ avoids $B$ and $C$ avoids $D$. Note that $B$ does not avoid
$C$, and $C$ does not avoid $B$. (At the last step, there is a faithful arc.) The statements confirm that
if $(p, q)  \not\leadsto (a, d)$, $(p, q)  \not\leadsto  (d, b)$, $(r, s)  \not\leadsto  (a, d)$, and $(r, s)  \not\leadsto  (b, a)$, 
then all other pairs of walks avoid each other, i.e., $A$ avoids $B, C, D$; $B$ avoids $A, D$; $C$ avoids $A, D$; and $D$ avoids $A, B, C$. 
In the first lemma, we assume all walks $A, B, C, D$ are congruent, while in the second lemma only $A,B$ are congruent, and $C, D$ are congruent; 
but on the other hand, all four walks are constricted and
have the same height, so we can replace them by their congruent embedded pre-images by Lemma
\ref{typer} (more exactly, by their congruent embedded pre-images, cf. the discussion after Lemma \ref{typer}).

\begin{lemma}\label{strong-core}
Let $A, B, C, D$ be four congruent walks in $H$, from $p, q, r, s$ to $a, b, b, d$  respectively,
such that $A$ avoids $B$ and $C$ avoids $D$. Suppose in $H^+$
\begin{enumerate}

\item $(p,q) \not\leadsto (a,d)$ and $(p,q) \not\leadsto (d,b)$, 
\item $(r,s) \not\leadsto (a,d)$ and  $(r,s) \not\leadsto (b,a)$.
\end{enumerate}
Then all pairs from $A, B, C, D$ avoid each other, except the pair $B, C$.
\end{lemma}

\begin{proof} Let $A$ be the walk $p=a_1,a_2,\dots,a_n=a$, $B$ the walk $q=b_1,b_2,\dots,b_n=b$, $C$ the walk 
$r=c_1,c_2,\dots,c_n=b$, and $D$ the walk $s=d_1,d_2,\dots,d_n=d$. Let $S_i$ denote the statement that all pairs from 
\begin{align*}
    A[a_{i+1},a], B[b_{i+1},b], C[c_{i+1},b], D[d_{i+1},d]
\end{align*}
avoid each other, except possibly $B[b_{i+1},b], C[c_{i+1},b]$. The Lemma claims that $S_0$ holds, while
$S_{n-1}$ holds vacuously. Therefore, let $i, 0 \le i \le n-1$ be the first index such that $S_i$ holds.

Note that $a_id_{i+1}$ is not a faithful arc. Otherwise, $(p,q) \leadsto (d,b)$ in $H^+$
by combining two walks in $H$, namely the walk $A[p,a_i]+D[d_{i+1},d]$, and the walk 
$B$. This implies that $b_id_{i+1}$ is also not a faithful arc, since otherwise 
$(p,q) \leadsto (a,d)$ by combining the walks $A$ and $B[q,b_i]+D[d_{i+1},d]$. (This uses the fact that
$a_id_{i+1}$ is not a faithful arc.) By a similar line of reasoning, we conclude that 
\begin{itemize}
\item
$c_ia_{i+1}$ is not a faithful arc (as otherwise $(r,s) \leadsto (a,d)$), and then $d_ia_{i+1}$ is not 
a faithful arc (otherwise $(r,s) \leadsto (b,a)$).

\end{itemize}

Now $d_ib_{i+1}$ is not a faithful arc. Otherwise $(p,q) \leadsto (a,d)$ in $H^+$ by 
combining two walks in $H$, namely the walk $A[p,a_i]+A^{-1}[a_i,a_{i+1}]+A[a_i,a]$ and  the walk 
$B[q,b_{i+1}]+b_{i+1}d_i+D[d_i,d]$. (This uses the fact that none of the $d_ia_{i+1}, a_id_{i+1}$ is a faithful arc. 
Similarly $a_ic_{i+1}$ is not a 
faithful arc (as otherwise $(r,s) \leadsto (a,d)$). 

Now $b_ia_{i+1}$ is not a faithful arc. Otherwise $(r,s) \leadsto (a,d)$ in $H^+$ by combining two walks in $H$, namely $C+B^{-1}[b,b_i]+A[a_{i+1},a]$ and 
the walk $D+D^{-1}[d,d_i]+D[d_i,d]$. (This uses the fact that none of the $d_ib_{i+1}$, $d_ia_{i+1}$ is a faithful arc.)
Similar argument implies that $d_ic_{i+1}$ is not a faithful arc.

Together with the fact that $a_ib_{i+1}$ and $c_id_{i+1}$ are not faithful arcs (corresponding to the assumption 
that $A$ avoids $B$ and $C$ avoids $D$), we obtain a contradiction with the minimality of $i$; therefore $i=0$,
and the lemma is proved.
\end{proof}
\subsection{Implication of four constricted walks}
A similar result applies to walks that are not all congruent, as long as they are constricted and have the same 
net length (Of course the pairs of walks that one avoids another one must be congruent by definition.)

\begin{lemma}\label{strong-core1}
Let  $A, B, C, D$ be four constricted walks of the same net length, from $p, q, r, s$ to $a, b, b, d$ respectively,
such that $A, B$ are congruent and $A$ avoids $B$, and $C, D$ are congruent and $C$ avoids $D$. Suppose in $H^+$
\begin{enumerate}
    \item $(p,q) \not\leadsto (a,d)$ and $(p,q) \not\leadsto (d,b)$,
    \item $(r,s) \not\leadsto (a,d)$ and $(r,s) \not\leadsto (b,a)$.
\end{enumerate}
Then there exist congruent walks $A', B', C', D'$ that are embedded pre-images of $A, B, C, D$ respectively, such that all 
pairs from $A', B', C', D'$ avoid each other, except the pair $B', C'$ and hence $A,B$ avoid each other and $C,D$ avoid each other.
\end{lemma}
\begin{proof} 
    Let $A$ be the walk $p=a_1,a_2,\dots,a_n=a$, $B$ the walk $q=b_1,b_2,\dots,b_n=b$, $C$ the walk
    $r=c_1,c_2,\dots,c_m=b$, and $D$ the walk $s=d_1,d_2,\dots,d_m=d$. We prove the lemma by induction
    on the sum of the lengths $m+n$. If $m+n=0$, i.e., $m=n=0$, this holds trivially.

    Suppose first that $A, B, C, D$ are strongly constricted from above (no prefix of $A$ has net length zero). This means that the first two arcs in
    each walk are forward arcs, and the walks $A-p, B-q, C-r, D-s$ are also constricted walks of the same 
    net length, with the first two congruent and the last two congruent. Moreover, in $H^+$, neither $(a,d)$ 
    nor $(d,b)$ is reachable from $(a_2,b_2)$, otherwise they would also be reachable from $(p,q)$ because 
    $A$ is assumed to avoid $B$. Similarly, neither $(a,d)$ nor $(d,b)$ is reachable from $(c_2,d_2)$. By the 
    induction hypothesis, $A-p, B-q, C-r, D-s$ have congruent embedded pre-images $A'', B'', C'', D''$ in which all 
    pairs except $B'', C''$ avoid each other. Noting that $A''$ starts in $a_2$, we let $A'$ consist of $p$
    concatenated with $A''$ (i.e., $A' = a_1a_2 + A''$), let $B'$ be $q$ concatenated with $B''$, and 
    similarly for $C'$ and $D'$. Since $A', B', C', D'$ are all congruent, we can apply Lemma \ref{strong-core},
    and conclude that all pairs avoid each other, except the pair $B', C'$.

    In the rest of the proof we will show that $B$ also avoids $A$, and that $D$ also avoids $C$; in other words, 
    that $A, B$ avoid each other and that $C, D$ avoid each other. By repeated application of Lemma \ref{typer} 
    we conclude that there exist congruent walks $A', B', C', D'$ from $p, q, r, s$ to $a, b, b, d$ that are embedded pre-images 
    of $A, B, C, D$ respectively. Since $A$ and $B$ are congruent and avoid each other, the walks $A', B'$ follow
    the same sequence of back and forth steps inside $A, B$, and also avoid each other. (Note that if $A', B'$ take
    a backward step along $A, B$ we can only conclude $A'$ avoids $B'$ if we know that also $B$ avoids $A$.) 
    Similarly, $C', D'$ also avoid each other. Therefore we can now apply Lemma \ref{strong-core} to $A', B', C', D'$ 
    and conclude that all pairs from $A', B', C', D'$ avoid each other, except the pair $B', C'$.

    Since we have already considered the case when all four walks $A, B, C, D$ are strongly constricted from 
    above, we may assume, up to symmetry, that $A, B$ are not strongly constricted from above, i.e., that there 
    exists a subscript $j>1$ such that $A[p,a_j]$ and $B[q,b_j]$ have net length zero. We take the subscript $j$ 
    is as large as possible, therefore $A[a_j,a], B[b_j,b]$ are strongly constricted from above and have the same 
    net length as $C, D$. We now apply the induction hypothesis to $A[a_j,a], B[b_j,b], C, D$ and conclude that 
    $A[a_j,a], B[b_j,b], C, D$ have congruent pre-images that pairwise avoid each other (except for the pre-images 
    of $B[b_j,b], C$). This implies that $A[a_j,a], B[b_j,b]$ also avoid each other, and $C, D$ also avoid each other. 
    If $C, D$ were also not strongly constricted from above, we could draw the similar conclusion that $A, B$ avoid 
    each other, as claimed. However, in general $C, D$ may happen to be strongly constricted from above, and 
    we proceed more carefully as follows: recall that our goal is to prove that $A$ and $B$ avoid each other. 
    Noting that by the definition of $j$ the arcs $a_{j-1}a_j$ and $b_{j-1}b_j$ are backward arcs; moreover, the
    first arcs $c_1c_2, d_1d_2$ of $C, D$ are forward arcs. Since $A$ avoids $B$ and $C$ and $D$ avoid each other,
    we can apply Lemma \ref{strong-core} to $A[a_{j-1},a], B[b_{j-1},b], c_2c_1 + C, d_2d_1 + D$ and conclude
    that $A[a_{j-1},a], B[b_{j-1},b]$ have embedded pre-images that avoid each other, and hence that $A[a_{j-1},a], B[b_{j-1},b]$ 
    also avoid each other. The idea of the proof is to continue this way backwards on $A, B$ until proving that they
    avoid each other in their entirety. Thus, let $i \le j$ be the minimum subscript such that :
    
    \begin{itemize}
        \item $A[a_i,a], B[b_i,b]$ avoid each other,
        \item there exists an $\ell$ 
    such that $A[a_i,a_j]$ has an embedded pre-image with a walk $W_C$ in $C^{-1}$ that starts in some vertex $c_{\ell}$ and ends 
    in $c_1$.
    \end{itemize}

\begin{figure}
\begin{center}
\includegraphics[height=6cm,width=14cm]{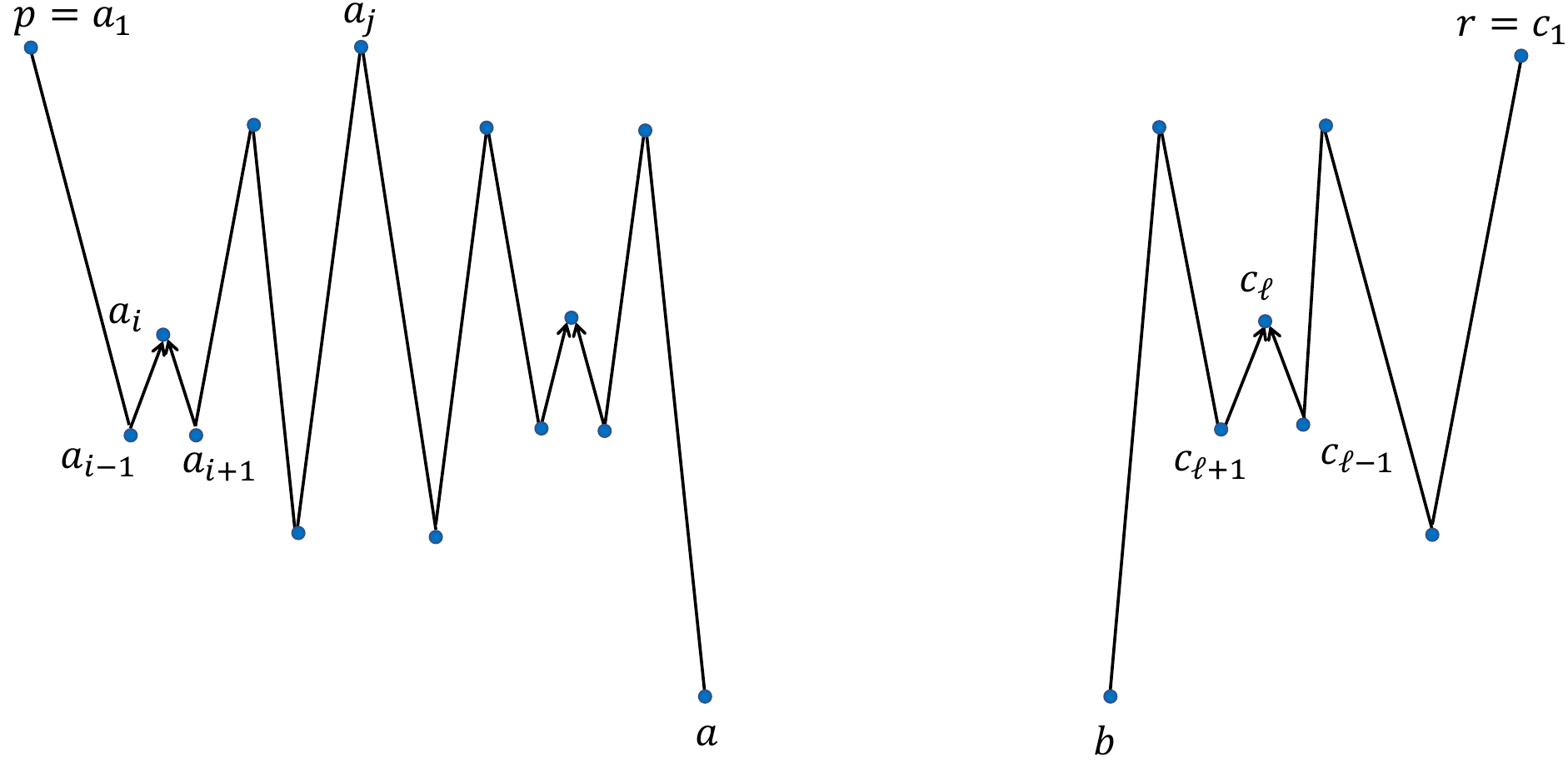}
\caption{The notation for Lemma \ref{strong-core1}; each straight segment represents a constricted walk. The figure shows the embedded pre-image of $A[a_i,a],C^{-1}[c_{\ell},c_1]+C$ (here $A[a_i,a]$ is a portion of $A$ from $a_i$ to $a$, here $C^{-1}$ is the reverse of $C$). Further assume that $a_ia_{i+1}$ is backward, i.e., $c_{\ell-1}c_{\ell} \in C$ is forward arc. Note that in general $A[a_1,a_{i-1}]$ may consists of several constricted segments. } \label{lemma34-fig1}
\end{center}
\end{figure}

Then $A[a_i,a]$ has an embedded pre-image with $W_C + C$ (see Figure \ref{lemma34-fig1}). Of course, this is also a pre-image of $B[b_i,b]$ 
and $W_D + D$ where $W_D$ is congruent to $W_C$ and starts in $d_{\ell}$ and ends in $d_1$.  

We have just shown (previous paragraph) that 
$i \leq j-1$ and for $i=j-1$ we can take $\ell=2$. We claim that $i=1$, which means in particular that $A, B$ avoid each other.
We proceed by contradiction.
    
Let $X, Y, Z, U$ be congruent walks that are embedded pre-images of $A[a_i,a], B[b_i,b], W_C + C, W_D + D$ respectively, 
and denote by $x_t, y_t, z_t, u_t$ the $t$-th vertices of these walks respectively.

Suppose first that $a_{i-1}a_i$ is a forward arc. We would like to show that $A[a_{i-1},a]$, $B[b_{i-1},b]$ avoid each other
and have embedded pre-image with $W'_C + C$ that starts in some $c_{\ell'}$. If $c_{\ell}c_{\ell + 1} \in C$ is also a backward 
arc, we can set $\ell' = \ell + 1$ and argue as above, adding $a_{i-1}a_i$ to $X$ and $c_{\ell+1}c_{\ell}$ (forward arc in $C^{-1}$) to $Z$. Moreover, if $a_ia_{i+1}$ is a backward arc ($c_{\ell-1}c_{\ell}$ is a forward arc in $C$, and backward arc in $Z$, since $X,Z$ are congruent)
then we can set $\ell'=\ell-1$ and argue as above, adding $a_{i-1}a_i$ to $X$ and $c_{\ell-1}c_{\ell}$ to $Z$. Otherwise (i.e., $a_ia_{i+1}$ is forward),
we claim there is another vertex $\ell''$ and another walk $W''_C$ in $C$ that starts in $c_{\ell''}$ and ends in $c_1$ 
and has an embedded pre-image with $A[a_i,a_j]$, and such that $c_{\ell" - 1}c_{\ell"}$ is a forward arc (see Figure \ref{lemma34-fig2}). Note that we have $x_1=a_i$; we let $t$ be the last subscript such that $X[x_1,x_t]$ is constricted from 
below and has net length zero ($t >1$ because $a_ia_{i+1}$ is a forward arc).

\begin{figure}
\begin{center}
\includegraphics[height=6cm, width=14cm]{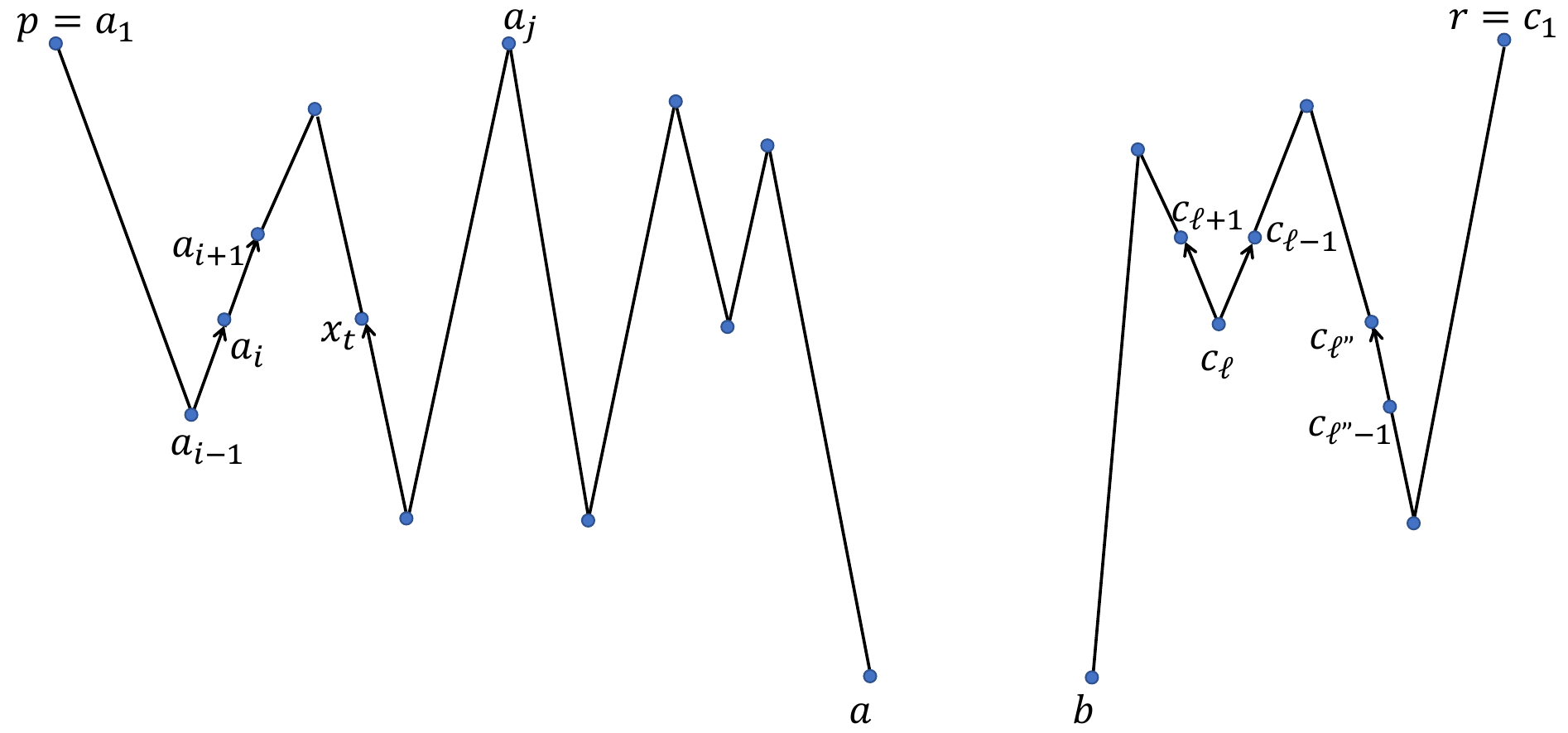}
\caption{The notation for Lemma \ref{strong-core1}; each straight segment represents a constricted walk. The figure shows the embedded  pre-image of $A[a_i,a],C^{-1}[c_{\ell},c_1]+C$. Further assume that $a_ia_{i+1}$ is forward, i.e., $c_{\ell-1}c_{\ell} \in C^{-1}$ is a forward arc.} \label{lemma34-fig2}
\end{center}
\end{figure}

Then it is easy to see that $x_tx_{t+1}$ is a backward arc. Indeed, the net length of 
$A[a_i,a_j]$ is strictly positive, as $a_{i-1}a_i$ is a forward arc, $A$ is constricted, and $A[a_1,a_j]$ has net length 
zero. Since $X$ and $Z$ are congruent, $z_tz_{t+1}$ is also a backward arc, and we can set $c_{\ell''}=z_t$. It remains 
to construct a walk $W''_C$ in $C^{-1}$, from $c_{\ell''}$ to $c_1$ that has an embedded pre-image with $A[a_i,a_j]$. Note that 
the walk $X[x_1,x_t]$ from $x_1=a_i$ to $x_t$ is congruent with the walk $Z[z_1,z_t]$ from $z_1=c_{\ell}$ to $z_t=c_{\ell''}$.
Both are constricted from below, and have the same maximum net length of a sub-walk; say $X[x_1,x_s]$ and $Z[z_1,z_s]$
are of maximum net length. Then applying Lemma \ref{typer} twice (once to $X[x_1,x_s], Z^{-1}[z_t,z_s]$ and once to 
$X^{-1}[x_s,x_1], Z[z_t,z_s]$, we obtain congruent walks $X^*, Z^*$ from $a_i$ to $a_i$ and from $c_{\ell''}$ to $c_{\ell}$ 
respectively. Then the concatenation $W''_C = Z^* + Z$ is a walk in $C^{-1}$ from $c_{\ell''}$ to $c_1$ that has a common 
pre-image $X^* + X$ with $A[a_i,a_j]$, as required. Notice that at this point we add $a_{i-1}a_i$ to $X^*+X+A[a_j,a]$ and add $c_{\ell''-1}c_{\ell''}$ into $W''_C+C$ and proceed as before. 

When $a_{i-1}a_i$ is a backward arc, the proof is similar. If $c_{\ell}c_{\ell + 1}$ is also a forward arc, we proceed as usual.
Otherwise, we let $t$ be the last subscript such that $X[x_1,x_t]$ is constricted from above and of net length zero. This again
means that $x_tx_{t+1}$ and hence also $z_tz_{t+1}$ is a forward arc, and we set $c_{\ell''}=z_t$. Then choosing $x_s$ so 
that the net length of $X[x_1,x_s]$ is minimized, and applying Lemma \ref{typer} twice -- to $X[x_1,x_s], Z^{-1}[z_t,z_s]$ and 
to $X^{-1}[x_s,x_1], Z[z_t,z_s]$, we obtain congruent walks $X^*, Z^*$ from $a_i$ to $a_i$ and from $c_{\ell''}$ to $c_{\ell}$ 
respectively, which yield the walk $Z^* + Z$ in $C^{-1}$ from $c_{\ell''}$ to $c_1$ that has an embedded pre-image $X^* + X$ 
with $A[a_i,a_j]$, as required.
\end{proof}

\section{Structure of Circuits in $H^+$ }\label{circuit-structure-property}

\subsection{M-Lemma}
We now single out a particular situation in which a circuit occurs in one strong component of the pair
digraph $H^+$.

\begin{theorem}\label{circuit-cycle}
Suppose $C$ is a closed walk in $H$ of net length greater than one, and $x, y$ are two
extremal vertices of $C$ such that the net length of $C[x, y]$ is positive. Let $P_x$ be the infinite walk
starting at $x$, obtained by continuously following the cycle $C$ in the positive direction and let $P_y$ 
obtained the same way starting at $y$. Suppose $H$ contains two congruent walks $X, Y$ such that 
$X$ avoids $Y$ and such that $X$ is an embedded pre-image of $P_x$ and $Y$ is an embedded pre-image of $P_y$. 
Then $H^+$ contains a strong circuit.
\end{theorem}
\begin{proof}
Let $f : X \rightarrow  P_x$ be a homomorphism taking the first vertex of $X$ to the first vertex of $X$, i.e., $x$, and similarly for $g : Y \rightarrow  P_y$. It is easy to see that $X$ contains vertices $x_0,x_1,...$ and $Y$ contains vertices $y_1, y_2, . . .$ such that (for all $i$)

\begin{itemize}
    \item $x_0$ is the first vertex of $X$, and $y_1$ is the first vertex of $Y$;
    \item $x_i$ is the vertex on $X$ corresponding to $y_{i+1}$ on $Y$
    \item $f(x_i) = g(y_i)$, and the vertex $v_i = f(x_i) = g(y_i)$ is extremal on $C$
    \item each segment $X[x_i, x_{i+1}]$ and $Y[y_i, y_{i+1}]$ has the same net length as $C[x, y]$.
\end{itemize}

Since $C$ has only finitely many extremal vertices, we must eventually have $f(x_i) = f(x_{i+j})$ for some positive $i$ and $j$. It is now clear that $(v_i,v_{i+1}),(v_{i+1},v_{i+2}),\dots,(v_{i+j},v_i)$ is a circuit in $H^+$, since $X[x_i, x_{i+1}] = C[v_i, v_{i+1}]$ avoids $Y[y_{i+1}, y_{i+2}] = C[v_{i+1}, v_{i+2}]$ (subscripts reduced modulo $j$).
\end{proof}
One particular situation is helpful to know.
\begin{corollary}\label{cyc}
If $H$ contains an induced cycle of net length greater than one, then a strong component of $H^+$ 
contains a circuit.
\end{corollary}

\vspace{2mm}
\begin{proof} Suppose $C$ is an induced cycle of net length $k > 1$. Recall that the cycle $C$ has $k$ extremal
vertices $v_0, v_1, \dots, v_{k-1}$ with each $C[v_i,v_{i+1}]$ (subscript addition modulo $k$) constricted
from below and of net length one.

We shall show that  $(v_0,v_1), (v_1,v_2), \dots,$ $(v_{k-1},v_0)$ belong to the same strong component 
of $H^+$. Indeed, for any $i=1, \dots, k-1$, we shall exhibit a directed path in $H^+$ from $(v_{i-1},v_i)$ 
to $(v_i,v_{i+1})$. These directed paths in $H^+$ will be constructed out of pairs of walks on the cycle $C$.

Assume first that the height of $C[v_{i-1},v_i]$ is at most the height of $C[v_i,v_{i+1}]$ (see Figure \ref{cycle-circuit} (left)).

\begin{figure}
\begin{center}
\includegraphics[height=6cm, width=13cm]{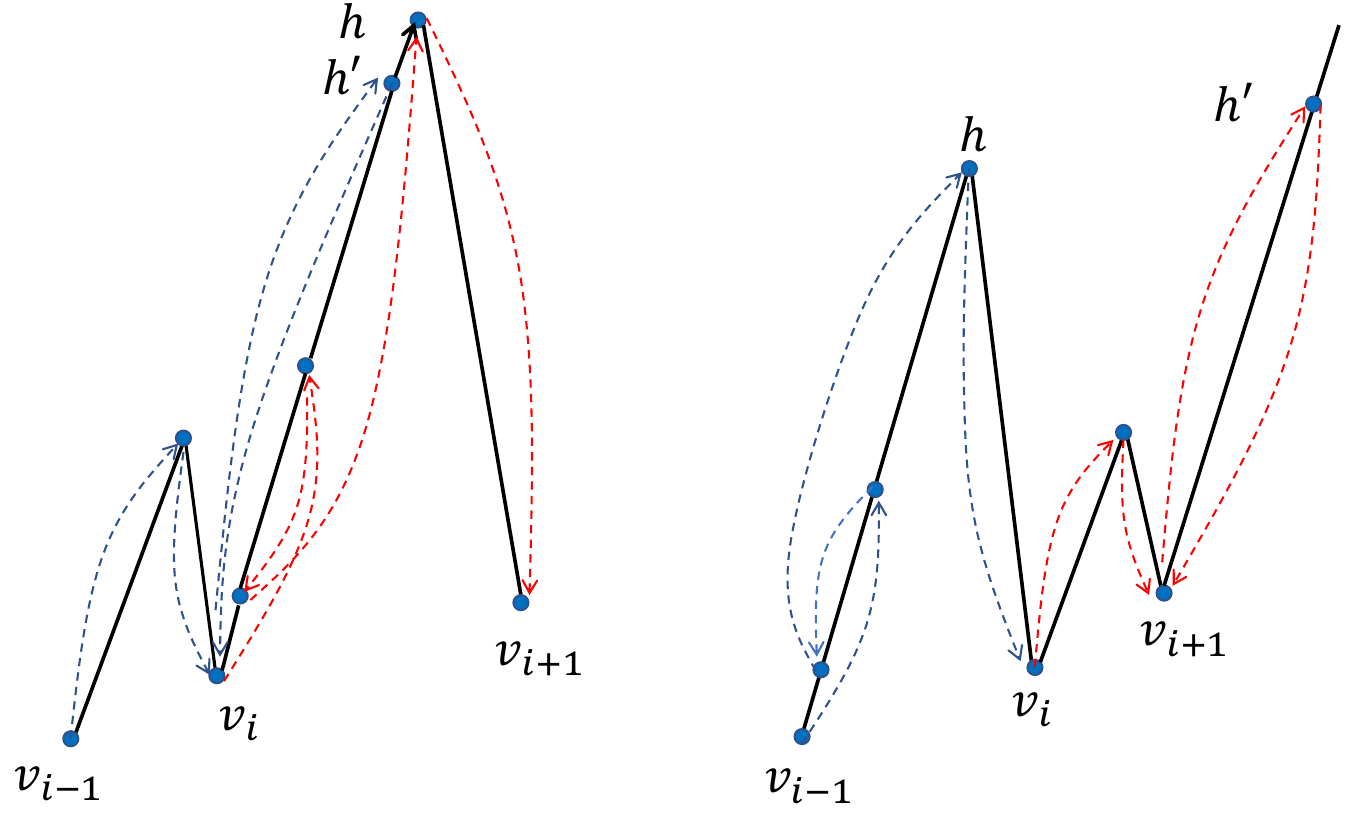}
\caption{Each (colored) black segment is an oriented path. For example, on the left figure, there is a constricted oriented path from $v_{i-1}$ to $h'$, and then there is an oriented path from $h$ to $v_{i+1}$ which is constricted. The dashed lines show the sequence of the vertices on the path to make congruent embedded pre-images.} \label{cycle-circuit}
\end{center}
\end{figure}

To be able
to use Lemma \ref{typer}, we consider the last vertex $h$ of $C[v_{i},v_{i+1}]$ maximizing the net length
of $C[v_{i},h]$, and the first vertex $h'$ of $C[v_{i},v_{i+1}]$ such that $C[h',h]$ has net length one. 

Now  $C[v_{i-1},h']$ and $C[v_i,h]$ are constricted and have the same net length. Thus by Lemma 
\ref{typer} they have a common pre-image $A$. Also $C^{-1}[h',v_i]$  and  $C[h,v_{i}]$  are 
constricted and have the same net length; thus they also have a common pre-image $B$. Let $X$
be the walk in $C$ from $v_{i-1}$ to $v_i$ corresponding to $A+B$, and let $Y$ be the walk in $C$ 
from $v_i$ to $v_{i+1}$ corresponding to $A+B$. We claim that $X$ avoids $Y$. Consider the $j$-th 
vertex $u$ of $X$, and the $(j+1)$-st vertex $v$ of $Y$. Note that the net lengths of $C[v_0,u]$ and
$C[v_0,v]$ differ by two; since $C$ is an induced cycle of net length greater than one, there can 
be no faithful arc between $u$ and $v$. This implies that there is a directed path in $H^+$ from 
$(v_{i-1},v_i)$ to $(v_i,v_{i+1})$.
 
If the height of $C[v_{i-1},v_i]$ is greater than the height of $C[v_i,v_{i+1}]$, we argue analogously.
We denote by $P$ the infinite walk obtained by continuously following $C$ in the positive direction.
Let $h$ be the last vertex of $C[v_{i-1},v_i]$ maximizing the net length of $C[v_{i-1},h]$, and let 
$h'$ be the first vertex of $P$ after $v_{i+1}$ such that $P[h,h']$ has net length one. Now Lemma
\ref{typer} can be applied to the walks $C[v_{i-1},h]$ and $C[v_{i+1},h']$, and to the walks $C[h,v_i]$
and $P^{-1}[h',v_{i+1}]$, yielding a common pre-image $A$ for the former pair and a common
pre-image $B$ for the latter pair. The walk $X$ in $C$ from $v_{i-1}$ to $v_i$ corresponding to $A+B$
again avoids the walk $Y$ in $P$ from $v_i$ to $v_{i+1}$ corresponding to $A+B$. 
\end{proof}

Here is a useful consequence of Theorem 8.2. We illustrate the notation in Figure 22.

\begin{lemma}[M-Lemma]\label{M}
Suppose that $n \geq t > 1, \ell > 0$ are integers, $(a_0,a_1), (a_1,a_2), \dots$ $, (a_n,a_0)$ is a circuit in $H^+$,
and, for each $i = 0, 1, \dots, t$, $p_i, q_i, g_i, h_i$ are vertices of $H$, and 
$A_i, A'_i, B_i, B'_i$ are walks in $H$, such that the following statements hold (subscript addition modulo $n+1$):
\begin{enumerate}
\item
$A_i$ is a constricted walk from $p_i$ to $h_i$ of net length $\ell$
\item
$B_i$ is a constricted walk from $q_i$ to $g_i$ of net length $\ell$
\item
$A'_i$ is a constricted walk from $h_i$ to $a_i$ of net length $-\ell$
\item
$B'_i$ is a constricted walk from $g_i$ to $a_i$ of net length $-\ell$
\item
$A_i + A'_i$ is congruent to and avoids $B_{i+1} + B'_{i+1}$
\item
if $k \neq j+1$, then $(p_i,q_{i+1}) \not\leadsto  (a_j,a_k)$  
\end{enumerate}

\begin{figure}
\begin{center}
\includegraphics[height=6cm,width=13cm]{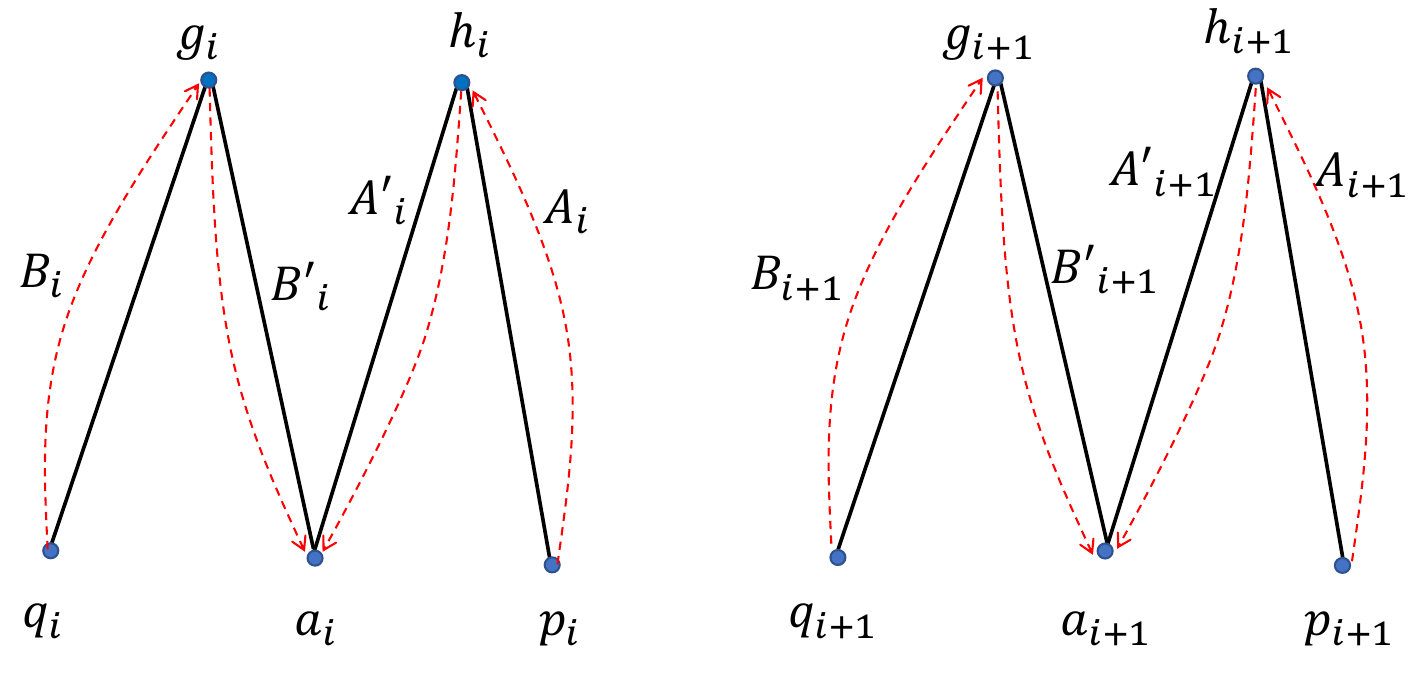}
\caption{The notation for Theorem \ref{M}; each straight segment represents a constricted walk. The dotted line shows the direction of the walks. Here, $A_i$ 
(form $p_i$ to $h_i$) avoids $B_{i+1}$(from $q_{i+1}$ to $g_{i+1})$, and $A'_i$ (from $h_i$ to $a_i$ ) avoids $A_{i+1}$ (from $g_{i+1}$ to $q_{i+1}$) } \label{M1}
\end{center}
\end{figure}

Let $C$ be any one of the walks $A'_i$ or $B'_i$ or $A^{-1}_i$ or of $B^{-1}_i$,
and let $D$ be any one of the walks $A'_j$ or $B'_j$ or $A^{-1}_j$ or of $B^{-1}_j$,
with $i \neq j$. 

Then $C, D$ have embedded pre-images that avoid each other. 
\end{lemma}


Another way to state the conclusion of the lemma is the following.

{\em There are embedded pre-images of all $A'_i, B'_i, A^{-1}_i, B^{-1}_i, i=0, 1, \dots, t$,
such that any two pre-images of walks with different subscripts avoid each other.}
\vspace{2mm}

The lemma will often be used for walks where $A'_i=A^{-1}_i$ and/or $B'_i=B^{-1}_i$ (or even 
$A'_i=A^{-1}_i=B'_i=B^{-1}_i$).
\begin{proof}
We prove the lemma with $t=n$, and it is easy to check that the proof allows any smaller $t, t \geq 2$.

We first prove that any $B'_i, B'_j$ with $i \neq j$ have embedded pre-images that avoid each other. We 
proceed by induction on $|j-i|$. If $|j-i|=1$, say $j=i+1$, we may apply Lemma \ref{strong-core1} to 
the walks $A'_{i-1}, B'_i, A'_i, B'_{i+1}$. Indeed, $A'_{i-1}$ avoids $B'_i$ and $A'_i$ avoids $B'_{i+1}$
by 5, and since the same condition also implies that   $(p_i,q_{i+1}) \leadsto (h_i,g_{i+1})$
in $H^+$, condition 6 implies that  $(h_i,g_{i+1}) \not\leadsto (a_{i-1},a_{i+1})$ and $(h_i,g_{i+1}) \not\leadsto (a_i,a_{i-1})$,
and similarly for $(h_{i-1},g_i)$. (Here we used the fact that $n>1$.) 

For the induction step, we again assume that $i < j$ and consider $B'_i, B'_j, B'_{j+1}$. By the
induction hypothesis, there are embedded pre-images of $B'_i, B'_j$ that avoid each other and also
embedded pre-images of $B'_j, B'_{j+1}$ that avoid each other. As noted earlier, we may assume that
all these pre-images are congruent to each other. Now assume there is a faithful arc from $B'_i$ to
$B'_{j+1}$, or from $B'_{j+1}$ to $B'_i$. It is easy to trace walks from $(a_i,a_j)$ on reverses of 
$B'_i, B'_j$ (that are known to avoid each other) up to the faithful arc, use the faithful arc, 
and then follow the walks $B'_{j+1}, B'_j$, also known to avoid each other to $(a_{j+1},a_j)$. This would imply that
$(a_j,a_{j+1}) \leadsto (a_j,a_i)$ which contradicts condition 7, and completes the induction proof.

A symmetric argument yields that any $A'_i, A'_j$ with $i \neq j$ have embedded pre-images that avoid 
each other.

It now follows that any $A'_i, B'_j, i \neq j,$ have embedded pre-images that avoid each other.
Indeed, $A'_i, B'_{i+1}$ are congruent by assumption, so it suffices to take the embedded pre-images
of $B'_{i+1}, B'_j$ that avoid each other, we have just constructed, and also use the same pre-image 
for $A'_i$. Then if there was a faithful arc between $A'_i$ and $B'_j$ (in either direction), we could 
use it to reach $(a_j,a_i)$ from $(p_i,q_{i+1})$, using the walks $A_i, B_{i+1}, A'_i,$ a portion of 
$B'_{i+1}$, the faithful arc, and a portion of $B'_j$. This contradicts condition 6 of the lemma. 
(We note that since $n>2$, we can always choose $i, j$ so that $j+1 \neq i$.)

Next we argue that for each $j \neq i$ there are embedded pre-images to $A_i, B_{i+1}, A_j,$ $B_{j+1}$, 
such that each pair except for the pre-images of $B_{i+1}$ and $A_j$ avoid each other. This will in 
particular imply that the pre-images of $A_i$ and $A_j$ avoid each other, and the pre-images of $A_i$ 
and $B_{j+1}$ avoid each other. It will also imply that $A_i, B_{i+1}$ avoid each other, and thus
$A_i, B_j$ avoid each other for all $j \neq i$.  This will imply the corresponding statements also
about their reverses. For any $i \neq j$, consider a new digraph $H^o$ obtained from $H$ by the 
addition of three new vertices $u, v, w$ and four new arcs $h_iu, g_{i+1}v, h_jv, g_{j+1}w$ (see Figure \ref{pfMs1}).  Then 
in $H^o$ we will apply Lemma \ref{strong-core1} to the walks $A_i + h_iu, B_{i+1} + g_{i+1}v$,
$A_j + h_jv, B_{j+1} + g_{j+1}w$, to conclude that $A_i, A_j$ as well as $A_i, B_{j+1}$ have embedded
pre-images that avoid each other. The assumptions of Lemma \ref{strong-core1} are easy to check
using the statements we have already proved. For instance, $(p_i,q_{i+1}) \not\leadsto (u,w)$
since otherwise we would have $(p_i,q_{i+1}) \leadsto (h_i,g_{j+1}) $. Since we have already
proved that $A'_i, B'_{j+1}$ avoid each other, this implies that $(p_i,q_{i+1}) \leadsto (h_i,g_{j+1})$, contradicting assumption 6 of the lemma.

\begin{figure}
\begin{center}
\includegraphics[height=6cm, width=15cm]{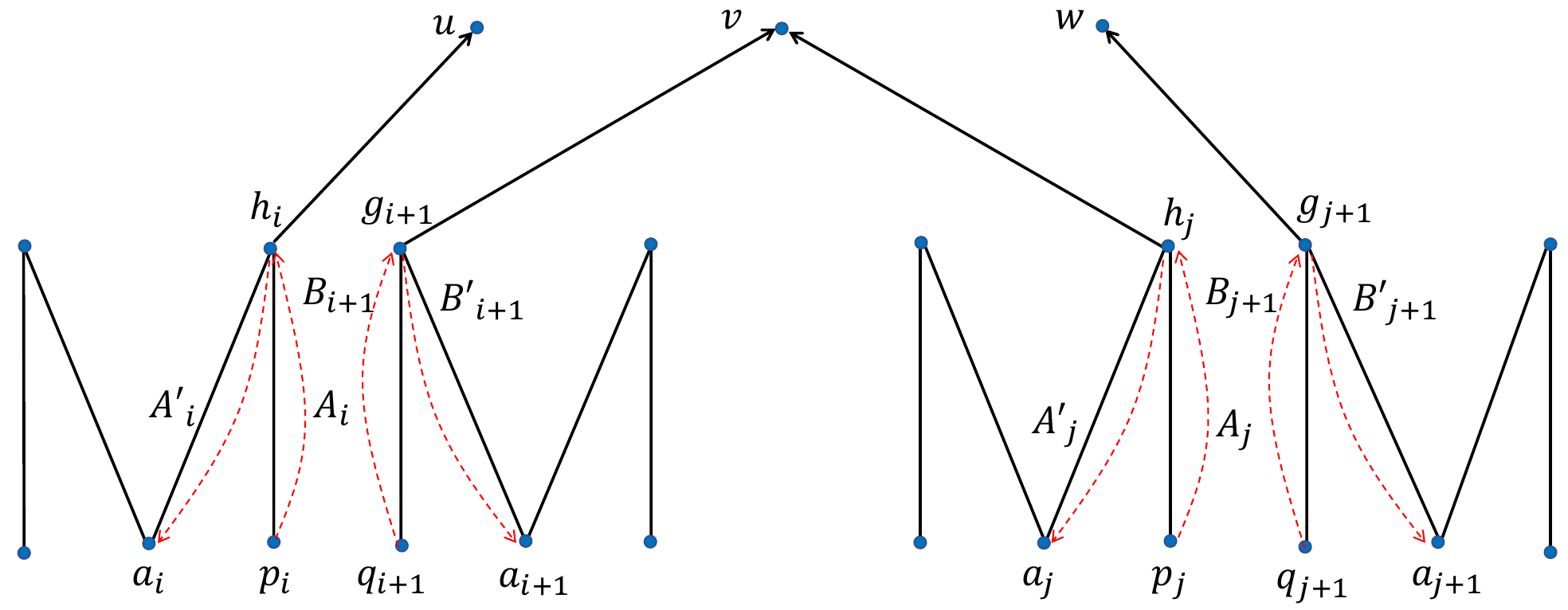}
\caption{Induction step in the proof of Lemma \ref{M}. The dashed arc show the direction of the oriented paths. Here $A_i+A'_i$ avoids $B_{i+1}+B'_{i+1}$, and $A_j+A'_j$ avoids $B_{j+1}+B'_{j+1}$. } \label{pfMs1}
\end{center}
\end{figure}

It remains to check the primed walks against the reverses of the none primed walks. The arguments are
symmetric, we focus on finding embedded pre-images of $A'_i, B'{i+1}, A^{-1}_j, B^{-1}_{j+1}$. We again
construct a new digraph $H^o$ with added vertices $u, v, w$ and arcs $a_iu, a_{i+1}v, p_jv, q_{j+1}w$.  
It is again easy to check, from the statements already proved, that Lemma \ref{strong-core1} applies
to the walks $A'_i + a_iu, B'_{i+1} + a_{i+1}v, A^{-1}_j + a_jv, B^{-1}_{j+1} + q_{j+1}w$ to imply that
$A'_i$ and $A^{-1}_j$ avoid each other, and that $A'_i$ and $B^{-1}_{j+1}$ as well as $A'_i$ and 
$B^{-1}_{i+1}$ also avoid each other.
\end{proof}
\subsection{Minimal circuits}
We now analyze a minimal circuit in $H^+$ under certain conditions and we derive properties of $H^+$.

\begin{definition}[closure-dual-free]\label{closure-dual-free}
A subset $T$ of $H^+$ is called closure-dual-free if for any $(x,y)\in T$, $(y,x)$ is not reachable from a pair in $T$, i.e., $(y,x) \not\in \widehat{T}$. 
\end{definition}

\begin{definition}[minimal circuit]\label{minimal-circuit-definition} 
Let $C : (a_0,a_1),(a_1,a_2),\dots,(a_n,a_0)$ be a circuit in $\widehat{T}$ where $T$ is a closure-dual-free set of pairs from $V(H^+)$ (i.e. $\widehat{T}$ is  dual-free). Let $S_0,S_1,\dots,S_n$ (not necessarily distinct) be the strong components in $T$ where $(a_i,a_{i+1})$ is in $\widehat{S_i}$, $0 \le i \le n$. We say $C$ is minimal if there is no other circuit $(a'_0,a'_1),(a'_1,a'_2),\dots,(a'_m,a'_0)$, $m<n$, where each $(a'_i,a'_{i+1})$ is in some $\widehat{S_j}$, $0 \le j \le n$.  
\end{definition} 

\begin{definition}[LL-pair]\label{LL-pair}
Let $S$ be a subset of $V(H^+)$. We say $(x,y)$ is an $LL$-pair (lower layer pair) with respect to $S$ if there exists a pair $(x',y') \in S$ that reaches $(x,y)$ via a directed path in $H^+$ which is constricted from below and has net value one.   
\end{definition}


To any $LL$-pair $(x,y)$ we can associate a directed walk $W_{x,y}$ from a pair in $S$ to $(x,y)$ with a positive net value; a suffix of such a walk ends at $(x, y)$, has net value 1 and is constricted below. We denote this suffix by $Z_{x,y}$. 

\begin{theorem}\label{itit}
Let $S$ be a set of pairs in $H^+$ and let   $\widehat{S}$ contains a minimal circuit $(a_0,a_1),(a_1,a_2),\dots,$ $(a_{n-1},a_n),  (a_n,a_0)$, $n>1$, 
such that each $(a_i,a_{i+1})$ is a $LL$-pair with respect to $\widehat{S}$. Then there exists another circuit $(a'_0,a'_1), (a'_1,a'_2), \dots, (a'_n,a'_0)$ of the pairs in $\widehat{S}$, and walks $P_i, Q_i, i=0, \dots, n$, in $H$, 
such that
\begin{enumerate}
    \item $P_i, Q_i$ are walks of net length one,
    \item $P_i, Q_i$ are constricted from below, $P_i$ from $a'_i$ to $a_i$, 
    $Q_i$ from $a'_{i+1}$ to $a_{i+1}$,
    \item $P_i$ and $Q_i$ are congruent and avoid each other.
\end{enumerate} 
\end{theorem}

\vspace{2mm}

\begin{proof} Let $Z_i$, $1 \le i \le n$, be a path from $(p'_i,q'_{i+1}) \in \widehat{S}$ to $(a_i,a_{i+1})$. Let $L_i$ be the height of $Z_i$.
We will find $n$ vertices $a'_i$ from amongst the $2n$ vertices $p'_i, q'_i$ which satisfy the conclusion. 
As an intermediate step, we will find $n$ vertices $a^*_i$ of the $2n$ vertices $p_i, q_i$ which also yield a
circuit in $\widehat{S}$. For any $i$, if $L_{i-1} < L_i$ we let $a^*_i = q_i$ and $a'_i = q'_i$ and if $L_{i-1} \geq L_i$ 
we let $a^*_i = p_i$ and $a'_i = p'_i$. We first prove that each pair $(a^*_i,a^*_{i+1})$ in the circuit 
$(a^*_0,a^*_1), (a^*_1,a^*_2), \dots, (a^*_n,a^*_0)$ can be reached from the corresponding pair $(a_i,a_{i+1})$.

\begin{figure}
\begin{center}
\includegraphics[height=6cm,width=14cm]{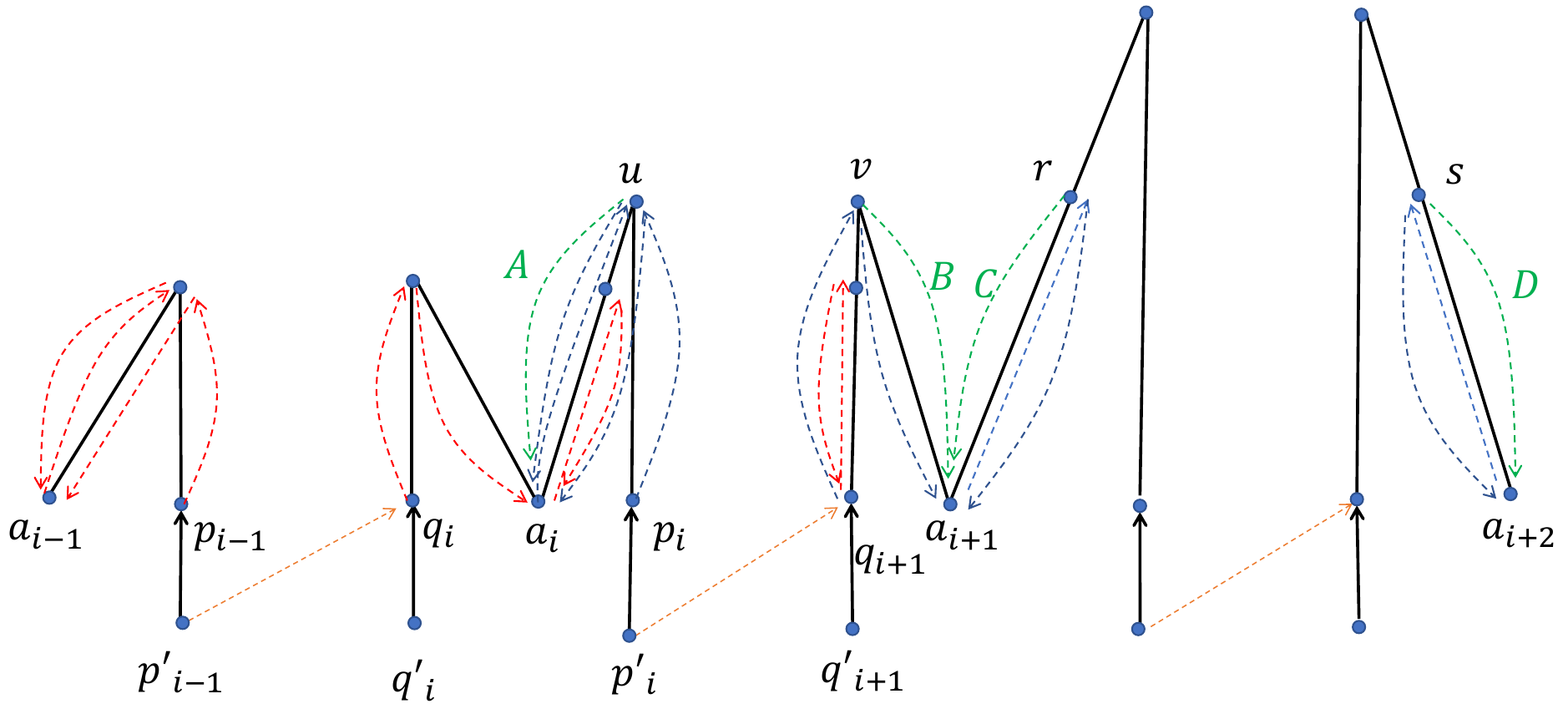}
\caption{Each straight segment represents a constricted walk. The dotted line shows the direction of the walks.walks $A,B$ avoid each other and walks $C,D$ avoid each other. The orange dashed arrow means missing arc in $H$.} \label{3Z}
\end{center}
\end{figure}

First consider the case that $L_{i-1} < L_i < L_{i+1}$, in which we have $(a^*_i,a^*_{i+1}) = (q_i,q_{i+1})$. 
We refer to Figure \ref{3Z} to summarize the steps of the proof. First of all, we find corresponding vertices 
$r, s$ so that the green walks $C$ from $r$ to $a_{i+1}$ and $D$ from $s$ to $a_{i+2}$ have net length $L_i$. 

Then Lemma \ref{strong-core1} is applied to the four green walks $A, B, C, D$, to conclude that $A, B$ avoid each other 
and $C, D$ also avoid each other. Moreover $A,C$ have a embedded pre-image that avoid each other and $B,D$ have embedded pre-image that avoid each other.  Now we can use Lemma \ref{M} on the suggested blue walks. Specifically, the blue walk from
$p_i$ to $u$, then taking $A$ to $a_i$ and then use $A^{-1}+A$ again; the blue walk from $q_{i+1}$ to $v$ followed by 
$B+C^{-1}+C$; and the blue walk $C^{-1}+C+C^{-1}+C$, and the blue walk $D^{-1}+D+D^{-1}+D$ (see Figure \ref{3Z}). 
We conclude that the walk $A^{-1}+A$ and the walk $B^{-1}$ concatenated with the walk from $v$ to $q_{i+1}$ have embedded pre-images that avoid each other. Thus 
$(a_i,a_{i+1}) \leadsto (a_i,q_{i+1})$ and $(a_{i+1},a_{i+2}) \leadsto (q_{i+1},a_{i+2})$. Then using similar arguments to 
the red walks of height $L_{i-1}$ we conclude that $(a_i,q_{i+1}) \leadsto (q_i,q_{i+1})$ and 
from $(a_{i-1},a_i) \leadsto (a_{i-1},q_i) $. This allows us to replace $a_i$ by $q_i$ and $a_{i+1}$ by $q_{i+1}$ in the circuit 
$(a_0,a_1), \dots, (a_n,a_0)$. (The proof in case $L_{i-1} < L_i < L_{i+1}$ is symmetric.)

If $L_i \geq L_{i-1}$ and $L_i \geq L_{i+1}$, then $(a^*_i,a^*_{i+1}) = (q_i,p_{i+1})$ and $(a'_i,a'_{i+1}) = (q'_i,p'_{i+1})$. 
Lemma \ref{M} can be similarly used to conclude that $(a_i,a_{i+1}) \leadsto (q_i,p_{i+1})$. Similarly by applying Lemma \ref{M}  
(on the blue walks in Figures \ref{plusminus}) we conclude that $(p_{i-1},p_{i+1}) \leadsto 
(a_{i-1},a_{i+1})$ and $(q_i,q_{i+2}) \leadsto (q_i,a_{i+2})$. 

It also follows that we can replace $a_i$ by $q_i$ and $a_{i+1}$ by $p_{i+1}$
in the circuit $(a_0,a_1), \dots, (a_n,a_0)$. Now $p'_{i+1}p_{i+1}$ is not an arc of $H$ as otherwise $(p'_{i-1},q_i)(p_{i+1},q_i)$ $\in A(H^+)$ and this contradicts the
minimality of the circuit. The absence of the arc $p'_{i+1}p_{i+1}$ would imply that $q'_ip_{i+1} \not\in A(H)$. 
Otherwise $(p'_{i-1},q'_i) (p_{i-1},p_{i+1}) \in A(H^+)$ and hence $(p'_{i-1},q'_i) \leadsto (a_{i-1},a_{i+1})$, contradicting the minimality of the circuit. 
We also note that $p'_{i+1}q_i \not\in A(H)$ as otherwise $(p'_{i+1},q'_{i+1})(q_i,p_{i+2}) \in A(H^+)$ and hence $(p'_{i+1},q'_{i+1}) \leadsto (q_i,a_{i+2})$, contradicting 
the minimality of the circuit.  Therefore $(a_i,a_{i+1}),$ $(p_i,q_{i+1}), (p'_i,q'_{i+1})$ are all reachable from each other.

\begin{figure}
\begin{center}
\includegraphics[height=6cm,width=14cm]{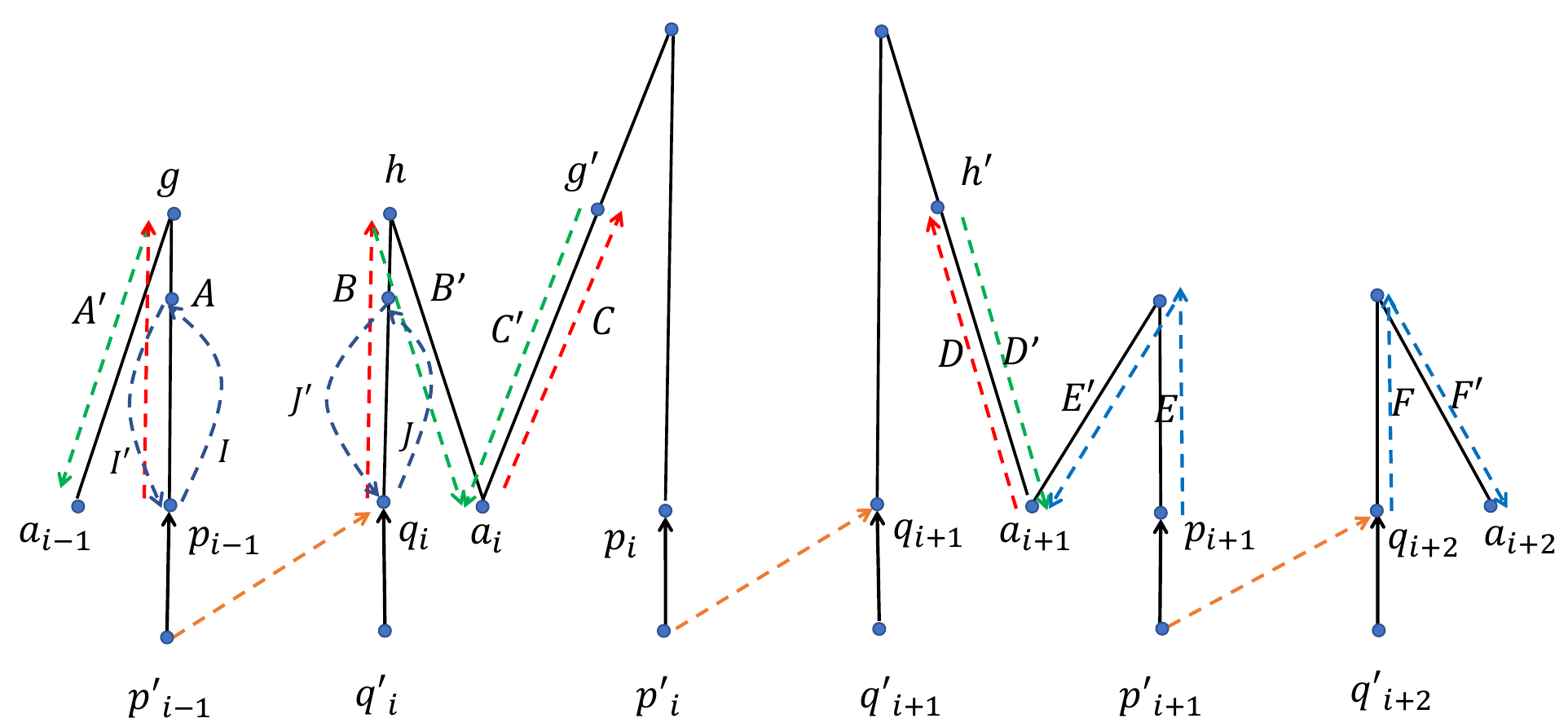}
\caption{Illustration of the proof of Lemma \ref{itit}. The orange dash arcs are the missing arcs. The red,green, blue dashed paths show the direction of the walks. } \label{plusminus}
\end{center}
\end{figure}

In case $L_i \leq L_{i-1}$ and $L_i \leq L_{i+1}$, we have $(a^*_i,a^*_{i+1}) = (p_i,q_{i+1})$, and by using Lemma \ref{M} in
a fashion similar to the above proofs, we conclude easily that there are the walks from $a_i$ to $a^*_i$ and from  $a_{i+1}$
to $a^*_{i+1}$ that avoid each other and are of net length zero, constricted from below. 

We now show that $a'_ia^*_{i+1}$ and $a'_{i+1}a^*_i$ are not arcs of $H$, completing the proof of the Lemma.
In fact, this has already been observed (in the previous case) for the $i$ which have $L_i \geq L_{i-1}$ and $L_i \geq L_{i+1}$.

If $L_{i-1} < L_i < L_{i+1}$. In fact, we assume $L_{i-1} < L_i < L_{i+1} < \dots L_{i+j} \geq L_{i+j+1}$ for some $j \geq 1$. 
Note that this means that 
\begin{align*}
    a^*_i=q_i, a^*_{i+1}=q_{i+1}, \dots, a^*_{i+j-1}=q_{i+j-1}, a^*_{i+j}=q_{i+j}, a^*_{i+j+1}=p_{i+j+1}.
\end{align*}
We first note that since $L_{i+j} \geq L_{i+j-1}$ and $L_{i+j} \geq L_{i+j+1}$ we already know that $p'_{i+j+1}q_{i+j}$ is not an 
arc of $H$.  By symmetry, $q'_{i+j}p_{i+j+1}$ is also not an arc of $H$. Next we argue that $p'_iq_{i+2}$ is not an arc of $H$ as 
otherwise $(p'_i,q'_{i+1}) \leadsto (a^*_{i+2},a^*_{i+1}) $ contradicting the minimality of the circuit. This implies that 
$q'_{i+1}q_{i+2}$ is not an arc of $H$, as otherwise $(p'_i,q'_{i+1}) \leadsto (p_i,q_{i+2}) $ and eventually 
$(q_i,q_{i+2})=(a^*_i,a^*_{i+2})$ (using Lemma \ref{M} on suitable portions of the walks). The same arguments imply
that $p'_{i+1}q_{i+3} \not\in A(H)$ and $q'_{i+2}q_{i+3} \not\in A(H)$, and so on until $p'_{i+j-2}q_{i+j} \not\in A(H)$ 
and $q'_{i+j-1}q_{i+j} \not\in A(H)$. (These will all be used later.) Now we proceed to show that $q'_{i+j}q_{i+j-1}$
is not arc of $H$, otherwise $(q'_{i+j},p'_{i+j+1})$ (which we have shown to be reachable from $(a^*_{i+j},a^*_{i+j+1})$)
can reach $(a^*_{i+j-1},a^*_{i+j+1})$ because $q'_{i+j}p_{i+j+1}=q'_{i+j}a^*_{i+j+1}$ is not an arc of $H$. This 
contradicts the minimality of the circuit. Note that now we have both $q'_{i+j}q_{i+j-1} \not\in A(H)$ and $q'_{i+j-1}q_{i+j} \not\in A(H)$. 
The first fact implies that $(a^*_{i+j-1},a^*_{i+j}) \leadsto (a'_{i+j-1},a'_{i+j}) $. The second fact implies that
we can repeat the argument to conclude that $q'_{i+j-1}q_{i+j-2} \not\in A(H)$, and continue the argument in this way, eventually
showing that $a'_{i+1}a^*_i=q'_{i+1}q_i \not\in A(H)$ and $a'_ia^*_{i+1}=q'_iq_{i+1} \not\in A(H)$.

It remains to consider those $i$ that have $L_i \leq L_{i-1}$ and $L_i \leq L_{i+1}$. In this case, $a^*_{i}=p_i$ and $a^*_{i+1}=q_{i+1}$.
If $L_{i+1} < L_{i+2}$, then $a^*_{i+2}=q_{i+2}$, and  we can use the previous argument to conclude that 
$q'_{i+1}q_{i+2}$ and $q'_{i+2}q_{i+1}$ are not arcs of $H$. This implies that $q'_{i+1}p_i$ is also not an
arc of $H$, otherwise the walk $q_{i+1}, q'_{i+1}, p_i$ avoids the walk $q_{i+2}, q'_{i+2}, q_{i+2}$ and hence 
the pair $(p_i,q_{i+2})$ $=(a^*_i,a^*_{i+2}) \in \widehat{S}$ 
, contradicting the minimality of the circuit. 

On the other hand, if $L_{i+1} \geq L_{i+2}$ we have $a^*_{i+2}=p_{i+2}$ and we use another previous case to conclude that
$q'_{i+1}p_{i+1}$ and $p'_{i+2}q_{i+1}$ are not arcs of $H$, and hence there is walk from $q_{i+1}$ to $p_i$ that avoids a
walk from $p_{i+2}$ to itself, thereby the pair $(p_i,p_{i+2})=(a^*_i,a^*_{i+2})$ is also in $\widehat{S}$, again yielding a contradiction.
\end{proof}

\section{Tools for the Proof of Correctness }\label{tools}
In this section we provide the tools needed to show the algorithm is correct.
We use the structural properties of walks in $H^+$ (Section~\ref{walks-structure}) and structural properties of circuits in $H^+$ (Section~\ref{circuit-structure-property}).

\subsection{Correctness of phase one}
In this subsection we prove a useful tool for proving the correctness of the first phase of the algorithm.

\subsubsection{Proof of Theorem~\ref{CLAIM-main-body}}

\begin{theorem}[Theorem~\ref{CLAIM-main-body}, repeated]\label{CLAIM}
Let $T$ be a closure-dual-free set of unbalanced components and assume that $\widehat{T}$ contains a minimal circuit $C$ with $n+1$ pairs. Then $n > 1$ and the following statements hold. 

\begin{enumerate}

\item There exists some minimal circuit (see Definition \ref{minimal-circuit-definition}) with extremal pairs \[(b_0,b_1),(b_1,b_2), \dots, (b_{n-1},b_n),(b_n,b_0)\] in $T$ such that the $i$-th, $0 \le i \le n$, pair in $C$ is in the same strong component as $(b_i,b_{i+1})$, and reachable
 from $(b_i , b_{i+1})$ by a symmetric walk of non-negative net value, and constricted from below. \label{CLAIM-1}

\item  For each $i$, $0 \le i \le n$, there exists an infinite walk $P_i$ that starts from $b_i$ and has unbounded positive net length. Furthermore, for every $i, j$, $0 \le i < j \le n$, $P_i$ and $P_j$ avoid each other. \label{CLAIM-2}

\item In statement \ref{CLAIM-1}, for a given $0 \le i \le n$, we can choose $(b_i , b_{i+1})$ to be any given extremal pair from its corresponding strong component. \label{CLAIM-3}

\item There is no directed path in $H^+$ from $(b_i,b_{i+1})$ to any of $(b_j,b_{j+1})$ 
$ i \ne j$, and to any of $(b_{j+1},b_j)$. \label{CLAIM-4}

\item There is no directed path in $H^+$ from any of $(b_{i+1},b_{i})$, $0 \le  i \le n$ to $(b_i,b_{i+1})$. \label{CLAIM-5}

\end{enumerate}
\end{theorem}

\begin{proof}
Recall that each unbalanced pair belongs to a strong component containing an unbalanced directed cycle. Plus, each unbalanced directed cycle contains an extremal pair. 
Let $C: (a_0,a_1),(a_1,a_2),\dots,(a_n,a_0)$. 
Now, for each $i$, let $C_i$ be the strong component of $H^+$ containing an extremal pair in $\widehat{T}$ where
$(a_i , a_{i+1} )$ is reachable from $C_i$, and let $D_i$ be the directed cycle in $C_i$ containing that extremal pair.

We first show that $n>1$. Otherwise, 
by definition $(x,y) \leadsto (a_0,a_1)$, and $(x',y') \leadsto (a_1,a_0)$ where $(x,y),(x',y') \in T$. Now by skew property we have $(a_0,a_1) \leadsto (y',x')$, and hence, $(x,y) \leadsto (y',x')$, a contradiction that $T$ is closure-dual-free.  \\

\noindent  \textbf{Proof of \ref{CLAIM-1}}. We first prove the following claim. 
\begin{claim}\label{bit}
Each $(a_i,a_{i+1})$ is an $LL$-pair with respect to $\widehat{T}$ (see definition \ref{LL-pair}).
\end{claim}

\begin{proof} 
First consider, a subscript $i$ such that $D_i$ has positive net value. Then, there is an infinite directed path $W$ 
continuously winding around $D_i$ in the positive direction which is constricted from below with unbounded net value. By following 
$W$ as far as necessary and then following a path that leads from $C_i$ to $(a_i,a_{i+1})$ (such a path exists both when $(a_i,a_{i+1})$
is in $C_i$ or reachable from $C_i$), we obtain a directed path $W_i$ in $H^+$ that is constricted from below (recall that when we say a directed path $X$ in $H^+$ is constricted from below, we mean the walks $A_1,A_2$ in $H$ corresponding to $X$ are constricted from below). We let $(p'_i,q'_{i+1})$ be 
the last vertex on $W_i$ such that the net value of $W_i[(p'_i,q'_{i+1}),(a_i,a_{i+1})]$ is one, and we set $Z_i = W_i[(p'_i,q'_{i+1}),(a_i,a_{i+1})]$. Let $(p_i,q_{i+1})$ be the second vertex of $Z_i$, i.e. the net value of $Z_i[(p_i,q_{i+1}),(a_i,a_{i+1})]$ is zero. 
Let $L_i$ be the maximum net value of a prefix of $Z_i$, i.e., of a directed path $Z_i[(p'_i,q'_{i+1}),(x,y)]$ for any $(x,y)$. Note that $L_i$ could be one, 
in case the $Z_i$ is also constricted from above, i.e., $Z_i$ is just one arc in $H^+$.

We emphasize for future reference that in this case the directed path $Z_i$ arises from $W_i$ that started on the cycle $D_i$.

A similar argument applies to a subscript $i$ such that $D_i$ has negative net value, but following the directed path 
$W$ discussed above (unbounded and non-positive) and then a path from $D_i$ to $(a_i,a_{i+1})$, we obtain a directed
walk $W_i$ in $H^+$ that is constricted from above but {\textbf { not} } constricted from below. Indeed, in 
such a case we can again let $(p'_i,q'_{i+1})$ be the last vertex on $W_i$ such that the net value of 
$W_i[(p'_i,q'_{i+1}),(a_i,a_{i+1})]$ is minus one and set $Z_i = W_i[(p'_i,q'_{i+1}),(a_i,a_{i+1})]$.

\begin{observation}
Since, the circuit is minimal, and $n>1$, there is no directed path in $H^+$ from a vertex on $Z_i$ to any $(a_r,a_s)$, $0 \le r \ne s-1,s \ne n$, as otherwise we get a shorter circuit. Thus in what follows we are able to apply  Lemma \ref{strong-core1} and Lemma \ref{M}. 
\end{observation}

Suppose next that there are two subscripts $i, i+1$ (addition modulo $n$) such that both $D_i$ and $D_{i+1}$ have 
negative net values, and both $W_i$ and $W_{i+1}$ are constricted, from some pairs $(u,v), (w,x) \in \widehat{T}$ 
to $(a_i,a_{i+1}), (a_{i+1},a_{i+2})$ respectively (see Figure \ref{D_iD_{i+1}-both-negative}). We may assume that the pairs $(u,v), (w,x)$ are on 
the cycles $D_i, D_{i+1}$, and that the net values of $W_i, W_{i+1}$ are the same and arbitrary (by choosing for their starting 
vertices a suitable extremal vertex on $D_i,D_{i+1}$). In this context, Lemma \ref{strong-core1} applies to the 
four walks $A_i, B_{i+1}, A_{i+1}, B_{i+2}$ in $H$ corresponding to $W_i=(A_i,B_{i+1}), W_{i+1}=(A_{i+1},B_{i+2})$, and we conclude that, in particular, $A_i, B_{i+1}$ 
avoid each other and $A_{i+1}, B_{i+2}$ avoid each other (see Figure \ref{D_iD_{i+1}-both-negative}). This implies that the reverse traversal of the cycles $D_i, D_{i+1}$ 
is also a cycle in $H^+$, of positive net value, and we can proceed as in the case when $D_i, D_{i+1}$ had positive net value.


\begin{figure}
\begin{center}
\includegraphics[scale=0.9]{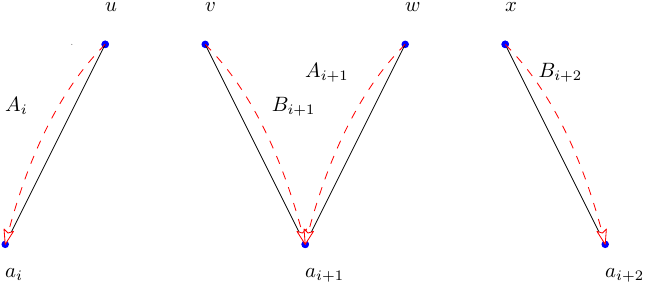}
\caption{Claim  \ref{bit} when $D_i,D_{i+1}$ have negative net value. $A_i$ avoids $B_{i+1}$, and $A_{i+1}$ avoids $B_{i+2}$, the dashed red lines show the direction of $A_i,A_{i+1},B_{i+1},B_{i+2}$.} 
\label{D_iD_{i+1}-both-negative}
\end{center}
\end{figure}

Thus, it remains to consider the case when $D_i$ has negative net value, and the value of $W_i$ is constricted (both from below and from above), and $D_{i-1}, D_{i+1}$ have positive net values. We illustrate this part of the 
proof in Figure \ref{D_{i-1}D_{i+1}-positive-D_i-negative}. Each directed path in $H^+$ is depicted as two walks in $H$ where the first avoids the 
second. We show in the Figure \ref{D_{i-1}D_{i+1}-positive-D_i-negative}  the starting vertex $(p'_{i-1},q'_i)$ 
of $Z_{i-1}$, its ending vertex $(a_{i-1},a_i)$, as well as the starting vertex $(p'_{i+1},q'_{i+2})$ of $Z_{i+1}$, 
and its ending vertex $(a_{i+1},a_{i+2})$. In the illustration we assume neither $Z_{i-1}$ nor $Z_{i+1}$ is constricted from above. (The proof in the cases where one or both are constricted is similar and easier.) 

Without 
loss of generality, we assume that the height of $Z_{i-1}$ is greater than or equal to the height of $Z_{i+1}$. 

As shown in the Figure  \ref{D_{i-1}D_{i+1}-positive-D_i-negative} , we let $(p_{i-1},q_i)$ be the
second vertex of $Z_{i-1}$ and $(p_{i+1},q_{i+2})$ the second vertex of $Z_{i+1}$. We also show a constricted directed
walk $W_i$ ending in $(a_i,a_{i+1})$. We assume $(g,h)$ is the last vertex on $Z_{i-1}$ that maximizes the net
value of the prefix $Z_{i-1}[(p'_{i-1},q'_{i-1}),(g,h)]$, and $(g',h')$ the last vertex of $W_i$ so that 
$Z_{i-1}[(g,h),(a_{i-1},a_i)]$ and $W_i[(g',h'),(a_i,a_{i+1})]$ have the same net values (note that both 
these directed paths are constricted). The directed path $Z_{i-1}$ in $H^+$ corresponds to two walks $p'_{i-1}+A_{i-1}+A'_{i-1},$
$q'_i+B_i+B'_i$ in $H$, as depicted, where the first avoids the second, in particular $A_{i-1}$ avoids $B_i$ and $A'_{i-1}$ avoids 
$B'_i$. ($A_{i-1}$ is the portion from $p_{i-1}$ to $g$ and $A'_{i-1}$ is the portion from $g$ to $a_{i-1}$, and similarly for $B_{i-1}$). 
Let $C$ be the walk from $g'$ to $a_{i}$, and $D$ the walk from $h'$ to $a_{i+1}$ so that $W_i[(g',h'),(a_i,a_{i+1})]=(C,D)$.
Notice that $C$ avoids $D$, and $A'_{i-1},C,B'_{i-1},D$ are constricted and have the same net length. Now we prove the following.

\begin{figure}
\begin{center}
\includegraphics[scale=0.8]{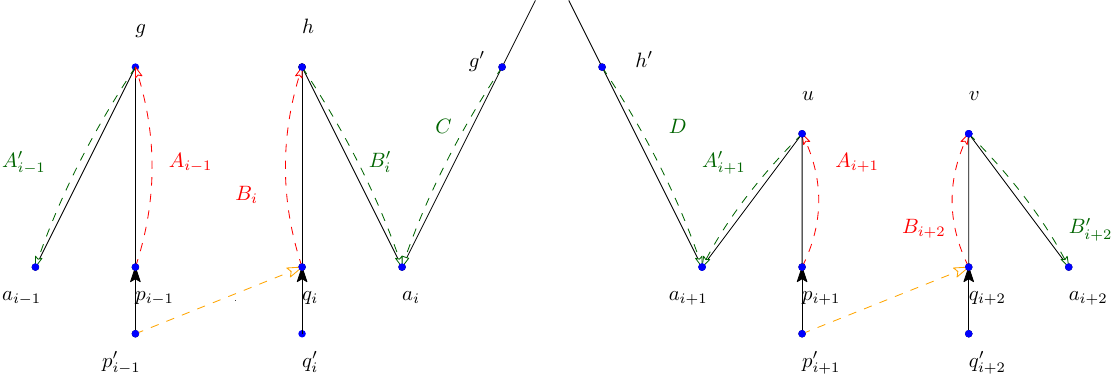}
\caption{Claim \ref{bit} when $D_{i-1},D_{i+1}$ have positive net value and $D_i$ has negative net value, $A_{i-1}$ avoids $B_i$, $A'_{i-1}$ avoids $B'_i$, and $C$ avoids $D$. Red and green dashed lines show the direction of the walks, and the orange arrows are the missing arcs.} 
\label{D_{i-1}D_{i+1}-positive-D_i-negative}
\end{center}
\end{figure}

\begin{enumerate}
\item[1.]
The walks $A'_{i-1}, B'_i, C, D$ have congruent embedded pre-images that avoid each other, except for the pre-images of $B'_i$ and $C$
\item[2.]
The walks $A^{-1}_{i-1}, A'_{i-1}, B^{-1}_i , B'_i$ have congruent embedded pre-images that avoid each other, except for the pre-images of $A^{-1}_{i-1},A'_{i-1}$ as well as $B^{-1}_i,B'$

\item[3.]
The pairs $(a_{i-1},a_i)$ and $(a_{i-1},q_i)$ are in the same strong component of $H^+$
\item[4.]
The walks $B_i+B'_{i}$ and $D^{-1}+D$ have congruent embedded pre-images that avoid each other. The walks $A_{i-1}+A'_{i-1}$ and $D^{-1}+D$ have congruent embedded pre-images that avoid each other
\item[5.]
The pairs $(a_i,a_{i+1})$ and $(q_i,a_{i+1})$ are in the same strong component of $H^+$. 
The pairs $(a_{i-1},a_{i+1})$ and $(p_{i-1},a_{i+1})$ are in the same strong component of $H^+$

\item [6.] The pairs $(a_{i},a_{i+2}),(a_i,q_{i+2})$ are in the same strong component of $H^+$. 
\end{enumerate}

Item 1 follows directly by Lemma \ref{strong-core1} because the minimality of the circuit $(a_0,a_1), (a_1,a_2),$ $\dots,(a_n,a_0)$ implies that $(a_{i-1},a_{i+1})$ or $(a_{i+1},a_i)$ or $(a_i,a_{i-1})$ can not be in $\widehat{T}$. Therefore, Lemma \ref{M} applies to the walks $A_{i-1},A'_{i-1},B_i,B'_i,C,C^{-1},D,D^{-1}$, using the same minimality arguments, verifying item 2. We also conclude that $D,B^{-1}_i,B'_i$ have congruent embedded
pre-images that avoid each other, except $B^{-1}_i,B_i$, as well as, $D,A^{-1}_{i-1},A'_{i-1}$ have congruent embedded pre-images that avoid each other, except $A^{-1}_{i-1},A'_{i-1}$. Now this implies that $B_i+B'_{i}$ and 
$D^{-1}+D$ have congruent embedded pre-images that avoid each other, and $A_{i-1}+A'_{i-1}$ and $D^{-1}+D$ have congruent embedded pre-images that avoid each other; verifying 4. As a consequence of 2, $A'^{-1}_{i-1}+A'_{i-1}$, and $B_i+B'_i$ have congruent embedded pre-images, implying that $(a_{i-1},a_i),(a_{i-1},q_i)$ are in the same strong component; implying 3. Using 4, we conclude that $(a_i,a_{i+1})$ and $(q_i,a_{i+1})$ are  in the same component, and $(a_{i-1},a_{i+1}),(p_{i-1},a_{i+1})$ are in the same strong component; implying 5. Analogous to 5, we derive 6. Items 3 and 5 imply that we can replace $a_i$ by $a'_i=q_i$ and obtain another circuit of pairs $(a_0,a_1), (a_1,a_2), \dots, (a_{i-1},a'_i), (a'_i,a_{i+1}), \dots, (a_n,a_0)$ in $\widehat{T}$. By similar argument on this new circuit (deducing 1,2,3,4,5,6 ) we may replace $a_{i+1}$ by $a'_{i+1}=p_{i+1}$. In the rest of the proof we assume that we have made the replacement, i.e., that $a_i=q_i, a_{i+1}=p_{i+1}$. Thus we continue with the circuit 
\[
X: (a_0,a_1), (a_1,a_2), \dots, (a_{i-1},q_i), (q_i,p_{i+1}),(p_{i+1},a_{i+2}), \dots, (a_n,a_0)
\]
in $\widehat{T}$. Note that $X$ is also minimal. Similar conclusion as in 1, 2, 3, 4, 5, 6 hold for circuit $X$.

We now show that, in the new circuit $X$, the path $Z_i$ actually exists, namely, that the (single-arc) walks $q'_iq_i$
and $p'_{i+1}p_{i+1}$ avoid each other, and hence the new $(a_i,a_{i+1})$ and $(q'_i,p'_{i+1})$ are
reachable from each other. First, we observe that $p'_{i-1}p_{i+1}$ is not an arc, otherwise 
$(p'_{i-1},q'_i) \leadsto (p_{i+1},q_i)$, contradicting the minimality of the circuit $X$ (because $(q_i,p_{i+1}) \in \widehat{T}$).   
Then $q'_ip_{i+1}$ is not 
an arc, otherwise,  $(p'_{i-1},q'_i) (p_{i-1},p_{i+1}) \in A(H^+)$, and according to item 5 for $X$, we have $(p_{i-1},p_{i+1}) \leadsto (a_{i-1},p_{i+1})$, a contradiction to minimality of $X$. 
Finally, $p'_{i+1}q_i$ is not an arc, otherwise, $(p'_{i+1},q'_{i+2})(q_i,q_{i+2})$ is an arc of $H^+$. By item 6 for  circuit $X$, $(q_i,a_{i+2}),(q_i,q_{i+2})$ are in the same strong component, and hence, $(p'_{i+1},q'_{i+2}) \leadsto (q_i,a_{i+2})$; contradicting the minimality of the circuit $X$. This would imply that $(a_i,a_{i+1})$ is also an $LL$-pair. 
\end{proof}

Since Claim \ref{bit} makes it possible to apply Theorem \ref{itit}, we can conclude that there exists another
circuit $(a'_0,a'_1), (a'_1,a'_2), \dots, (a'_n,a'_0)$ in $\widehat{T}$ with the corresponding walks $P_i, Q_i$
from Theorem \ref{itit}. We can repeat the argument obtaining at the $i$-th step an ordered sequence of vertices
$a^i_0, a^i_1, \dots, a^i_n$ such that $(a^i_0,a^i_1)$, $(a^i_1,a^i_2), \dots, (a^i_n,a^i_0)$ is a circuit in $\widehat{T}$
with each $(a^{i}_0,a^i_1)$ is reachable from $(a^{i+1}_0,a^{i+1}_1)$ by a symmetric walk of net value minus one. Thus there exist $r \neq s$ such that $a^r_0=a^s_0, a^r_1=a^s_1, \dots, a^r_n=a^s_n$. 
This circuit has each pair $(a^i_k,a^i_{k+1})$ extremal; and hence, we set $b_k=a^i_k$, $0 \le k \le n$. \\

\noindent{\textbf{Proof of \ref{CLAIM-2}}}. Continuing the last sentence from the proof of 1, let $W_i$, $0 \le i \le n$ be a closed walk from $b_i$ to $b_i$ of net length $|s-r|$, and constricted from below; let $W'_i$ be the walk from $b_{i+1}$ to $b_{i+1}$ where $W_i,W'_i$ avoid each other.  For $0 \le i \le n$, let $Q_i$, be the walk that winding around $W_i$, $a$ times for some positive integer $a$, and let $Q'_i$ be the walk obtained by winding around the closed walk $W'_i$, $a$ times. Let $h_i \in Q_i$ such that $R_i=Q_i[b_i,h_i]$ is constricted and have net length $a|r-s|$ (notice that $h_i$ could be $b_i$). Let $h'_i \in Q'_i$ be the corresponding vertex to $h_i$ and let $R'_i=Q'_i[b_{i+1},h'_i]$. Notice that since $R_i,R'_i$ avoid each other and are constricted, $(R_i)^{-1},(R'_i)^{-1}$ also avoid each other. Now one  can apply the Lemma \ref{M} on $R_0,(R'_0)^{-1},R_1,(R'_1)^{-1},\dots,R_n,(R'_n)^{-1}$ and conclude that $R_1,R_2,\dots,R_n$ have congruent embedded pre-images $P_1,P_2,\dots,P_n$ that all avoid each other. Notice that $P_i$ starts at $b_i$, $0 \le i \le n$. This proves (2). \\

\noindent{\textbf{Proof of \ref{CLAIM-3}}}. 
By statement (\ref{CLAIM-2}) of Theorem \ref{CLAIM}, there exist infinite walks $P_i$, $0 \le i \le n$, starting at $b_i$ with unbounded positive net length. Moreover, all pairs $P_i,P_j$, $0 \le i < j \le n$ avoid each other.

We prove the statment for $i=0$ (the other cases are similar). Now let $(x,y)$ be an arbitrary extremal pair in the  component $C_0$ containing $(b_0,b_1)$. 
We may assume there exists a directed path $W$ from $(x,y)$ to $(b_0,b_1)$ which is  constricted and has positive net value. This can be done by going around a directed cycle, in $C_0$ containing $(x,y)$, in the positive direction as many times as needed. If the cycle has negative net value then similar argument is applied. This means that $W=(X,Y)$ where $X$ is constricted and has positive net length and avoids $Y$. Note that the net length of $X$ could be arbitrary large. 

Let $b'_i$, $0 \le i \le n$ be a vertex on $P_i$ such $P_i[b'_i,b_i]$ is constricted and has the same net length as $X$. Now by applying the Lemma \ref{M} on $X,Y,P_1[b'_1,b_1],P_2[b'_2,b_2]$, and on $P_n[b'_n,b_n],P_0[b'_0,b_0],X,Y$ we  conclude that $X,Y$ have congruent embedded pre-images $X',Y'$ that avoid each other. Moreover, we can choose $Q_0$ from $x$ to $b_0$, $Q_1$ from $y$ to $b_1$, $Q_2$ from $b'_2$ to $b_2$, and $Q_n$ from $b'_n$ to $b_n$ in such a way that $Q_0,Q_1,Q_2,Q_n$ are all congruent and all avoid each other. Note that $Q_0,Q_1,Q_2,Q_n$ are congruent embedded pre-images of $X',Y',P_2[b'_2,b_2],P_n[b'_n,b_n]$ respectively.    
This means $(x,y),(y,b'_2),(b'_2,b'_3),...,\\(b'_n,x)$ is also a circuit, and each $(b'_i,b'_{i+1})$ is an extremal pair.  Now it is easy to see that there exist $P'_0,P'_1,\dots,P'_n$ starting at $x,y,b'_2,b'_3,\dots,b'_n$ (respectively) so that $P'_i,P'_j$ avoid each other. \\

\noindent {\textbf {Proof of \ref{CLAIM-4}}}.
Suppose there exists a directed path $W$ in $H^+$ 
from $(b_i,b_{i+1})$ to $(b_j,b_{j+1})$. 
We may assume $W$ has non-positive net value (the argument for the other case is similar). 
Now define $W'$ to be a walk in $H^+$ starting at $(b_i,b_{i+1})$ and then following $W$ to $(b_j,b_{j+1})$ and then following $W_j$ in negative direction sufficiently many times such that $W'^{-1}$ is constricted from below. Recall that by (3) $W_j$ is the directed cycle in $H^+$ containing 
$(b_j,b_{j+1}$).

Let $W'=(X_1,X_2)$ and observe that $X_1$ is a walk in $H$ from $b_i$ to $b_j$ and $X_2$ is 
a walk from $b_{i+1}$ to $b_{j+1}$ and $X_1$ avoids $X_2$. This implies that $(b_i,b_{i+1})$ and $(b_j,b_{j+1})$ are in the same strong component of $H^+$.  

Let $b'_0,b'_1,\dots, b'_n$ be the vertices on $P_0,P_1,\dots,P_n$ such that for every $0 \le r \le n$, $r \ne j+1,j+2$, $P_r[b'_r,b_r]$ has the same net length as $X_1^{-1}$ (reverse of $X_1$). Set $b'_j=b_j$, $b'_{j+1}=b_{j+1}$. Observe that $(b'_r,b'_{r+1}),(b_r,b_{r+1})$ are in the same strong component of $H^+$ because $P_r,P_{r+1}$ avoid each other. This would imply that $(b'_0,b'_1),(b'_1,b'_2),\dots, (b'_n,b'_0)$ is also a circuit in $\widehat{T}$. Consider two vertices $p \in P_{j+1}$ and $q \in P_{j+2}$ where
\[
A=P_{j+1}[b_{j+1},p]P_{j+1}^{-1}[p,b_{j+1}]
\]
\[
B=P_{j+2}[b_{j+2},q]P_{j+2}^{-1}[q,b_{j+2}]
\]
are congruent (avoid each other) and have the same height as $X_1^{-1}$. Now by Lemma \ref{M} on $Y_1=P_{i}[b'_{i},b_i]X_1$, $Y_2=P_{i+1}[b'_{i+1},b_{i+1}]X_2$, and $A,B$ we may assume that $Y_2,B$ have congruent embedded  pre-images $Y'_2,B'$ that avoid each other.  However, there exists a path $W_1$ in $H^+$ from $(b'_{j+1},b'_{j+2})$ to $(b'_i,b'_{j+2})$, implying a shorter circuit in $\widehat{T}$. 
Note that $W_1$ consists of two walks $Y_2'^{-1}$ and $B'^{-1}$ that avoid each other, and hence, $W_1$ is indeed in $H^+$.  \\

\noindent{\textbf{Proof of 5}} Assume without loss of generality that $i = n$, the other cases are symmetric. Using
 \ref{CLAIM-1}, we may also suppose that $(b_0,b_n)$ is an extremal pair. Suppose there is a directed path in $H^+$ from 
$(b_0,b_n)$ to $(b_n,b_0)$. Using \ref{CLAIM-3}, there is a circuit $(d_0,d_1), \dots , (d_{n-1},d_n ),(d_n,d_0)$ with $(d_n,d_0 )=(b_0,b_n)$ where  $(d_i,d_{i+1}) \leadsto (b_i,b_{i+1} )$ (via a symmetric directed path according to \ref{CLAIM-1}) and $(d_n,d_0)$ is in the same strong component as 
$(b_n,b_0)$. This implies that $(b_n,b_0),(b_n,a_0)$ is a shorter circuit, a contradiction to the minimality of the circuit.
\end{proof}   

%

Recall that we have denoted by $C_i$ the (strong) component of $H^+$ containing the pair $(b_i,b_{i+1})$.
The following proposition is not used anywhere in the paper but it is interesting to know. 

\begin{proposition} \label{all-avoid-in-Di}
Let $T$ be a closure-dual-free set of unbalanced pairs, and assume that $\widehat{T}$ contains a circuit. 
Let $(b_0,b_1),(b_1,b_2),\dots,(b_{n},b_0)$ be any minimum circuit in $\widehat{T}$.  Then each component $C_i$ has all arcs symmetric.
\end{proposition}
\begin{proof} We will show that every directed path $W'$ in $C_i$ from some $(c,d)$ to $(b_i,b_{i+1})$ consists of two walks, $X$ 
from $c$ to $b_i$ and $Y$ from $d$ to $b_{i+1}$, that avoid each other.  Since every arc of $C_i$ lies on  such 
a walk, this proves the Proposition. By Theorem \ref{CLAIM} \ref{CLAIM-1}, we may assume $(b_i,b_{i+1})$ is an extremal pair in component $C_i$, $0 \le i \le n$.  

Recall that $(b_i,b_{i+1})$ lies on $D_i$ (which is a closed walk of positive net 
value). 
We also note that, by Theorem \ref{CLAIM} (\ref{CLAIM-2}) that $P_i,P_{i+1}$ are obtained by repeatedly following cycle $D_i$ in positive direction. 

Consider a directed path $W$ from $(b_i,b_{i+1})$ to $(c,d)$ in $C_i$ that has a negative net value and it is constricted from above. Such a directed path is obtained 
by starting at $(b_i,b_{i+1})$ and going around the cycle $D_i$ in negative direction sufficiently many times and then going to $(c,d)$. Now consider 
the directed path $W'$ going from $(c,d)$ to $(b_i,b_{i+1})$ and then following the directed path $W''$ around the cycle $D_i$ in the negative direction, so that 
$WW'W''$ is constricted and has negative net value, (again this can be obtained by going around $D_i$ in negative direction sufficiently many times). 
The directed path $WW'W''$ gives two constricted walks $A, B$ from some $b_i, b_{i+1}$ to $b_i, b_{i+1}$ respectively, where $A$ avoids $B$. Now let $C, D$ 
be two walks from $p, q$ to $b_{i+1}, b_{i+2}$ respectively, that avoid each other and have the same negative net value as $A$. (We may assume $p \in P_{i+1}$ and $ p \in P_{i+2}$ ). Now by Lemma \ref{strong-core1} we conclude that $A, B$ avoid each other, and 
hence, the walks $X$ and $Y$ constituting $W'$ avoid each other. This implies that $C_i$ has all arcs symmetric.  
\end{proof}

Following the proof of the Theorem \ref{itit}  one can obtain the following corollary. 

\begin{corollary}\label{on-the-same-level-circuit}
 
Let $T$ be a set of pairs in $H^+$ that is closure-dual-free.  Suppose $\widehat{T}$ contains a circuit. Let $(a_0,a_1),(a_0,a_1),(a_1,a_2),\dots,(a_{n},a_0)$  be any minimal circuit in $\widehat{T}$. 
  
If $(a_i,a_{i+1})$ is a $LL$-pair with respect to $\widehat{T}$ then let $(p_i,q_{i+1})$ be the second vertex on $Z_i$( a constricted walk from below with net value one that ends at $(a_i,a_{i+1})$) otherwise let $X_i$ 
be a constricted directed path from below of net value zero from $(p_i,q_{i+1}) \in \widehat{T}$ to $(a_i,a_{i+1})$ ($X_i$ could be just a path in strong component containing $(a_i,a_{i+1})$.  

Then there exists another circuit $(a''_0,a''_1), (a''_1,a''_2), \dots,  (a''_n,a''_0)$ of pairs, and walks $P'_i, Q'_i, i=0, \dots, n$,
in $H$, such that $P'_i, Q'_i$ are walks of net length zero, constricted from below, $P'_i$ from $a''_i$ to $a_i$, 
$Q_i$ from $a''_{i+1}$ to $a_{i+1}$, and such that $P'_i$ and $Q'_i$  avoid each other. Here each $(a''_i,a''_{i+1})$ is either $(p_i,q_{i+1})$ or $(q_{i},p_{i+1})$ or $(q_{i},q_{i+1})$ or $(p_i,p_{i+1})$. 
\end{corollary}

\begin{corollary}\label{not-imply-two}  
Let $T$ be a set of pairs. Suppose $T$ is closure-dual-free and assume that $\widehat{T}$ contains a circuit. Let $(a_0,a_1),(a_1,a_2),\dots,(a_{n},a_0)$  be any minimal circuit in $\widehat{T}$. Let $W_1$ be a directed path from $(p,q) \in T$ to $(a_i,a_{i+1})$ of net value zero and let $W_2$ be a directed path from $(p,q)$ to $(a_{i+1},a_{i+2})$ of net value zero.  Then at least one of the $W_1,W_2$ is not constricted from below. 
\end{corollary}

\begin{proof} For contradiction $W_1$ and $W_2$ both are constricted from below. 
We may assume $L_1$, the height of $W_1$ is at most $L_2$, the height of $W_2$. Now by Corollary \ref{on-the-same-level-circuit} and the proof of the Theorem \ref{itit}  we have $(a_{i},a_{i+1}),(q,q)$ are in the same strong component, a contradiction. Similarly if  
$L_2 < L_1$ then $(a_{i},a_{i+1})$ and $(p,p)$ are in the same strong component of $H^+$, a contradiction.  
\end{proof}

\subsection{Correctness of phase two} 

The goal of this subsection is to show that after adding pair $(p,r)$ (line \ref{line18}) into $V_c$ and computing $Tr(V_c)$ we don't encounter a circuit. To show this, we assume a minimal circuit occurs, and obtain some properties of such a circuit, and finally derive a contradiction. 


\begin{definition}[original pair, 1-implied pair, chain]
Pair $(p,q)$ in the Algorithm \ref{alg-main} line \ref{line17}  is called an original pair. Any pair $(u,v)$ that is reachable from an original pair is called 
{\em $1$-implied} pair. We say a pair $(u,v) \in V_c$ is by transitivity if there exist 
\begin{align*}
    (u,u_1),(u_1,u_2),\dots,(u_{m-1},u_m),(u_m,v) \in V_c. 
\end{align*}

We say pairs $(x_1,y_1),(x_2,y_2),...,(x_n,y_n)$ form a chain of pairs (between $x_1$, $x_{n+1}$) or simply a chain when $y_i=x_{i+1}$ for every $1 \le i \le n$. 

\end{definition}
\begin{definition}[depth of a pair]
Let $Ch : (a_0,a_1),(a_1,a_2),\dots,(a_{n-1},a_n)$ be a chain  where each $(a_i,a_{i+1}) \in V_c$. 
The {\it depth} of a pair $(a_i,a_{i+1})$ is defined as follows. Before handling the pairs on layer $k$, the depth of the existing pairs in $V_c$ is zero. If $(x,y) \leadsto (x',y')$ then the depth of $(x,y),(x',y')$ are the same. If $(x_0,x_{n})$ is by transitivity on $(x_0,x_1),(x_1,x_2),\dots,(x_{n-1},x_n)$ then the depth of $(x_0,x_n)$ is one plus the maximum depth of $(x_j,x_{j+1})$ where $(x_j,x_{j+1})$, for some $0 \le j \le n-1$, has the maximum depth among $(x_0,x_1),(x_1,x_2),\dots,(x_{n-1},x_n)$.

\end{definition}

\begin{definition}(minimal chain)
Let $Ch : (a_0,a_1),(a_1,a_2),\dots,(a_{n-1},a_n)$, $n>1$ be a chain where each $(a_i,a_{i+1}) \in V_c$.  We say $Ch$ is a {\em minimal chain} if

\begin{itemize}
    \item  no pair $(a_i,a_{i+1})$, $0 \le i \le n-1$ is by transitivity.
    \item the depth of $Ch$ is minimum; the depth of $Ch$ is the maximum depth of $(a_i,a_{i+1})$, $0 \le i \le n$. 
\end{itemize}

\end{definition}

\begin{lemma} \label{at-most-1-implied}
Let $Ch: (y_0,y_1),(y_1,y_2),\dots,(y_{m-1},y_m)$, $m>1$ be a minimal chain in $V_c$ which is currently circuit free. For each pair $(y_i,y_{i+1})$, $0 \le i \le m-1$, let $X_i=(E_i,F_{i+1})$ be a directed path in $H^+$ with net value zero from a pair $(p_i,q_{i+1}) \in V_c$ to $(y_i,y_{i+1})$ where $E_i$ is a constricted from below walk with net length zero from $p_i$ to $y_i$,  and avoiding $F_{i+1}$ which is a walk from $q_{i+1}$ to $y_{i+1}$.   
Let $h(X_i)$ denote the height of $X_i$. 
Then the following hold.
\begin{itemize}
    \item [1.] Suppose $(y_i,y_{i+1})$ is an $1$-implied pair reachable from an original pair $(p_i,q_{i+1})$. Then $h(X_i) > h(X_j)$, $ j \ne i$. \label{1-lemma-chain}
    \item [2.] At most one of the $(y_i,y_{i+1})$, $0 \le i \le m$ is an $1$-implied pair. \label{2-lemma-chain}

   \item [3.] Suppose $(y_i,y_{i+1}),(y_{i+1},y_{i+2})$, $1 < i \le m-2$ are $LL$-pairs with respect to $V_c$ (see Definition~\ref{LL-pair}) such that $h(X_i) \le  h(X_{i+1})$. Then $(y_{i-1},y_i)$ is also an $LL$-pair, and $h(X_{i-1}) < h(X_i)$. \label{3-lemma-chain}
   
   \item [4.] Suppose $(y_i,y_{i+1}),(y_{i+1},y_{i+2})$, $1 < i \le m-2$ are $LL$-pairs (with respect to $V_c$) such that $h(X_i) \ge  h(X_{i+1})$. Then $(y_{i-1},y_i)$ is also an $LL$-pair, and $h(X_{i+1}) > h(X_{i+2})$. \label{4-lemma-chain}
   
   \item [5.] Suppose $(y_i,y_{i+1}),(y_{i+2},y_{i+3})$ are $LL$-pairs (with respect to $V_c$). Then  
   \[\min\{h(X_{i}),h(X_{i+2})\} < h(X_{i+1}) < \max \{h(X_i),h(X_{i+2})\},\] 
   and $(y_{i+1},y_{i+2})$ is also an $LL$-pair. \label{5-lemma-chain}
\end{itemize}
\end{lemma}
\begin{proof} {\em Proof of 1.}
Since $V_c$ does not have a circuit, $(p_j,q_{j+1}) \not\leadsto (y_{i+1},y_i)$ and $(p_i,q_{i+1}) \not\leadsto (y_{j+1},y_j)$ (Figure \ref{h(xi)>h(xj)}). Moreover, since $Ch$ is a minimal chain, $(p_j,q_{j+1}) \not\leadsto (y_i,y_{j+1})$, and $(p_i,q_{i+1}) \not\leadsto (y_i,y_{j+1})$. 
For contradiction assume $h(X_i) \le h(X_j)$. Let $E_i=A_i+A'_{i}$, ($+$ is the concatenation) where $A_i$ is walk from $p_i$ to $g_i$, and $A'_{i}$ is a walk from $g_i$ to $y_i$; $g_i$ is a vertex on $E_i$ with a maximum height and $(A'_{i})^{-1}$ is constricted. Let $F_{i+1}=B_{i+1}+B'_{i+1}$ where $A_i$ avoids $B_{i+1}$ and $A'_i$ avoids $B'_{i+1}$, and let $h_{i+1}$ be the corresponding vertex to $g_i$ on $F_{i+1}$. Let $C=A'_{j}[g,y_j]$ so that $C$ is constricted and have the same net length as $A'_i$, and let $D=B'_{j+1}[h,y_{j+1}]$ where $C$ avoids $D$. 

Now, by applying Lemma \ref{strong-core1} on walks $A'_i,B'_{i+1},C,D$ (see Figure \ref{h(xi)>h(xj)}), we conclude that $C,D$ have  embedded pre-images that avoid each other, and hence, $C,D$ avoid each other. Now by applying Lemma \ref{M} on walks $A'_i,A^{-1}_i,B'_{i+1},(B_{i+1})^{-1},C,C^{-1},D,D^{-1}$, we conclude that $E_i,F_{i+1}$ avoid each other. Moreover, $(p_i,y_{j+1}) \leadsto (y_i,y_{j+1})$. We note that $(y_{j+1},y_i) \not\in V_c$. Notice that $(y_{j+1},p_i) \not\in V_c$, as otherwise, since $(y_{i+1},p_i) \leadsto  (y_{j+1},y_j)$ we would have $(y_{j+1},y_j) \in V_c$; a circuit in $V_c$. 
Now according to the rules of the Algorithm \ref{alg-main} lines \ref{line17}, \ref{line18}, $(p_i,y_{j+1}) \in V_c$ should have been added into $V_c$ before $(p_j,q_{j+1})$, contradicting the minimality of the chain $Ch$; unless $(y_{j+1},p_i)$ is already in $V_c$ which is not possible. Notice that when $p_i=p_j$ again $(p_i,y_{j+1})$ is a pair that should be added into $V_c$ (because we consider any circuit after running line \ref{cotinue-with-p0}) which gives a shorter chain). The argument for the case $h(X_j) < h(X_i)$ is analogous. This proves the first premise of the lemma.  \\

\begin{figure}
\begin{center}
\includegraphics[height=6cm,width=14cm]{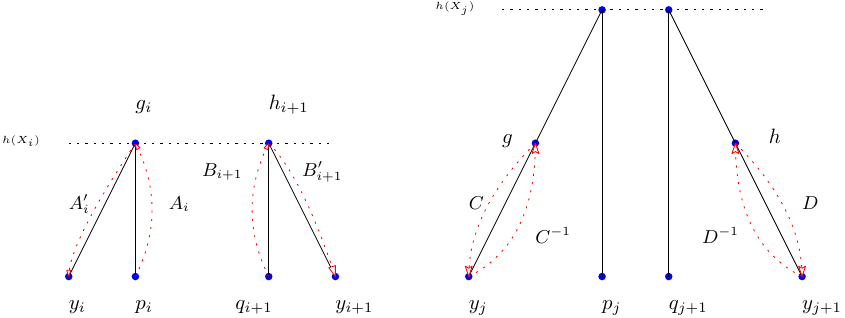}
\caption{In the proof of Lemma \ref{at-most-1-implied} (1) assuming $h(X_i) < h(X_j)$, $A_i$ avoids $B_{i+1}$, $A'_i$ avoids $B'_{i+1}$. $C$ avoids $D$ and $C,A'_i$ are constricted and have the same net length.} 
\label{h(xi)>h(xj)}
\end{center}
\end{figure}

\noindent {\it {Proof of 2.}} According to (1) we must have $h(X_i) > h(X_j)$. Now since $(y_j,y_{j+1})$ is also an $1$-implied pair, we have $h(X_j) > h(X_i)$, a contradiction. \\

\noindent {\it {Proof of 3.}} 
For $j=i,i+1$, let $(p'_j,q'_{j+1}) \in V_c$ such that $(p'_j,q'_{j+1})(p_j,q_{j+1})$ is an arc of $H^+$, and $p'_jp_{j},q'_{j+1}q_{j+1} \in A(H)$ (see Figure \ref{leftright1}) . For contradiction first  assume that  $(y_{i-1},y_i)$ is an $1$-implied pair. Now by (1) we have $h(X_{i-1}) > h(X_i),h(X_{i+1})$. Let $X_j=(E_j,F_{j+1})$, $j=i-1,i,i+1$, where $E_j$ avoids $F_{j+1}$. Since $Ch$ is a minimal chain, similar to the proof of (1), by applying Lemma \ref{strong-core1} and Lemma \ref{M} on appropriate portion of $E_{i-1},F_{i},E_{i+1},F_{i+2}$ together with $E_i,F_{i+1}$ (see Figure \ref{leftright1}) we conclude that :

\begin{itemize}
    \item $E_i,F_{i+1}$ avoid each other,
    \item $E_{i+1},F_{i+2}$ avoid each other,
    \item $(p_i,q_{i+2}),(y_i,y_{i+2})$ are in the same strong component of $H^+$,
    \item $(p_i,q_{i+1}),(p_i,p_{i+1}),(y_i,y_{i+1})$ are in the same strong component $H^+$,
    \item $(q_{i+1},q_{i+2}),(y_{i+1},y_{i+2})$ are in a same strong component of $H^+$. 
\end{itemize}

(1) $p'_{i+1}p_i \not\in A(H)$ as otherwise, $(p'_{i+1},q'_{i+2})(p_i,q_{i+2})$ is an arc of $H^+$ and since $(p_i,q_{i+2}),(y_i,y_{i+2})$ are in the same strong component, we have $(p'_{i+1},q'_{i+2}) \leadsto (y_i,y_{i+2})$, a contradiction to the minimality of $Ch$ (see Figure \ref{leftright1}).  

(2) Note that $q'_{i+2}p_i \not\in A(H)$. Otherwise, $(p'_{i+1},q'_{i+2})(p_{i+1},p_i)$ is an arc of $H^+$  because $p'_{i+1}p_i \not\in A(H)$, and hence,  $(p'_{i+1},q'_{i+2}) \leadsto (y_{i+1},y_i)$ ( because $(p_i,p_{i+1}),(y_i,y_{i+1})$ are in the same strong component); yielding a circuit in $V_c$.  

(3) We observe that $p'_iq_{i+2} \not\in A(H)$, as otherwise, $(p'_i,q'_{i+1})(q_{i+2},q_{i+1}) \in A(H^+)$, and hence, $(p'_i,q'_{i+1}) \leadsto (y_{i+2},y_{i+1})$ ( because $(p_i,p_{i+1}),(y_i,y_{i+1})$ are in the same strong component); yielding a circuit in $V_c$.  

Observe that $(p'_i,q'_{i+2})(p_i,q_{i+2})$ is an arc of $H^+$. Now $q'_{i+2}q_{i+1} \not\in A(H)$, as otherwise, $(y_i,y_{i+1}),$ $(p_i,q_{i+1})$ ,$(p'_i,q'_{i+2})$ are in the same strong component, and hence, $(p'_i,q'_{i+1}) \leadsto (p_i,q_{i+1}) \leadsto (p'_i,q'_{i+2}) \leadsto (p_i,q_{i+2}) \leadsto (y_i,y_{i+2})$. This is a contradiction to the minimality of the chain $Ch$. Moreover, $q'_{i+1}q_{i+2} \not\in A(H)$, as otherwise,   $(p'_i,q'_{i+1}) \leadsto (p_i,q_{i+2}) \leadsto (y_i,y_{i+2}) \in V_c$; a contradiction to the minimality of the chain $Ch$.

Now $(q_{i+1},q_{i+2}),(q'_{i+1},q'_{i+2})$ are in the same strong component. However, since \[(p'_{i+1},q'_{i+2}) \leadsto (y_{i+1},y_{i+2}) \leadsto (q_{i+1},q_{i+2}) \leadsto (q'_{i+1},q'_{i+2}),
\]
we conclude that $(q'_{i+1},q'_{i+2})$ is in $V_c$ and on a lower layer. Therefore, by transitivity rule of the algorithm, we should have $(p'_i,q'_{i+2}) \in V_c$, and hence, $(p_i,q_{i+2}) \in V_c$, and consequently $(y_i,y_{i+2}) \in V_c$. This is a contradiction to the minimality of the chain $Ch$. \\

\begin{figure}
\begin{center}
\includegraphics[height=5cm]{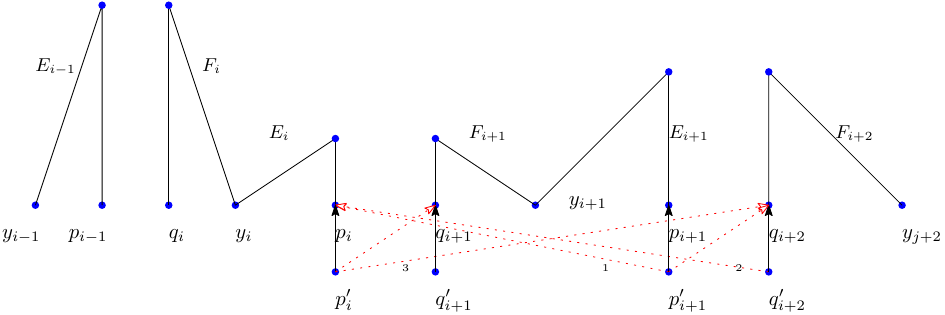}
\caption{Lemma \ref{at-most-1-implied} (3) suppose for contradiction $h(X_{i-1})> h(X_{i}), h(X_{i+1})$. $E_{i-1}$ avoids $F_i$, $E_i$ avoids $F_{i+1}$, and $E_{i+1}$ avoids $F_{i+2}$. The dash arcs are missing. We first show $p'_{i+1}p_i$ is a missing arc, then $q'_{i+1}p_i$ is a missing arc, and finally $p'_iq_{i+2}$ is a missing arc.} 
\label{leftright1}
\end{center}
\end{figure}

\noindent {\it {Proof of 4.}} It is analogous to proof of 3.\\

\noindent {\it {Proof of 5.}} 
For $j =i,i+2$, let $(p'_j,q'_{j+1}) \in V_c$ such that $(p'_j,q'_{j+1})(p_j,q_{j+1})$ is an arc of $H^+$. For contradiction, assume $h(X_{i+1}) > h(X_{i+2}),h(X_i)$ as depicted in figure \ref{leftright}.

First assume $h(X_i) \ge h(X_{i+2})$. By applying Lemma \ref{strong-core1} and Lemma \ref{M}, on appropriate portion of $E_i,F_{i+1},E_{i+1},F_{i+2},E_{i+2},F_{i+3}$ we conclude that $E_i,F_{i+1}$ avoid each other, and $E_{i+2},F_{i+3}$ avoid each other. Moreover,   $(q_{i+1},p_{i+2}),(y_{i+1},y_{i+2})$ are in the same strong component of $H^+$,  $(p_i,p_{i+2}),(y_i,y_{i+2})$ are in the same strong component of $H^+$, and $(q_{i+1},p_{i+2}),(y_{i+1},y_{i+2})$ are in the same strong component of $H^+$. 

(1) $p'_ip_{i+2} \not\in A(H)$, as otherwise, $(p'_i,q'_i)(p_{i+2},q_{i+1})$ is an arc of $H^+$, and hence $(p'_i,q'_{i+1})$ $\leadsto$ $(p_{i+2},q_{i+1})$ $\leadsto$  $(y_{i+2},y_i)$, implying a circuit in $V_c$ (see Figure \ref{leftright}).
\begin{figure}
\begin{center}
\includegraphics[height=6cm,width=14cm]{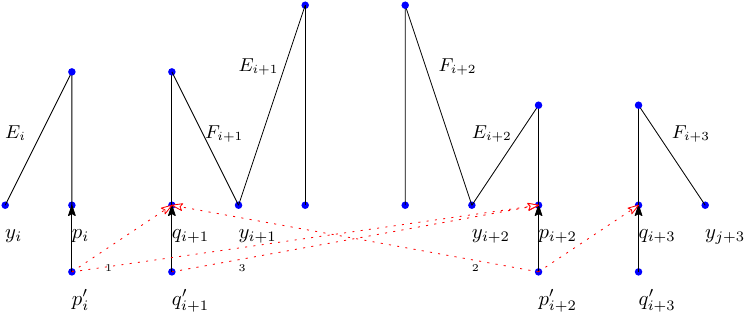}
\caption{
Lemma \ref{at-most-1-implied} (5) suppose for contradiction $h(X_i),h(X_{i+2}) < h(X_{i+1})$.  $E_{i}$ avoids $F_{i+1}$, $E_{i+1}$ avoids $F_{i+2}$, and $E_{i+2}$ avoids $F_{i+3}$. The dash arcs are missing. We first show $p'_ip_{i+2}$ is a missing arc, then $p'_{i+2}q_{i+1}$ is a missing arc, and finally $q'_{i+1}p_{i+2}$ is a missing arc.} 
\label{leftright}
\end{center}
\end{figure}
Notice that since $(q_{i+1},y_{i+2}),(y_{i+1},y_{i+2})$ are in the same strong component, we observe that $(y_0,y_1),\dots,(y_i,q_{i+1}) \\,(q_{i+1},y_{i+2}),(y_{i+2},y_{i+3}), \dots,(y_{m-1},y_m)$ is also a chain in $V_c$. However, by applying Lemma \ref{M} on this new circuit, we conclude that $(q_{i+1},q_{i+2})$, $ (q_{i+1},y_{i+3})$ are in the same strong component.  

(2) $p'_{i+2}q_{i+1} \not\in A(H)$, as otherwise, $(p'_{i+2},q'_{i+3})(q_{i+1},q_{i+3}) \in A(H^+)$, and hence, $(p'_{i+2},q'_{i+3}) \leadsto (q_{i+1},y_{i+3})$, yielding  a shorter chain. 

(3) $q'_{i+1}p_{i+2} \not\in A(H)$, as otherwise, $(p'_i,q'_{i+1})(p_i,p_{i+2})$, and hence, $(p'_i,q'_{i+1}) \leadsto (p_i,p_{i+2}) \leadsto (y_i,y_{i+2})$, a shorter chain. 

Now $(q'_{i+1},p'_{i+2}),(y_{i+1},y_{i+2})$ are in the same strong component, and hence, $(q'_{i+1},p'_{i+2}) \in V_c$. Since $(p'_i,q'_{i+1}),(q'_{i+1},p'_{i+2}) \in V_c$ and both on a lower layer, $(p'_i,p'_{i+2}) \in V_c$. However, because $(p'_i,p_{i+2}) \leadsto (p_i,p_{i+2}) \leadsto (y_i,y_{i+2})$ we get a shorter chain, a contradiction.

The same argument is applied when $ h(X_i) < h(X_{i+2})$. 
\end{proof}

\begin{lemma}\label{chain-structure}
Let $(a_i,a_{i+1}) \in V_c$ be a pair reachable from $(y_0,y_m) \in V_c$ via a symmetric directed path is constricted from below and has net value zero. 
Let $Ch: (y_0,y_1),(y_1,y_2),\dots,(y_{m-1},y_m)$, $m>1$, be a minimal chain in $V_c$ where $V_c$ is circuit free. Then $(a_i,a_{i+1})$ is by transitivity on the pairs $(a_i,b_1),(b_1,b_2),\dots,(b_{r-1},b_r),(b_r,a_{i+1})$ in $V_c$ where each of them is either an $LL$-pair or $1$-implied pair. 
\end{lemma}
\begin{proof}
We use induction on the depth of $(a_i,a_{i+1})$. For each pair $(y_i,y_{i+1})$, $0 \le i \le m-1$, let $X_i$ be a
directed path in $H^+$ from $(p_i,q_{i+1}) \in V_c$ to $(y_i,y_{i+1})$ which is constricted from below and has net value zero. Let $h(X_i)$ denote the height of $X_i$. We denote the $X_i=(E_i,F_{i+1})$ where $E_i$ is a constricted from below walk from $p_i$ to $y_i$, with net length zero, and avoiding $F_{i+1}$.\\ 

\noindent {\textbf{Base of Induction:}} First assume that each pair in $Ch$ is either an $LL$-pair or is $1$-implied pair. In other worlds, the depth of $(a_i,a_{i+1})$ is $2$. When $(y_i,y_{i+1})$, $0 \le i \le m-1$, is an $LL$-pair then let $(p'_i,q'_{i+1}) \in V_c$ such that $(p'_i,q'_{i+1})(p_i,q_{i+1})$ is an arc in $H^+$ ($p'_ip_i,q'_{i+1}q_{i+1} \in A(H)$, $p'_iq_{i+1} \not\in A(H)$).  
Let $X : (y_0,y_m) \leadsto (a_i,a_{i+1})$ be the symmetric path as stated in the Lemma. Then the following hold.

\begin{enumerate}
    \item $(y_m,y_0) \not\leadsto (y_1,y_0)$, as otherwise, $(y_0,y_1) \leadsto (y_0,y_m)$ and this contradicts the minimality of the chain $Ch$.
   
   \item $(y_m,y_0) \not\leadsto (y_m,y_1)$, as otherwise, $(y_1,y_m) \leadsto (y_0,y_m) \leadsto (a_i,a_{i+1})$, and this contradicts the minimality of the chain, and our assumption about $(a_i,a_{i+1})$. 
   
   \item $(p_0,q_1) \not\leadsto (y_m,y_1)$, as otherwise, we would get a circuit $(y_1,y_2),\dots,(y_{m-1},y_m),(y_m,y_1)$ in $V_c$, a contradiction.  
   
   \item $(p_0,q_1) \not\leadsto (y_0,y_m)$, as otherwise, it contradicts the minimality of the chain $Ch$.
  
    \item $(y_{m-1},y_m) \not\leadsto (y_{m-1},y_0)$, as otherwise, $(y_0,y_{m-1}) \leadsto (y_m,y_{m-1})$, and we get a circuit in $V_c$, a contradiction.   
   
    \item $(y_{m-1},y_m) \not\leadsto (y_0,y_m)$, as otherwise, $(y_m,y_0) \leadsto (y_m,y_{m-1})$, and 
    we get a circuit in $V_c$ , a contradiction. 
    
    \item $(p_{m-1},q_{m}) \not\leadsto (y_0,y_m)$, as otherwise,  it contradicts the minimality of $Ch$.
    
     \item $(p_{m-1},q_{m}) \not\leadsto (y_{m-1},y_0)$, because of the minimality of the chain $Ch$.

\end{enumerate}

\begin{figure}
\begin{center}
\includegraphics[scale=0.8]{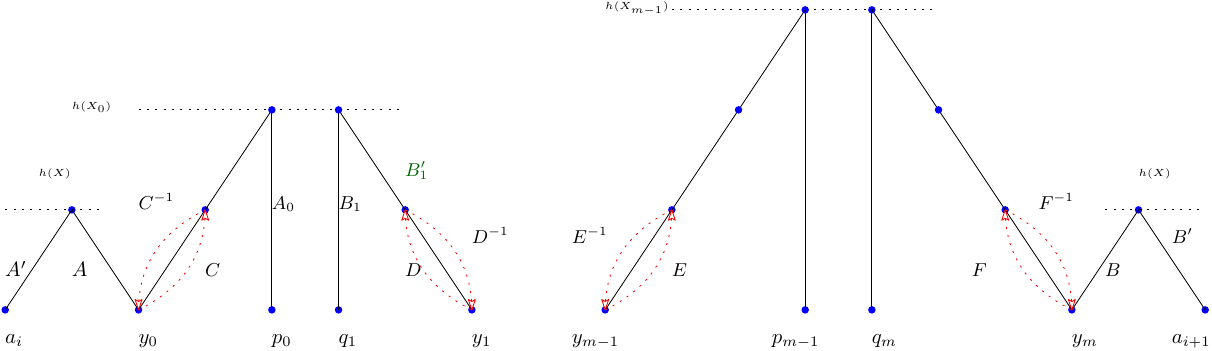}
\caption{In Lemma \ref{chain-structure}, assume $h(X_0),h(X_{m-1}) > h(X)$, here $X=(A+A',B+B')$ where $A+A'$, and $B+B'$ avoid each other. $A,B$, $A'^{-1},B'^{-1},C,D,E,C^{-1},D^{-1},E^{-1},F^{-1}$ are constricted and have the same net length. $C$ avoids $D$, and $E$ avoids $F$.} 
\label{chain-basic-pairs}
\end{center}
\end{figure}

First suppose $h(X_{m-1}) \ge h(X)$ and $h(X_0) \ge h(X)$. According to (1,2,3,4) by Lemma \ref{strong-core1}, and Lemma \ref{M} on the four walks inside $X_0,X$ (in Figure \ref{chain-basic-pairs}, $A,A',B,B',C,C^{-1},D,D^{-1}$), we conclude  that $(y_0,y_1) \leadsto (a_i,y_1)$. Similarly by considering (5,6,7,8) and applying Lemmas \ref{strong-core1}, \ref{M} on the walks inside $X_{m-1},X$ (in the Figure \ref{chain-basic-pairs}, $A,A',B,B', E,E^{-1},F,F^{-1}$), we conclude that $(y_{m-1},y_m) \leadsto (y_{m-1},a_{i+1})$. Thus, we obtain the chain $(a_i,y_1),(y_1,y_2), \dots,(y_{m-2},y_{m-1}),\\ (y_{m-1},a_{i+1})$, and the lemma holds. 

Next we continue by assuming that $\min \{h(X_0),h(X_{m-1})\}  < h(X)$. We prove the lemma when $h(X_0) \le h(X_{m-1})$ (the argument for the other case is similar). This assumption together with Lemma \ref{at-most-1-implied} (1,2) imply that $(y_0,y_1)$ is not an original pair, and hence, it is an $LL$-pair.

\begin{claim}\label{1-claim}
We show that $m=2$, and $(y_1,y_2)$ is an $1$-implied pair. 
\end{claim}
\begin{proof}
First suppose $(y_1,y_2)$ is an original pair. Now according to (1) we have $h(X_1) >h(X_0)$, moreover, by similar argument as  in Lemma \ref{at-most-1-implied}(3), if $(y_2,y_3)$ exits then it is not an $LL$-pair, and according to item (1) of the Lemma \ref{at-most-1-implied}, $(y_2,y_3)$ is not an $1$-implied pair. Therefore, $m=2$ and $(y_1,y_2)$ is an $1$-implied pair, and the claim is proved in this case.  

Thus, we continue by assuming $m>2$ and that $(y_1,y_2)$ is an $LL$-pair. 
Let $1< j \le m-1$ be the smallest subscripts such that $(y_j,y_{j+1})$ is an $1$-implied pair. Suppose such a $j$ exists. Now according to (1,2,3,4) from Lemma \ref{at-most-1-implied} we conclude that $h(X_0)< h(X_1) < \dots < h(X_{m-1})$ and all the pairs $(y_0,y_1),\dots,(y_{m-2},y_{m-1})$ are $LL$-pairs and $(y_{m-1},y_m)$ is an $1$-implied pair.  If $j$ doesn't exist then all the pairs $(y_0,y_1),(y_1,y_2),\dots,(y_{m-1},y_m)$ are $LL$-pairs, and $h(X_0)<h(X_1) \dots < h(X_{m-1})$.  

In any case by the argument in Lemma \ref{at-most-1-implied} (3) we conclude that $h(X) < h(X_1)$, and hence, $h(X) < h(X_{m-1})$. Now in this case (as seen before) by applying the Lemmas \ref{strong-core1}, \ref{M} on appropriate segments of the walks inside $X_0,X_1,X$, we conclude that $(y_0,y_1)$ and $(a_i,y_1)$ are in the same strong component of $H^+$. Moreover, by applying the Lemmas \ref{strong-core1}, \ref{M} on appropriate segments of the walks inside  $X,X_{m-1},X_{m-2}$, we conclude that $(y_{m-1},y_m)$ and $(y_{m-1},a_{i+1})$ are in the same
strong component of $H^+$. Therefore, $(a_i,y_1),(y_1,y_2),\dots,(y_{m-1},a_{i+1})$ is the required chain in the lemma and the lemma holds. Thus, we may assume that $m=2$, and hence, the claim is proved. 
\end{proof}

We continue by assuming $m=2$. First assume that $h(X_1) > h(X)$. Now again similar to the argument in the proof of the Claim \ref{1-claim}, by applying Lemma \ref{strong-core1} and Lemma \ref{M} on the walks inside $X_0,X_1,X$, we conclude that $(y_0,y_1),(a_i,y_1)$ are in the same strong component of $H^+$, and $(y_1,y_2),(y_1,a_{i+1})$ are in the same strong component of $H^+$; a contradiction to minimality of the chain $Ch$. 
Therefore, $h(X) > h(X_1)$. Now in this case again using the same application of Lemma \ref{strong-core1}, \ref{M}, we conclude that $(y_0,y_1)$ and $(y_0,p_1)$ are in the same strong component of $H^+$. This would imply that $(y_0,p_1)$ is an $LL$-pair because $(y_0,y_1)$ is an $LL$-pair. However, this is a contradiction to the choice of $p_1$, as it implies that $(y_0,p_1) \in V_c \cap L_k$. \\

\begin{remark}
By applying Lemma \ref{M} on $X_i,X_{i+1}$ when $h(X_i) \le h(X_{i+1})$, we conclude that $X_i$ is a symmetric path. Similarly if $h(X_i) \ge h(X_{i+1})$ then $X_{i+1}$ is symmetric. 
\end{remark}

\noindent \textbf{Induction hypothesis: }
Suppose some $(y_i,y_{i+1})$ is not an original pair. If $h(X_i) \le h(X_{i+1})$ or $h(X_i) < h(X_{i-1})$ then $X_i$ is symmetric, and hence, by induction hypothesis $(y_i,y_{i+1})$ is by transitivity on pairs where each is either an original pair or is an $LL$-pair. Otherwise, suppose $h(X_i) > h(X_{i+1}),h(X_{i-1})$. We show that this is not possible as follows. Note that $X_{i-1},X_{i+1}$ are symmetric, and by induction hypothesis, 
$(y_{i-1},y_i)$ is by transitivity on original pairs or $LL$-pairs. So we may assume each of the $(y_{i-1},y_i)$ and $(y_{i+1},y_{i+2})$  is either an $LL$-pair or is an original pair. Note that by Lemma \ref{at-most-1-implied} none of the $(y_{i-1},y_i)$ and $(y_{i+1},y_{i+2})$ is an original pair, and now this is a contradiction according to Lemma \ref{at-most-1-implied} (3).  
\end{proof}

\begin{lemma}\label{special-M}
Let $X,Y$ be two directed paths that are constricted from below with net value zero in $H^+$. Suppose $X$ starts from $(p,q)$ and reaches $(a,b)$ where $p=b$ or $q=a$. Suppose $Y$ starts from $(r,s)$ and reaches $(b,c)$. Furthermore, assume that $h(X) \leq h(Y)$ (here $h(X)$ is the height of $X$). Then one of the following occurs. 
\begin{enumerate}
    
     \item $(r,s) \leadsto (a,c)$; via a directed path that is constricted from below and has net value zero. 
     
     \item $(r,s) \leadsto (b,a)$; via a directed path that is  constricted from below and has net value zero,
     
    \item $(p,q) \leadsto (a,c)$; via a directed path that is  constricted from below and has net value zero, 
    \item $(p,q) \leadsto (c,b)$; via a directed path that is  constricted from below and has net value zero.

\end{enumerate}
\end{lemma}
\begin{proof}
Let $(g_1,h_1)$,  be a vertex on $X$ with the maximum height, and let $(g_2,h_2)$, be a vertex on $Y$ with the maximum height. 
Let $X=(A+A',B+B')$ where $A$ starts from $p$ and ends at $g_1$, and $A'$ starts from $g_1$ and ends at $a$. $B$ starts from $q$ and ends at $h_1$ ($h_1$ corresponding to $g_1$), and $B'$ starts from $h_1$ and ends at $b$. Notice that $A+A'$ avoids $B+B'$.

Let $Y=(C+C',D+D')$ where $C$ starts from $r$ and ends at $g_2$, and $C'$ starts from $g_2$ and ends at $b$. $D$ starts from $s$ and ends at $h_2$ ($h_2$ corresponding to $g_2$), and $D'$ starts from $h_2$ and ends at $c$. Note that $A,C$ are constricted and $(A')^{-1},C'^{-1}$ are also constricted. 

By assumption of the lemma we have $h(X) \leq h(X)$ so, let $(g,h)$ be a vertex on $Y$ such that $E=C'[g,b]$ is constricted and have the same net length as $A'$, and $F=D'[h,c]$ is constricted and have the same net length as $B'$ (see Figure \ref{7.15}). 

\begin{figure}
\begin{center}
\includegraphics[scale=1.0]{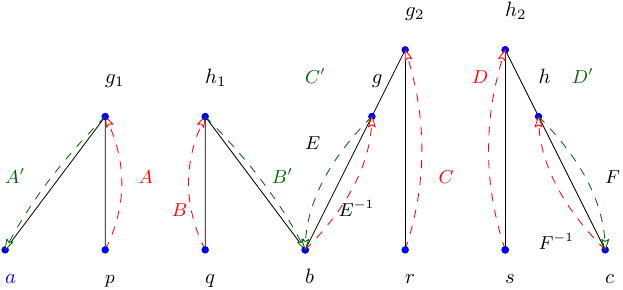}
\caption{In Lemma \ref{special-M}, assume $h(X) \le h(Y)$; height of $X$ is smaller than the height of $Y$ which is the same as the height of $A$. Here $X=(A+A',B+B')$, where $A+A'$ avoids $B+B'$, and $Y=(C+C',D+D')$ where $C+C'$ avoids $D+D'$,  $A',B', A^{-1},B^{-1},E,F,E^{-1},F^{-1}$ are constricted, and $E$ avoids $F$.} 
\label{7.15}
\end{center}
\end{figure}

Suppose none of the 1,2,3,4 occurs. Now we can apply Lemma \ref{strong-core1} on walks  $A',B',E,F$, and hence, conclude that $E,F$ have congruent embedded pre-images that avoid each other. Now Lemma~\ref{M} is applied to the walks $A,A',B,B',E,E^{-1},F,F^{-1}$ (in Figure \ref{7.15}), and hence, we conclude that $(B)^{-1}$ and $A'$ have (congruent) embedded pre-images that avoid each other, and $B',(A)^{-1}$ have (congruent) embedded pre-images that avoid each other. But, this is not possible because when $p=b$, the endpoint of $B'$ and $(A)^{-1}$ are the same and they cannot avoid each other. Similarly, we get a contradiction when $q=a$. Therefore, one of the (1), (2), (3), (4) should occur. Notice that the existence of a directed path constricted from below with net value zero, follows from using some faithful arc between one of the pairs of walks ($A+A'$, $E^{-1}+E$), ($A+A'$, $F^{-1}+F$), and ($B+B'$, $F^{-1}+F$) in the proof of the Lemma ~\ref{strong-core}.
\end{proof}

\begin{lemma}\label{special-M1}
Let $X,Y$ be two constricted from below with net value zero in $H^+$. Suppose $X$ starts from $(p,q)$ and reaches $(a,b)$. Suppose $Y$ starts from $(r,s)$ and reaches $(b,c)$ where $r=c$ or $s=b$. Furthermore, assume that $h(Y) \leq h(X)$. 

Then one of the following occurs. 
\begin{enumerate}
     \item $(r,s) \leadsto (a,c)$; via a directed path that is  constricted from below and has net value zero. 
    
     \item $(r,s) \leadsto (b,a)$; via a directed path that is  constricted from below and has net value zero,
    
     \item $(p,q) \leadsto (a,c)$; via a directed path that is constricted from below and has net value zero,

    \item $(p,q) \leadsto (c,b)$; via a directed path that is  constricted from below and has net value zero,

\end{enumerate}
\end{lemma}

\subsubsection{Proofs of Lemmas~\ref{p-line17-main-body} and \ref{p-line18-main-body}}

\begin{lemma}[Lemma~\ref{p-line17-main-body}, paraphrased]\label{p-line17}
Suppose $V_c$ does not contain a circuit before executing Line \ref{line17}, and furthermore,  $L_{k} \cap R$ is not empty (line \ref{line16}). Then there exists a vertex $p$ in line \ref{line17} and there exists a pair $(p,r)$ on line \ref{line18}.  


\end{lemma}
\begin{proof}
We construct digraph $G'=(V,A)$ as follow:
\begin{itemize}
    \item $V(G')= \{p \in V(H) \mid (p,x) \in R \cap L_k \}$
    \item $A(G')=\{ xp \mid (x,p) \in V_c \cap L_k \text{ or } (p,x) \in L_k \text{ with }
    (p,x) \leadsto (x,p) \}$.
\end{itemize}

If there exists a vertex $p$ in $G'$ with in-degree zero then $p$ is the desired vertex. Otherwise, there exists a directed cycle $v_0,v_1,v_2,\dots,v_n,v_0$ in $G'$.
Now this means there exists a circuit $C_1:(v_0,v_1),(v_1,v_2),\dots,(v_n,v_0)$ so that each  $(v_i,v_{i+1})$ is in $V_c \cap L_k$ or $(v_{i+1},v_i) \in L_k$ with $(v_{i+1},v_i) \leadsto (v_i,v_{i+1})$. Now we further relax the conditions on the pairs of $C_1$ and we may assume there exists a circuit $C:(x_0,x_1),(x_1,x_2),\dots,(x_n,x_0)$ such that each $(x_i,x_{i+1})$ is either a pair in $V_c$ or $(x_i,x_{i+1}) \not\in V_c$ and $(x_{i+1},x_i) \leadsto (x_{i},x_{i+1})$ and $(x_{i+1},x_i) \in  L_k$. Notice that if $(x_i,x_{i+1}) \in V_c$ then $(x_{i+1},x_i)$ in $V_d$, and hence, $(x_{i+1},x_i) \not\in R$. Let $X_i$ be a directed path in $H^+$ from $(x_{i+1},x_i)$ to $(x_i,x_{i+1})$ (if exists). Notice that by the definition of the layers and because $(x_{i+1},x_i) \in L_k$, $X_i$ is constricted from below. Otherwise, consider a  pair $(a,b)$ on $X_i$, where $X[(x_{i+1},x_i),(a,b)]$ has net value less than zero. Notice that $X_i : (x_{i+1},x_i) \leadsto (a,b) \leadsto (x_i,x_{i+1})$, and by skew property, we have $(x_{i+1},x_i) \leadsto (b,a) \leadsto (x_i,x_{i+1})$. Now since $(a,b)$ is on a lower layer, either $(a,b)$ or $(b,a)$ has been placed in $V_c$, and hence $(x_i,x_{i+1})$ is already in $V_c$, a contradiction to our assumption.

Now suppose $X_i$ has positive net value. Similarly, if $X_i$ has net value greater than zero, then by skew property, the reverse of $X_i$, $X^{-1}_i$, which is also a path from $(x_{i+1},x_i)$ to $(x_i,x_{i+1})$ has negative net value, and hence, there exists some $(a,b) \in X^{-1}_i$, so that $(a,b)$ placed on a lower layer than $(x_{i+1},x_i)$. This means either $(a,b) \in V_c$ or $(b,a) \in V_c$. Now again this means $(x_i,x_{i+1})$ should be in $V_c$, and $(x_{i+1},x_i) \in V_d$ and not in $R$. 

We further relax the conditions on $C$, and we may assume $C$ has minimum length among all the circuits where each pair on the circuit is either a pair in $V_c$ or is a pair $(x,y)$ such that $(y,x) \leadsto (x,y)$ via a directed path that is constricted from below and has net value zero. In other words, we may assume we cannot short cut $C$. However, Claim~\ref{all-V_c-pairs} shows that the pairs on circuit $C$ must be in $V_c$, a contradiction that $V_c$ is circuit free. Before we proceed we observe the following. 

\begin{observation}\label{obs1} Since $C$ has a minimum length then the following occur. 
\begin{enumerate}
    \item $(x_{i+2},x_i) \not\leadsto (x_i,x_{i+2})$, via a directed path that is  constricted from below and has net value zero. Otherwise, we replace $C$ by $(x_0,x_1),\dots,(x_{i-1},x_i),(x_i,x_{i+2}) ,\dots, (x_n,x_0)$, a contradiction to $C$ having minimum length.   
    
    \item If $(x_i,x_{i+1}) \not\in V_c$ then $(x_{i+2},x_i) \not\leadsto (x_{i+1},x_{i})$ (i.e.  $(x_i,x_{i+1}) \not\leadsto (x_i,x_{i+2})$). Otherwise, by skew property, $(x_{i+2},x_i) \leadsto (x_{i+1},x_i) \leadsto (x_i,x_{i+1}) \leadsto (x_i,x_{i+2})$,  contradiction to (1).
    
     \item Similar to (2), if $(x_{i+1},x_{i+2}) \not\in V_c$ then $(x_{i+2},x_i) \not\leadsto (x_{i+2},x_{i+1})$ (i.e. $(x_{i+1},x_{i+2}) \not\leadsto (x_i,x_{i+2})$ )
 
\end{enumerate} 
\end{observation}

\begin{claim}\label{all-V_c-pairs}
Every $(x_i,x_{i+1})$, $0 \le i \le n$,  is in $V_c$.
\end{claim}
\begin{proof}
Suppose there exists $0 \le i \le n$ such that $(x_i,x_{i+1}) \not \in V_c$. Thus according to our assumption, we have $(x_{i+1},x_i) \leadsto (x_i,x_{i+1})$ by a directed path, say $X_i$, constricted from below and with net value zero. Note that $n>1$ as otherwise, we must have  $(x_0,x_1) \leadsto (x_1,x_0)$, and $(x_1,x_0) \leadsto (x_0,x_1)$, and hence, a strong circuit of length 2 in $H^+$, a contradiction.

\noindent 
\textbf{Case 1. $(x_{i+1},x_{i+2}) \not \in V_c $. } 
This means, $(x_{i+2},x_{i+1}) \leadsto (x_{i+1},x_{i+2})$ by a directed path, say $X_{i+1}$, in $H^+$; constricted from below and with net value zero.

\begin{observation} \label{obs2} Since $C$ has a minimum length then the following hold. 
\begin{enumerate}
    \item At most one of the $X: (x_{i+2},x_i) \leadsto (x_i,x_{i+1})$, $Y : (x_{i+2},x_i) \leadsto (x_{i+1},x_{i+2})$ exists (here $X,Y$ are constricted from below and with net-value zero).  Otherwise, by Observation \ref{obs1} (1, 2, 3), $(x_{i+2},x_i) \not\leadsto \{(x_i,x_{i+2}), (x_{i+1},x_{i}),(x_{i+2},x_{i+1})\}$. However, this is a contradiction to Lemmas \ref{special-M}, \ref{special-M1} for $X,Y$.
    
     \item At most one of the $X: (x_{i+2},x_i) \leadsto (x_i,x_{i+1})$, $Y : (x_{i},x_{i+1}) \leadsto (x_{i+1},x_{i+2})$ exists (here $X,Y$ are constricted from below and with net-value zero). Otherwise, by Observation \ref{obs1} $(x_i,x_{i+1}) \not\leadsto (x_i,x_{i+2})$. Moreover, $(x_i,x_{i+1}) \not\leadsto (x_{i+1},x_i)$, as otherwise, we have $(x_{i+1},x_i) \leadsto (x_i,x_{i+1}) \leadsto (x_{i+1},x_i)$, a strong circuit of length 2 in $H^+$. Finally, by Observation \ref{obs1} (1, 3) $(x_{i+2},x_i) \not\leadsto \{(x_i,x_{i+1}),(x_{i+2},x_{i+1})\}$. However, this is a contradiction to Lemmas \ref{special-M}, \ref{special-M1} for $X,Y$.
     
     \item At most one of the $X: (x_{i+1},x_{i+2}) \leadsto (x_i,x_{i+1})$, $Y : (x_{i},x_{i+1}) \leadsto (x_{i+1},x_{i+2})$ exists (here $X,Y$ are constricted from below and with net-value zero). Otherwise, by Observation \ref{obs1}(2) $(x_i,x_{i+1}) \not\leadsto (x_i,x_{i+2})$. Moreover, $(x_i,x_{i+1}) \not\leadsto (x_{i+1},x_i)$, as otherwise, a strong circuit of length 2 in $H^+$. Analogously, $(x_{i+1},x_{i+2}) \not\leadsto \{(x_i,x_{i+2}),(x_{i+2},x_{i+1})\}$. However, this is a contradiction to Lemmas \ref{special-M},\ref{special-M1} for $X,Y$.
     
     \item At most one of the $X: (x_{i+2},x_{i}) \leadsto (x_{i+1},x_{i+2})$, $Y : (x_{i+1},x_{i+2}) \leadsto (x_{i},x_{i+1})$ exists (here $X,Y$ are constricted from below and with net-value zero). This proof is analogous to (2). 
     
\end{enumerate}
\end{observation}

\noindent Now according to Lemma \ref{special-M} (when $h(X_i) <h(X_{i+1})$) or Lemma \ref{special-M1} (when $h(X_i) \ge h(X_{i+1})$) one of the following occurs. 

\begin{itemize}
    \item[1.] $(x_{i+2},x_{i+1}) \leadsto (x_i,x_{i+2})$;  via a directed path which is constricted from below and has net value  zero, 
    
    \item[2.] $(x_{i+2},x_{i+1}) \leadsto (x_{i+1},x_i)$; via a directed path which is constricted from below and has net value  zero. 
    
     \item[3.] $(x_{i+1},x_{i}) \leadsto (x_{i},x_{i+2})$; via a directed path which is constricted from below and has net value  zero,
    
    \item[4.] $(x_{i+1},x_i) \leadsto (x_{i+2},x_{i+1})$; via a directed path which is constricted from below and has net value  zero,

\end{itemize}

In what follows, we show that none of the 1,2,3,4 (above) occurs; yielding a contradiction. \\

\noindent \textbf{For contradiction, suppose 1 occurs}, i.e. $(x_{i+2},x_{i+1}) \leadsto (x_i,x_{i+2})$. By skew property, $X'_{i+1} :(x_{i+2},x_i) \leadsto (x_{i+1},x_{i+2})$. Now consider the directed paths $X_i,X'_{i+1}$ (see Figure \ref{1-1-Lemma18}) (recall : $X_i : (x_{i+1},x_i) \leadsto (x_i,x_{i+1})$).

\begin{figure}
\begin{center}
\includegraphics[scale = 0.9]{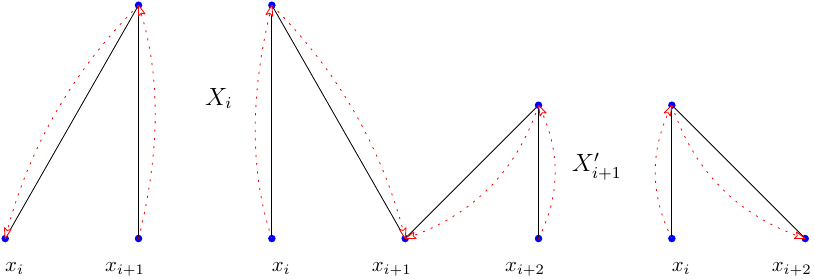}
\caption{Assuming $(x_{i+2},x_i) \leadsto (x_{i+1},x_{i+2})$}
\label{1-1-Lemma18}
\end{center}
\end{figure}

\noindent 1.1. By Observation \ref{obs1} (1) $(x_{i+2},x_i) \not\leadsto (x_i,x_{i+2})$.\\ 

\noindent 1.2. By Observation \ref{obs1} (2) $(x_{i+2},x_{i}) \not\leadsto (x_{i+1},x_i)$.\\

\noindent 1.3.  $(x_{i+1},x_i) \not\leadsto (x_{i},x_{i+2})$. Otherwise, $(x_{i+2},x_i) \leadsto (x_i,x_{i+1})$ and because $(x_{i+2},x_i) \leadsto (x_{i+1},x_{i+2})$ we get a contradiction by Observation \ref{obs2} (1).

\noindent 1.4.  $(x_{i+1},x_i) \not\leadsto (x_{i+2},x_{i+1})$. Otherwise, by skew property $(x_{i+1},x_{i+2}) \leadsto (x_i,x_{i+2})$, and because $(x_{i+2},x_i) \leadsto (x_{i+1},x_{i+2})$ we get a contradiction by Observation \ref{obs2} (2). 

\noindent Since none of the 1.1, 1.2, 1.3, and 1.4 occurs, we get a contradiction by Lemmas \ref{special-M}, or \ref{special-M1} for walks $X_i,X'_{i+1}$. Therefore, 1 does not occur. \\

By analogous argument for the proof of (1), we can show that 3 does not occur. We show that 4 does not occur, and analogously, 2 can not occur. \\

\noindent \textbf{For contradiction, suppose 4 occurs}. Thus, by skew property, we have $X'_i : (x_{i+1},x_{i+2}) \leadsto (x_i,x_{i+1})$; via  a directed path which is constricted from below and have net value zero.  Now consider the directed paths $X'_i,X_{i+1}$ (see Figure \ref{4-Lemma18})

\begin{figure}
\begin{center}
\includegraphics[scale = 0.9]{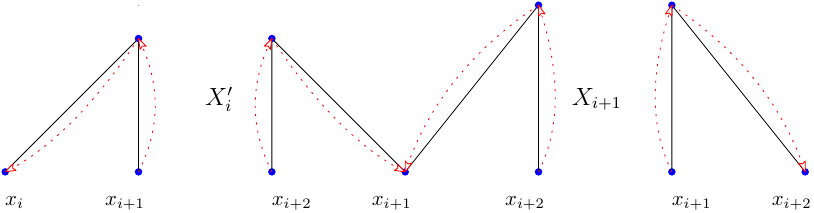}
\caption{Assuming $(x_{i+1},x_{i+2}) \leadsto (x_{i},x_{i+1})$}
\label{4-Lemma18}
\end{center}
\end{figure}

\noindent 4.1. $(x_{i+1},x_{i+2}) \not\leadsto (x_{i+2},x_{i+1})$, as otherwise, 
 since $(x_{i+2},x_{i+1}) \leadsto (x_{i+1},x_{i+2})$, we get a strong circuit of length 2; a contradiction.  

\noindent 4.2. $(x_{i+1},x_{i+2}) \not\leadsto (x_{i},x_{i+2})$. Otherwise, by skew property we have $(x_{i+2},x_i) \leadsto (x_{i+2},x_{i+1})$, a contradiction by Observation \ref{obs1} (3). 

\noindent 4.3. $(x_{i+2},x_{i+1}) \not\leadsto (x_{i},x_{i+2})$. This follows from Observation \ref{obs2} (4). 

\noindent 4.4. $(x_{i+2},x_{i+1}) \not\leadsto (x_{i+1},x_{i})$. This follows from Observation \ref{obs2} (3).\\

\noindent Since none of the 4.1,4.2,4.3, and 4.4 is satisfy for walks for $X'_i,X_{i+1}$, we get a contradiction according to Lemmas \ref{special-M}, \ref{special-M1}. Therefore, we conclude that (4) does not occur. \\

Therefore, we conclude that none of the conditions 1,2,3, and 4 satisfied for the walks $X_i,X_{i+1}$ which is a contradiction according to Lemma \ref{special-M} or Lemma \ref{special-M1}. \\

\noindent \textbf{Case 2. $(x_{i+1},x_{i+2})\in V_c $.} 
When $(x_j,x_{j+1}) \in V_c$, $0 \le j \le n$, then let $X_j$ be a directed path  constricted from below and with net value zero in $H^+$, from $(p_j,q_{j+1}) \in V_c$ to $(x_j,x_{j+1})$. Note that $(x_j,x_{j+1})$ could be an $LL$-pair with respect to $V_c$ and in this case $X_j$ is a portion of a path from a vertex in $V_c$, on a lower layer, to $(x_j,x_{j+1})$, and if this is not the case then $X_j$ is from a pair $(p_j,q_{j+1}) \in L_k \cap V_c$ to $(x_j,x_{j+1})$ where $(p_j,q_{j+1})$ is by transitivity on some pairs already placed in $V_c$. Recall that when $(x_j,x_{j+1}) \not\in V_c$ then let $X_j$ be a directed path constricted from below and with net value zero in $H^+$, from $(x_{j+1},x_j)$ to $(x_j,x_{j+1})$.
Observe that the following hold (see Figure \ref{Case2-Lemma18}).

\begin{figure}
\begin{center}
\includegraphics[scale = 0.9]{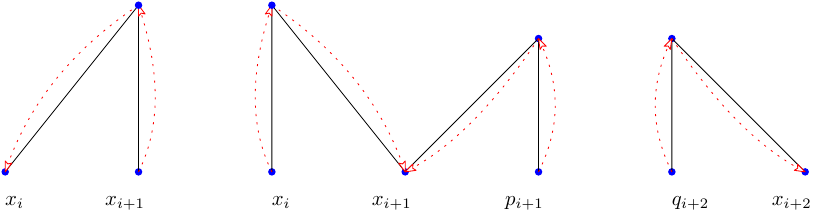}
\caption{Assuming $(x_{i+1},x_{i+2})$ is in $V_c \cap L_k$}
\label{Case2-Lemma18}
\end{center}
\end{figure}

\begin{itemize}
    \item [1.] $(p_{i+1},q_{i+2}) \not\leadsto (x_i,x_{i+2})$, as otherwise, $(x_i,x_{i+2})$ is a pair in $V_c$, contradiction to our assumption about $C$. 
    
    \item [2.] $(p_{i+1},q_{i+2}) \not\leadsto (x_{i+1},x_i)$ as otherwise, $(p_{i+1},q_{i+2}) \leadsto (x_{i+1},x_i)\leadsto (x_i,x_{i+1})\leadsto (q_{i+2},p_{i+1});$ a circuit in $V_c$.
    
    \item [3.] $(x_{i+1},x_i) \not\leadsto (x_{i+2},x_{i+1})$ as otherwise, $(x_{i+1},x_{i+2}) \leadsto (x_i,x_{i+1})$ implying that $(x_i,x_{i+1})$ is in $V_c$; a contradiction to our assumption about $(x_i,x_{i+1})$. 
    
    \item [4.] $(x_{i+1},x_i) \not\leadsto (x_i,x_{i+2})$.     Otherwise, let $X'_i=(x_{i+2},x_i) \leadsto (x_i,x_{i+1})$. Now by 1,2, $(p_{i+1},q_{i+2}) \not\leadsto (x_i,x_{i+2})$, and $(p_{i+1},q_{i+2}) \not\leadsto (x_{i+1},x_i)$. By Observation \ref{obs1} (1) $(x_{i+2},x_i) \not\leadsto (x_i,x_{i+2})$. Moreover, $(x_{i+2},x_i) \not\leadsto (x_{i+2},x_{i+1})$, as otherwise, by skew property  $(x_{i+1},x_{i+2})$ $\leadsto (x_i,x_{i+2})$, implying $(x_i,x_{i+2}) \in V_c$, a contradiction to the minimality of $C$. However, this is a contradiction according to Lemmas \ref{special-M}, \ref{special-M1} for the paths $X'_i,X_{i+1}$.
\end{itemize}

Now since none of the (1,2,3,4) occurs, by Lemma \ref{special-M} we have $h(X_i) > h(X_{i+1})$ and as a conclusion of Lemmas \ref{special-M},\ref{special-M1}, and \ref{M} it is easy to show that $X_{i+1}$ is a symmetric path. 

Note that by the argument in  Case (1), $(x_{i-1},x_i)$ must be a pair in $V_c$. 
Thus, again by a similar argument (using Lemma \ref{special-M1}) we conclude that $h(X_{i-1}) < h(X_i)$, and $X_{i-1}$ is symmetric. Now if both $(x_{i-1},x_i),(x_{i+1},x_{i+2})$ are $LL$-pair with respect to $V_c$ then according to the argument in Lemma \ref{at-most-1-implied} (5) we get a contradiction (i.e. $(x_i,x_{i+1})$ is in $V_c$). So, we may assume that at least one of the $(x_{i-1},x_i),(x_{i+1},x_{i+2})$ is not an $LL$-pair, and hence, one of the  $(p_{i-1},q_i),(p_{i+1},q_{i+2})$ is by transitivity.

We may assume that $(p_{i+1},q_{i+2})$ is by transitivity. We may assume that $(p_{i+1},q_{i+2})$ is a minimal chain. According to the rules of the algorithm, we start selecting original pairs on layer $L_k$ after identifying the vertex $p$ on line \ref{line17}. Thus,  at the beginning no pair is an original pairs. However, according to Lemma \ref{at-most-1-implied} the chain $(p_{i+1},q_{i+2})$ is replaced by the $LL$-pairs. This would allow us without loss of generality to assume $(p_{i+1},q_{i+2}$ is an $LL$-pair, and similarly, $(p_{i-1},q_i)$ is also an $LL$-pair (with respect to $V_c$). Thus again, according to the argument in Lemma \ref{at-most-1-implied} (5) we get a contradiction (i.e. $(x_i,x_{i+1})$ is in $V_c$). 
\end{proof}
Note that it is clear that $(p,q)$ on line \ref{line17} satisfies the condition of line \ref{line18} of Algorithm \ref{alg-main}. 
\end{proof} 

\begin{lemma}[Lemma~\ref{p-line18-main-body} repeated]\label{balanced-correctness}
By adding pair $(p,r)$ on line \ref{line18} of Algorithm~\ref{alg-main}, and computing $Tr(V_c)$, there will not be a circuit in $Tr(V_c)$. 
\end{lemma}
\begin{proof}
Suppose by adding $(p,r)$ 
we close a minimal circuit $C: (a_0,a_1),\dots,(a_{n-1},a_n),(a_n,a_0)$. We show that $n>1$. Otherwise, this means $(p,r) \leadsto (a_0,a_1)$, and also $(p,r) \leadsto (a_1,a_0)$. Now by skew property, we have $(p,r) \leadsto (a_1,a_0) \leadsto (r,p)$; contradiction to the choice of $(p,r)$. 

Notice that $C$ is minimal if every sub-chain of it is minimal. Thus, by Lemma \ref{chain-structure}, we may assume that each $(a_i,a_{i+1})$ is either $1$-implied or is $LL$-pair. Observe that, there is at least one $1$-implied pair. Suppose $(a_i,a_{i+1})$ is an $1$-implied pair. Note that by Lemma \ref{at-most-1-implied}(2) there is no any other $1$-implied pair. Thus the rest of the pairs are $LL$-pairs.
For each pair $(y_i,y_{i+1})$, $0 \le i \le n$, let $X_i$ be a constricted path of net value zero from $(p_i,q_{i+1}) \in V_c$ to $(y_i,y_{i+1})$. Let $h(X_i)$ denote the height of $X_i$. Now, by Lemma \ref{at-most-1-implied} (1) we must have $h(X_i) > h(X_{i+1})$ and $h(X_i) > h(X_{i-1})$ but this is a contradiction according to Lemma \ref{at-most-1-implied} (3).
\end{proof}

\section{Implementation and Time Complexity}\label{sec-time-complexity}
Let $n$ denote the numbers of vertices, and $m$ denote the numbers of arcs of $H$. We may assume that $H$ is weakly connected, and hence, $n \in 
\mathcal{O}(m)$. We note that, $H^+$ has $n^2$ vertices and $\mathcal{O}(m^2)$ arcs. 
To see that, for each arc $xx'$ of $H$, a particular pair $(x,y)$ (for some $y$ ) would appear 
in $d^+(y)$ pairs of $H^+$, and hence, it would count in $d^+(y)$ of the arcs in $H^+$. Therefore, for each arc $xx'$ of $H$, we would have at most 
$\sum_{y \in V(H)} d^+(y) = m$ arcs of $H^+$. Overall, after summing up, there are at most $O(m^2)$ arcs in $H^+$, and hence,  the construction of $H^+$ takes $\mathcal{O}(m^2)$. Deciding whether a strong component $S$ of $H^+$ contains a circuit can be done in 
$\mathcal{O}(m^2)$. To see this, we  create a new digraph $G$ whose vertices are $V(H)$, and there is an 
 arc from $x \in G$ to $y \in G$ whenever $(x,y) \in S$. If $G$ contains a directed cycle then $H^+$ contains a strong circuit, and hence, $H$ does not admit a min ordering. 
Note that $G$ has at most $O(n^2)$ arcs, and checking whether $G$ contains a cycle, takes $O(n^2)$ (running a BFS).    
Checking for circuit in all of the strong components of $H^+$ takes $\mathcal{O}(m^2+n^2)=\mathcal{O}(m^2)$ because $H^+$ has at most $\mathcal{O}(m^2)$ arcs. 

At the first stage of the algorithm; handling the unbalanced components, we maintain a digraph $G$ whose vertices are $V(H)$ where for each pair 
$(x,y) \in V_c$ we add an arc from $x$ to $y$ in $G$. $G$ is updated, each time a strong unbalanced component $S$ is added into $V_c$. 
Checking whether by adding $\widehat{S}$ into $V_c$, we close a circuit in $V_c$ takes, $\mathcal{O}(n^2)$. Again overall, it takes 
$\mathcal{O}(m^2+n^2)$ to handle all the strong unbalanced components. 

At the second stage of the algorithm (balanced components), we need to do extra work, and that is finding the layers $L_0,L_1,\dots,$. The layering can be done this way. Start with a pair $(x,y)$ that does not have an incoming negative arc into it in $H^{\#}$ (balanced pair). Now we do a BFS in $H^{\#}$ and give label zero to $(x,y)$. If $(u,v)$ has label $i$ and $e=(u,v)(u',v') \in A(H^{\#})$ is a positive arc, i.e., $uu',vv' \in A(H)$ and $uv' \not\in A(H)$, then $(u',v')$ gets label $i+1$ otherwise if $e$ is a negative arc then $(u',v')$ gets a label $i-1$. Now $L_0$ would be the set of all pairs with label zero. Note that this process takes $\mathcal{O}(m^2)$ because $H^+$ has $\mathcal{O}(m^2)$. Once we identify the layers, then we process each strong component $S$ and its $\widehat{S}$. In order to check whether $V_c$ contains a circuit, again we maintain the digraph $G$ as explain in the unbalanced case. We also need to take the transitive closure of $V_c$, and again this takes $O(n^2)$ where $n$ is the number of arcs in $G$. Notice that when we take the transitive closure some of the  arcs of $H^{\#}$ are going to be chosen and places into $V_c$. Therefore, taking transitive closure in all the steps of the algorithm would take at most $\mathcal{O}(n^2)$. The overall process takes $\mathcal{O}(m^2)$ because at each layer we deal with some of the pairs of $H^+$ and not all of them. Therefore, we have the following Lemma. 

\begin{lemma} \label{algorithm-complexity}
The running time of Algorithm $\ref{alg-main}$ is $\mathcal{O}(|A(H|)^2)$.  
\end{lemma}

\section{k-arc Digraphs and k-min Ordering } \label{sec-k-min}
Here, we discuss a generalization to $k$-arc digraphs, and the corresponding extension of min ordering, which has a similar obstruction characterization
and recognition algorithm. This class of digraphs is the core class in the study of list homomorphism and approximation of $H$-coloring. When $H$ is a $k$-arc digraph then finding a list homomorphism from an input digraph $G$ (together with the lists) to $H$ is polynomial time solvable (see Lemma 6.3 in \cite{arxiv-approximation}).  Moreover, the minimum cost homomorphism problem is approximable within a constant factor for this class of digraphs~\cite{RRS}.
 
 A $k$-{\em min ordering} of a digraph $H$ is a partition of $V(H)$ into $k$ subsets 
$V_0, V_1, \dots,$ $V_{k-1}$, and a linear ordering $<$ of each of these subsets $V_i$, such that each arc of 
$H$ belongs to some $V_i \times V_{i+1}$, $0 \le i \le k-1$, and $u < w, z < v$ and $uv, wz \in A(H)$ imply 
that $uz \in A(H)$ for any $u, w \in V_i$, $v, z \in V_{i+1}$, with all subscript addition modulo $k$. Theorem 3 in \cite{mfcs2018} 
can be extended to $k$-min orderings as follows. A $k$-{\em arc representation} of a digraph $H$ on 
a circle $C$ with $2k$ special points ({\em poles} $N_0, N_1, \dots, N_{k-1}, S_0, S_1, \dots, S_{k-1}$ (in 
this clockwise order) consists of intervals $I_v$, $J_v, v \in V(H)$ consistent as before, now each $I_v$ 
containing $S_{i+1}, S_{i+2},\dots,S_{k-1},$ $N_0, N_1, \dots, N_i$, for some $0 \le i \le k-1$, and no 
other poles, and each $J_v$ containing $N_{i+1}, N_{i+2}, \dots, N_{k-1}, S_0, S_1, \dots, S_i$, for some 
$0 \le i \le k-1$, and no other poles, such that $uv \in A(H)$ if and only if $I_u$ and $J_v$ are disjoint.

\begin{theorem}[\cite{arxiv-approximation}] 
A digraph $H=(V,A)$ is a k-arc digraph if and only if it admits a $k$-min ordering.
\end{theorem}

In some cases when min orderings do not exist, there may still exist extended
min orderings, which is sufficient for the polynomial solvability
of LHOM$(H)$ \cite{soda}. We denote by $\vec{C}_k$ the directed 
cycle on vertices $0, 1, \dots, k-1$. We shall assume in this section
that $H$ is weakly connected. This assumption allows us to conclude that any two homomorphisms 
$\ell, \ell'$ of $H$ to $\vec{C}_k$ define the same partition of $V(H)$
into the sets $V_i = \ell^{-1}(i)$, and we will refer to these sets without
explicitly defining a homomorphism $\ell$. 
Thus suppose $H$ is homomorphic to $\vec{C}_k$, and let $V_i$ be the 
partition of $V(H)$ corresponding to all such homomorphisms. 

Note that any $H$ is homomorphic to the one-vertex digraph with a loop
$\vec{C}_k$, and a $1$-min ordering of $H$ is just the usual min
ordering. Also note that a min ordering of a digraph $H$ becomes a 
$k$-min ordering of $H$ for any $\vec{C}_k$ that $H$ is homomorphic
to. However, there are digraphs homomorphic to $\vec{C}_k$ which have a 
$k$-min ordering but do not have a min ordering, for instance 
$\vec{C}_k$ (with $k>1$). 

We observe for future reference that an unbalanced digraph $H$ has
only a limited range of possible values of $k$ for which it could be
homomorphic to $\vec{C}_k$, and hence a limited range of possible 
values of $k$ for which it could have a $k$-min orderings. It is easy
to see that a cycle $C$ admits a homomorphism to $\vec{C}_k$ only if
the net length of $C$ is divisible by $k$ \cite{homobook}. Thus any cycle
of net length $q > 0$ in $H$ limits the possible values of $k$ to the divisors
of $q$. If $H$ is balanced, it is easy to see that $H$ has a $k$-min
ordering for some $k$ if and only if it has a min ordering.

For a digraph $H$ homomorphic to $\vec{C}_k$ we shall consider the 
following version of the pair digraph. The digraph $H^{(k)}$ is the subgraph 
of $H^+$ induced by all ordered pairs $(x,y)$ where $x,y$ belong to the same set
$V_i$, $0 \le i \le k-1$.  We say that $(u,v)$ is a {\em symmetrically $k$-invertible pair} in 
$H$ if $H^{(k)}$ contains a directed path joining $(u,v)$ and $(v,u)$. Thus a 
symmetrically $k$-invertible pair is a symmetrically invertible pair in 
$H$ in which $u$ and $v$ belong to the same set $V_i$. Note that $H$ 
may contain symmetrically invertible pairs, but no symmetrically 
$k$-invertible pair. Consider, for instance the directed hexagon 
$\vec{C}_6$. The pair $0, 3$ is symmetrically invertible and
symmetrically $3$-invertible, but not symmetrically $6$-invertible. The extended version of our main theorem follows. 
\begin{theorem}\label{main-k}
The following statements are equivalent for a weakly connected digraph $H$.

\begin{enumerate}
\item
$H$ admits a $k$-min ordering.
\item
there exists a positive integer $k$ such that $H$ is homomorphic to $\vec{C}_k$
and no component of $H^{(k)}$ contains a circuit.
\end{enumerate}
\end{theorem}
\begin{proof} We shall in fact prove that the following statements are equivalent
for a positive integer $k$ such that $H$ is homomorphic to $\vec{C}_k$:

\begin{enumerate}
\item
$H$ admits a $k$-min ordering
\item
no component of $H^{(k)}$ contains a circuit
\end{enumerate}

Suppose $H$ admits linear orderings $<$ of sets $V_i$ satisfying 
the Min property between consecutive sets $V_i, V_{i+1}$. Any
circuit $(x_0,x_1), (x_1,x_2), \dots, (x_n,x_0)$ in $H^{(k)}$ must have
all vertices $x_0, x_1, \dots, x_n$ in the same set $V_i$, and hence 
if all the pairs $(x_i,x_{i+1})$ were in the same component of $H^{(k)}$ 
we would obtain the same contradiction with transitivity of $<$ as 
above the statement of Theorem \ref{inv}. This proves that 1 implies 2.

Now we prove that 2 implies 1. Thus assume that $H$ is
homomorphic to $\vec{C}_k$ and no strong component of $H^{(k)}$ contains a circuit. 
We shall construct a $k$-min ordering 
of $H$. We have again the components of $H^{(k)}$ in dual pairs 
$C, C'$, where $C'$ consists of the reverses of the pairs in $C$, 
and we can proceed with a similar algorithm as before. 
At each stage of the algorithm, some component of $H^{(k)}$ is chosen 
and its dual component discarded. We again choose a component $X$ according to the rules in Algorithm \ref{alg-main}. 

The proof of correctness is analogous to  the proof of Theorem \ref{unbalanced-correctness}, and Lemma \ref{balanced-correctness}. 
\end{proof}

We again note that the theorem implies a polynomial time
algorithm to test whether an input digraph $H$ has a $k$-min
ordering. As noted above, it suffices to 
check for each component of $H$ separately, so we may 
assume that $H$ is weakly connected. If $H$ is balanced, 
we have already observed this is only possible if $H$ has 
a min ordering, which we can check in polynomial time. 
Otherwise we find any unbalanced cycle in $H$, say, of net 
length $q$, and then test for circuits in components $H^{(k)}$
for all $k$ that divide $q$.

\section{Connection to other Polymorphisms}\label{connection-other-polymorphism}

We first draw a comparison between the obstruction to conservative majority and conservative Maltsev and conservative semilattice. A conservative \emph{majority} polymorphism $\mu$ of $H$ is a ternary polymorphism such that  $\mu(x,x,y)=\mu(x,y,x)=\mu(y,x,x)=x$ for all $x,y\in V(H)$. 

 A conservative \emph{Maltsev} polymorphism $h$ of $H$ is a ternary polymorphism such that  $h(x,y,y)=h(y,y,x)=x$ for all $x,y\in V(H)$. 

\begin{definition}
Let $H$ be a digraph. Define $H^{+k}$ to be the digraph with the vertex set $\{ (a_1,a_2,\dots,a_k) | a_i \in V(H), 1 \le i \le k\}$  and the arc set 
\begin{align*}
A(H^{+k})=&\{ 
(a_1,a_2,\dots,a_k)(b_1,b_2,\dots,b_k) | a_ib_i (b_ia_i ) \in A(H),  1 \le i \le k,\\  
 &a_1b_j  (b_ja_1) \not\in A(H) \ \ \forall j,  2 \le j \le k  \}.
\end{align*} 
\end{definition} 

When $k=2$, then we get the usual $H^+$ defined in the previous sections. 

A {\it permutable triple} is three vertices $a,b,c$ together with vertices $\alpha_a,\alpha_b,\alpha_c, \beta_{ab},\beta_{bc},\beta_{ca}$ together with three directed paths $P_1,P_2,P_3$ in $H^{+3}$  such that $P_1 : (a,b,c) \leadsto (\alpha_a,\beta_{bc},\beta_{bc})$, $P_2 :(b,c,a) \leadsto (\alpha_b,\beta_{ca},\beta_{ca})$ and finally $P_3 : (c,a,b) \leadsto (\alpha_c,\beta_{ab},\beta_{ab})$. 

\begin{theorem}\cite{soda} 
A digraph $H$ admits a conservative majority if and only if $H$ does not admit a permutable triple. 
\end{theorem}

We say  $a,b,c \in V(H)$ is a {\it Maltsev triple} if there exist vertices $\alpha_a,\alpha_c,\beta_{ab},\beta_{bc}$ in $H$ such that $(\alpha_a,\beta_{bc},\beta_{bc}) \leadsto (a,b,c)$ in $H^{+3}$ and $(\alpha_c,\beta_{ab},\beta_{ab}) \leadsto (c,b,a)$ in $H^{+3}$
\begin{theorem}\cite{soda} 
A digraph $H$ admits a conservative Maltsev polymorphism if and only if $H$ does not admit a Maltsev triple. 
\end{theorem}
\subsection{Strong circuit implies composable closed walk in H}
We continue to consider other interesting polymorphisms. Note that the arguments in Theorem \ref{CLAIM} and lemmas and corollaries in the previous sections can be used even when we analyze a minimal circuit in a strong component of $H^+$ as long as $H^+$ does not contain an invertible pair. 

\begin{theorem} \label{before-last}
If there exists a circuit in a strong component of $H^+$ then either $H^+$ contains an invertible pair or there exists a closed walk $W$ in $H$ composed 
of walks $W[v_0,v_1]$, $W[v_1,v_2]$,$\dots,W[v_r,v_0]$ with the
following properties: 
\begin{enumerate}
    \item each $W[v_i,v_{i+1}]$ is constricted from below,
    \item each $W[v_i,v_{i+1}]$ has a positive net length $\ell$ (all have the same net length $\ell$),
    \item $W[v_i,v_{i+1}], W[v_{j},v_{j+1}]$ 
avoid each other for every $0 \le i < j \le r$ ($v_{r+1}=v_0$).
\end{enumerate}   
\end{theorem} 
\begin{proof} 
Suppose $C: (a_0,a_1), (a_1,a_2), \dots, (a_n,a_0)$ is a circuit in a strong
component $S$ of $H^+$. We may assume $n>1$, as otherwise, there exists an invertible pair and we are done. We first show the following. 

\begin{claim}
$C$ is minimal and $n>1$. Moreover $S$ is an unbalanced component. 
\end{claim} 
\begin{proof} 

If $S \leadsto S'$ then we consider a circuit in $S'$ and observe that in this case $S' \not\leadsto S$, as otherwise, $S=S'$, and hence, there exists an invertible pair in $S$. So without loss of generality we may assume $S \not\leadsto S'$.\\

\noindent \textbf{Case 1. $S$ is unbalanced} 
Let $C' : (a'_0,a'_1),(a'_1,a'_2),\dots,(a'_m,a'_0)$ be a minimal circuit in $\widehat{S}$. By the above assumption $m>1$, and hence, according to Theorem \ref{CLAIM} (2), there exists another minimal circuit $(b'_0,b'_1),\dots,(b'_m,b'_0)$ all belong to $S$. \\

\noindent \textbf{Case 2. $S$ is balanced} 
Consider the circuit $C$ inside $S$. Let $(x,y)$ be a pair on the lowest layer of $S$. Now consider a directed path $X_i=(A_i,B_i)$ from $(x,y)$ to $(a_i,a_{i+1})$, $0 \le i \le n$. We may assume that each walk has the maximum height. 

According to this assumption, there is no path from a vertex on a path $X_i$ to $(a_{i+1},a_i)$ as otherwise, this would imply that $(x,y) \leadsto (a_{i+1},a_i)$ as well as $(x,y) \leadsto (a_i,a_{i+1})$, and hence, $(x,y) \leadsto (y,x)$, a contradiction to $S \not\leadsto S'$. 
When considering the directed paths $X_0,X_1,\dots,X_n$ then one of the following happens.

\begin{itemize}
    \item there is no path from a vertex on a path $X_i$ to some $(a_r,a_s)$, $r \ne s-1,s$. This means we would be able to apply the Lemma \ref{M}.
    
    \item there is a path from a vertex on $X_i$ to some  $(a_r,a_s)$, $r \ne s-1,s,s+1$
    ($s+1$ is not possible as otherwise it would mean $(x,y) \leadsto (y,x)$), with the same net-value as $X_i$. In this case we get a shorter circuit of length greater than $1$ in $\widehat{S}$.   
\end{itemize}

As a conclusion of the above observations, we may assume there exists a minimal circuit $C'$ that has length  $m>1$, and each pair $(a'_i,a'_{i+1})$, $0 \le i \le m$ of $C'$ is reachable from $(x,y)$ via $X'_i$ that is a constricted path from below and has  non-negative net value. Now by Corollary \ref{not-imply-two} at most one $X'_i$, $0 \le i \le m$ has net value zero. If every $X'_i$ has net value more than zero then by applying Lemma \ref{itit} there exists another circuit $C'_1: (a^1_0,a^1_1),(a^1_1,a^1_2),\dots, (a^1_n,a^1_0)$ where the net-value of a path from $(x,y)$ to $(a^1_i,a^1_{i+1})$ is one less than the net value of $X'_i$, $0 \le i \le m$. By repeatedly applying the Lemma \ref{M}, we may assume that $X'_i$ for some $i$ has net value zero. Therefore, $(a'_{i-1},a'_i)$ and $(a'_{i+1},a'_{i+2})$ are $LL$-pairs. We may assume that $X'_i$ has a maximum height (max height in component $S$); this can be done by going through a vertex with max height from $(x,y)$ and back to $(x,y)$ and add this path as a prefix to $X'_i$. Now by a similar argument as the one in the  proof of Lemma \ref{at-most-1-implied} (5), we conclude that $(a_i,a_{i+1})$ is an $LL$-pair, a contradiction (see Figure \ref{leftright}). This proves the claim. 
\end{proof}

We continue by assuming that $S$ is unbalanced. By Theorem  \ref{CLAIM} we may assume that  
$(b_i,b_{i+1})=(a_i,a_{i+1})$, $0 \le i \le n$, and $(a_i,a_{i+1})$ is extremal and all lie on one directed cycle $(X,Y)=D$ in $S$. 
This can be done because all the pairs lie on the same strong component $S$. 
We also assume that the net value of such $D$ is minimum. We may assume that the net value of a directed path $W$ in $S$ 
from $(a_i,a_{i+1})$ to $(a_{i+1},a_{i+2})$ is not zero. Otherwise, in this case, we know $W$ is constricted from below  
because $(a_i,a_{i+1})$ is an extremal pair, and hence, $W=(A_i,B_i)$ where $A_i$ and $B_i$ avoid each 
other ($A_i$ is part of $P_i$ and $B_i$ is part of $P_{i+1}$). 
Now it is easy to see that there is a path from $(a_i,a_{i+1})$ to $(a_i,a_{i+2})$ 
(a walk from $a_i$ to a vertex with the maximum height on $A_i$ and then back to $a_i$ and a walk on $B_i$ from $a_{i+1}$ to $a_{i+2}$ would give 
to a path in $H^+$ from $(a_i,a_{i+1})$ to $(a_i,a_{i+2})$ in $S$) and hence we get a shorter circuit. 

Consider the walks $P_i$, $0 \le i \le n$ for circuit $C$ according to Theorem \ref{CLAIM} (4). 
Each walk $P_i$ starts at $a_i$ and it is constricted and has unbounded positive net length. 
Every $P_i,P_j$, $0 \le i < j \le n$ avoid each other. Moreover $P_i$ is obtained by walking around the closed walk $X$.  
Thus, without loss of generality let the net value of portion of $D$ from $(a_0,a_1)$ to $(a_1,a_2)$ 
be the smallest positive net value $\ell$, and let the net value of $D$ be $m$ where $m$ is minimum. 
Observe that $(n+1) \ell \le m$. 
 
Let $X'$ be the 
closed walk starting at $a_0$ corresponding to $P_0$. Observe that $X'$ is also a closed walk starting at $a_1$ corresponding to $P_1$.  

Now consider the following walks : $W_0=X'[a'_0,a'_1]$, where $a'_0=a_0$ and $a'_1=a_1$, and 
$W_j=X'[a'_j,a'_{j+1}]$; $j=1,2,\dots$ where $a'_{j+1}$ is a extremal vertex on $X'$ and $W_j$ has net length $\ell$. 
Observe that $W_i,W_{i+1}$, $1 \le i$ avoid each other since $P_0,P_1$ avoid each other. 
Now at some point we must have $a'_{r}=a'_k$. Without loss of generality we may assume that $a'_0=a'_k$. 
Note that $r \le m$ because $X'$ has at most $m$ extremal vertices (where the net length between any two of them is not zero), and the net value of $D$ is $(r+1)\ell$. 

At this point $(a'_0,a'_1),(a'_1,a'_2),\dots,(a'_r,a'_0)$ is a circuit in $S$. We show that $W_i$ and $W_j$ avoid each other for 
every $0 \le i < j \le r$. Observe that if there is a faithful arc from the $q$-th vertex of $W_i^{-1}$ to the $(q+1)$-th vertex of $W_{i+2}^{-1}$ then we would have $(a'_{i},a'_{i+1}) \leadsto (a'_{i+2},a'_{i+1})$ (using the faithful arc at index $q$) and because $W_i,W_{i+1}$ avoid each other and $W_{i+1},W_{i+2}$ avoid each other we conclude that $(a'_{i},a'_{i+1})$ and $(a'_{i+2},a'_{i+1})$ are both in $S$. 
On the other hand, both  $(a'_{i},a'_{i+1}),(a'_{i+1},a'_{i+2})$ are in $S$, and 
hence, $(a'_{i+1},a'_{i+2}),(a'_{i+2},a'_{i+1})$ are in $S$, a contradiction. 
Therefore, for every $i$ we have $W_i,W_{i+2}$ avoid each other. Now if $r=2$ then we are done so we may assume that $r \ge 3$. 

Let $d$ be the smallest integer 
such that $W_j$ and $W_{j+d}$ do not avoid each other ($d \ge 2$). Without loss of generality, we may assume that $d \le r/2$, and hence, we assume that $W_0, W_d$ do not avoid each other and $d \le r/2$. By considering the last faithful arc on $W_0^{-1}$ to $W_d^{-1}$ we conclude that  $X_1 : (a'_{d-1},a'_d) \leadsto (a'_{d},a'_1)$ has net value $\ell$, and since $W_{d-1},W_{d}$ avoid each other, and 
$W_0,W_{d-1}$ avoid each other, $(a'_{d-1},a'_d),(a'_d,a'_1)$ are in $S$. We also note that using the faithful arc from $W_0^{-1}$ to $W_d^{-1}$ we have $(a'_0,a'_1) \leadsto (a'_d,a'_1)$ which is  a directed path of net value zero, and hence, 
$X_2 : (a'_d,a'_1) \leadsto (a'_0,a'_1) \leadsto (a'_1,a'_2)$ has net value $\ell$. 
Now we have a circuit $(a'_1,a'_2),(a'_2,a'_2), \dots,(a'_{d-1},a'_d),(a'_d,a'_1)$, and a closed walk that 
consists of $a'_1,a'_2,\dots,a'_d,a'_1$  and  has net length 
$(d-1) \ell +\ell+\ell=(d+1)  \ell < (r+1) \ell$.  This is a contradiction to the assumption about the net value of $D$. Therefore, $W_i$ and $W_j$ avoid each other for every $0 \le i < j \le r$. This proves the theorem. 
\end{proof}

\subsection{Collapse: CSL = CST = Conservative Cyclic of all arities}
A polymorphism $f$ of $H$ of arity $k$ is {\em totally symmetric} if $f(x_1,x_2,\dots,x_k)=$
$f(y_1,y_2,\dots,y_k)$ whenever the sets $\{y_1,y_2,\dots,y_k\}$ and $\{x_1,x_2,\dots,x_k\}$ are the same.
A {\em set polymorphism} of $H$ is a mapping $f$ of the non-empty subsets of $V(H)$ to $V(H)$,
such that $f(S)f(T) \in A(H)$ whenever $S, T$ are non-empty subsets of $V(H)$ with the property 
that for each $s \in S$ there is a $t \in T$ with $st \in A(H)$ and also for every $t \in T$
there is an $s \in S$ with $st \in A(H)$. It is easy to see, cf. \cite{benoit,fv}, that $H$
has a conservative set polymorphism if and only if it has conservative totally symmetric (CTS)
polymorphisms of all arities $k$. 
A polymorphism $f$ of arity $k$ on digraph $H$ is called {\em cyclic} if 
$f(x_1,x_2,\dots,x_k)=f(x_2,x_3,\dots,x_k,x_1)$ for all $x_1,x_2,\dots,x_k \in V(H)$.

We note that a digraph $H$ that admits a CSL polymorphism also admits CTS polymorphisms of all arities: the conservative set function that assigns to each set $S$ the minimum under the min ordering. Moreover, a CTS polymorphism applies to all arities, including arity two, whence it implies a CC polymorphism. Thus, the class of digraphs with a min ordering is included in the class of digraphs with a conservative set polymorphism, which is included in the class of digraphs with a CC polymorphism. 
\begin{theorem}\label{last}
A digraph $H$ admits a CSL polymorphism if and only if it admits a conservative set polymorphism.
\end{theorem}
\begin{proof} 
Since a min ordering allows to define a conservative set polymorphism as the minimum,
it suffices to show that a digraph that does not have a min ordering also
cannot have a conservative set polymorphism. We show this by showing that a circuit
in one component of $H^+$ means that $H$ does not have a conservative set polymorphism.

So suppose $(a_0,a_1), (a_1,a_2), \dots, (a_n,a_0)$ is a circuit in a strong
component $C$ of $H^+$. Then, by Theorem~\ref{before-last}, we have that either there exists an invertible pair in $H^+$, and hence, there is no CC polymorphism, or there exists a closed walk $W$ composed of walks $W[v_0,v_1],$ $W[v_1,v_2],\dots,W[v_r,v_0]$ with the
following properties: 
\begin{itemize}
    \item each $W[v_i,v_{i+1}]$ is constricted from below,
    \item each $W[v_i,v_{i+1}]$ has a positive net length $\ell$, \item $W[v_i,v_{i+1}]$ and $W[v_{j},v_{j+1}]$ avoid each other for every $0 \le i < j \le r$ ($v_{r+1}=v_0$).
\end{itemize}   
Now for any conservative set polymorphism $f$, we must have
$f(v_0,v_1,\dots,v_r)=f(v_1,v_2,\dots,$ $v_{r},v_0)$,
but since the walks $W[v_i,v_{i+1}],W[v_{j},v_{j+1}]$, $0 \le i \ne r$ avoid each other, we apply the polymorphism definition on the vertices of the walks $W[v_0,v_1],W[v_1,v_2],\dots,W[v_r,v_0]$ and conclude that, if $f(v_0,v_1,\dots,v_r)=v_i$  then $f(v_1,v_2,\dots,v_r)$ must be $v_{i+1}$, a contradiction.
\end{proof}
We remark that in the proof we have only used the fact that $H$ does not have a conservative cyclic polymorphism.
Thus, we have actually proved the following 
\begin{theorem}
The class of bi-arc digraphs coincides with each of the following classes of digraphs: 
\begin{enumerate}
\item digraphs with a CSL polymorphism,
\item digraphs with a conservative set polymorphism,
\item digraphs with CTS polymorphisms of all arities, and
\item digraphs with conservative cyclic polymorphisms of all arities.
\end{enumerate}
\end{theorem}

\section{NP-complete Cases and a Dichotomy Classification}\label{np-complete}
\subsection{NP-completeness for arity 3}
For positive integer $n$, let $I_n=\{1,2,\dots,n\}$. 
Let $B$ be an instance of the betweenness problem. We are given a set $U$, and a 
subset $S$ from set $\{ (i,j,k) | i,j,k \in I_n\}$. The goal is to find an ordering $u_1 < u_2 < \dots <u_n$ of the vertices in $U$
such that for every $(i,j,k) \in S$, either $u_i < u_j < u_k$ or $u_k< u_j < u_i$. It is known that the betweenness problem in NP-complete \cite{betweenness}. 

\begin{theorem}
Deciding if a ternary relation admits a CSL polymorphism is NP-complete.
\end{theorem}
\begin{proof}
We start from an instance of the betweenness $B=(U,S)$, and construct two digraphs $H_1$ and $H_2$ and from these two digraphs we construct a ternary relation. Let $U=\{u_1,u_2,\dots,u_n\}$. The vertex set of $H_1$, consists of $r=|S|+1$ copies of $U$, here $|S|$ is the number of triples in $S$. Formally, we have:
\begin{itemize}
    \item $V(H_1)=\{a_{i,j} \mid 1 \le i \le |S|+1, \text{ and } 1 \le j \le n, \text{  $a_{i,j}$ corresponds to $u_j$ in copy $i$ of $U$} \}$,
    \item  $A(H_1)=\{ a_{t ,i}a_{t, j},a_{t, j}a_{t, k} \mid (i,j,k)  \text { is the $t$-th element of $S$}  \}$.
\end{itemize}

The vertex set of $H_2$, $V(H_2)=V(H_1)$, and the arc set of 
$H_2$, $A(H_2)=\{ a_{i,j}a_{i+1,j} | 1 \le i \le r, \text{ and } 1 \le j \le n\} $. In other words, the arc set of $H_2$ is $n$ copies of induced directed path of length $r$. 

\begin{claim}\label{NP-cl1}
There is an ordering of $U$ satisfying the betweenness condition if and only if there is an ordering of $V(H_1)$ that is a min ordering with respect to both $H_1$ and $H_2$.
\end{claim}
\begin{proof}
If $U$ has an ordering $<$ consistent with all the triples, then we can order the vertices of $H_1$ by taking this ordering on all copies of $U$, and put all the vertices of the $i$-th copy before all the vertices of the $(i+1)$-th copy. It is easy to see that the resulting ordering is a min ordering. Conversely, if $<$ is a min ordering of $H_1,H_2$ simultaneously, then the arcs in $A_2$ ensures that all copies are ordered in the same way, i.e., if $x$ precedes $y$ in some copy then it also precedes it in next copy and hence in all the copies of $U$.  
This means that there is an ordering $<$ of $U$ corresponding to all of them.
The arcs in  $A_1$ ensures that each triple is consistent with respect to $<$. This is because when $ab,bc$ are arcs in $A_1$, and $H_1$ has a min ordering then either $a<b<c$ or $c < b <a$ in the ordering. 
\end{proof}

 Let $R$ be a ternary relation constructed as follows. The ground set of $R$ is $V(H_1)$. The tuples of $R$ are $(a_{t,i},a_{t,j},a_{t+1,j+1})$ where $a_{t,i}a_{t,j} \in A(H_1)$. In other words, the tuples of $R$ are $(x,y,z)$ where $xy$ is an arc of $H_1$ and $yz$ is an arc of $H_2$. 

$\Rightarrow)$ Suppose there is an ordering $<$ of the vertices of $H_1$ that is a min ordering for both $H_1$ and $H_2$. According to the proof of the Claim \ref{NP-cl1}, in this ordering the vertices of each copy appear together and according to the ordering of the elements of $U$ in the betweenness instance (when all the triples are satisfied). More precisely, suppose $u_1< u_2 < \dots < u_n$ is the ordering of $U$, where for each triple $(i,j,k) \in S$ either $u_i<u_j<u_k$ or $u_k < u_j < u_i$. Then 
$a_{t,1} < a_{t,2},\dots, a_{t,n} < a_{t+1,1} < a_{t+1,2} < \dots < a_{t+1,n}$, for $1 \le t \le r$ is an ordering of $V(H_1)$ which is a min ordering for both $H_1,H_2$. 

Now define the semilattice function $f(a,b)=f(b,a)=a$ if $a<b$ in the ordering, i.e., $f(a,b)=\min \{a,b\}$. 
We show that $f$ is a semilattice for $R$. We need to show that $R$ is closed under $f$. 
Suppose $(a,b,c) \in R$, and $(a',b',c') \in R$. Thus, $ab,a'b' \in A(H_1)$, 
and $bc,b'c' \in A(H_2)$. Since $<$ is a min ordering for both $H_1,H_2$, 
$\min \{a,a'\} \min \{b,b'\}$ is an arc of $H_1$ and $\min \{b,b'\} \min \{c,c'\}$ is an arc of $H_2$. This means $f(a,a')f(b,b') \in A(H_1)$, and $f(b,b')f(c,c') \in A(H_2)$, and hence, by definition of $R$, $(f(a,a'),f(b,b'),f(c,c'))$ is in $R$. Note that since min ordering is commutative and associative, $f$ is a semilattice.

$\Leftarrow)$ Conversely, suppose $f$ is a semilattice on $R$. We will define an ordering on $V(H_1)$ that is a min ordering on both $H_1,H_2$. 
In order to find such a min ordering, we are going to modify $f$ first. We first obtain $f_1$ from $f$ as follows. 
For every $a_{t,i},a_{s,j}$, $t<s$, 
$f_1(a_{t,i},a_{s,j})=f_1(a_{s,j},a_{t,i})=a_{t,i}$. In any other case $f_1(x,y)=f(x,y)$. In other words, $f_1$ and $f$ have the same outcome on the vertices inside each copy of $U$. Clearly by definition $f_1(x,y)=f_1(y,x) \in \{x,y\}$. Thus, $f_1$ is a CC operation. 
Now we show that $f_1(x,f_1(y,z))=f_1(f_1(x,y),z)$ for every $x,y,z \in V(H_1)=V(H_2)$. 

If all $x,y,z$ belong to the same copy of $U$, then because $f$ has associative property, $f_1$ would be associative. Suppose $x=a_{t,i}$, $y=a_{s,j}$, $z=a_{p,k}$. We may assume $|\{t,s,p\}| >1$, otherwise, since $f$ is associative, $f_1$ is also associative. First suppose $t < s,p$. Then by definition  $f_1(a_{t,i},f_1(a_{s,j},a_{p,k}))=a_{t,i}$, and $f_1(a_{t,i},a_{s,j})=a_{t,i}$ and since $f_1(a_{t,i},a_{p,k})=f_1(a_{t,i},a_{s,j})=a_{t,i}$, we have
$f_1(a_{t,i},f_1(a_{s,j},a_{p,k}))=f_1(f_1(a_{t,i},a_{s,j}),a_{p,k})$. So we may assume $t \ge \min \{s,p\}$. If $t > s,p$ then $f_1(a_{t,i},f_1(a_{s,j},a_{p,k}))=f_1(a_{s,j},a_{p,k})$, and $f_1(f_1(a_{t,i},a_{s,j}),a_{p,k})=f_1(a_{s,j},a_{p,k})$, and $f_1(a_{t,i},f_1(a_{s,j},a_{p,k}))=f_1(f_1(a_{t,i},a_{s,j}),a_{p,k})$. Third, assume $t=s < p$ (or $t=p < s$). In this case $f_1(a_{t,i},f_1(a_{s,j},a_{p,k}))=f_1(a_{t,i},a_{s,j})=f(a_{t,i},a_{s,j})$, and $f_1(f_1(a_{t,i},a_{s,j}),a_{p,k})=f_1(f(a_{t,i},a_{s,j}),a_{p,k})=f(a_{t,i},a_{s,j})$. The other case can be treated similarly. Therefore, $f_1$ is associative. Next, we show that $R$ is closed under $f_1$. Suppose 
\[(a_{t,i},a_{t,j},a_{t+1,j}),(a_{s,i'},a_{s,j'},a_{s+1,j'}) \in R,\]
and let  \[\mu=(f_1( a_{t,i}, a_{s,i'}),f_1( a_{t,j}, a_{s,j'}), f_1(a_{t+1,j},a_{s+1,j'})).\] First suppose $t<s$. Then $\mu=(a_{t,i},a_{t,j},a_{t+1,j}) \in R$. Similarly, if $t>s$ then $\mu \in R$. So we continue by assuming $t=s$. In this case according to the construction of $R$, we have $i'=j$, $j'=\ell$ such that $a_{t,i}a_{t,j},a_{t,j}a_{t,\ell} \in A(H_1)$. Thus,  $\mu=(f(a_{t,i},a_{t,j}),f(a_{t,j},a_{t,\ell}), f(a_{t+1,j},a_{t+1,\ell}))$ and hence, $\mu \in R$. 

By the above discussion, without loss of generality, we continue by assuming that 
$f$ has the following property. For every $a_{t,i},a_{s,j}$, $t<s$, $f(a_{t,i},a_{s,j})=a_{t,i}$.

Suppose $a_{t,i}a_{t,j},a_{t,j}a_{t,\ell}$ are arcs of $H_1$. Then, $(a_{t,i},a_{t,j},a_{t+1,j}), (a_{t,j},a_{t,\ell},a_{t+1,\ell}) \in R$, and since $f$ is a semilattice we have  $(f(a_{t,i},a_{t,j}),f(a_{t,j},a_{t,\ell}),f(a_{t+1,j},a_{t+1,\ell})) \in R$. Moreover, because the arcs of $H_1$ are among each copy of $U$, one of the following must hold. 
\begin{itemize}
    \item $f(a_{t,i},a_{t,j})=a_{t,i}, f(a_{t,j},a_{t,\ell})=a_{t,j},$ $f(a_{t+1,j},a_{t+1,\ell})=a_{t+1,j}$, and $f(a_{t,i},a_{t,\ell})=a_{t,i}$ (because $f$ is associative) 
    \item $f(a_{t,i},a_{t,j})=a_{t,j}, f(a_{t,j},a_{t,\ell})=a_{t,\ell},$  $f(a_{t+1,j},a_{t+1,\ell})=a_{t+1,\ell}$, and $f(a_{t,i},a_{t,\ell})=a_{t,\ell}$ (because $f$ is associative).  
\end{itemize}
Thus, we would have the following observation. 
\begin{observation} \label{obs-np}
Notice that at this point the restriction of $f$ on the arcs of $H_1$, is a CSL. In other words, if $a_{t,i}a_{t,j},a_{t,j}a_{t,\ell}$ are arcs of $H_1$, then $f(a_{t,i},a_{t,j})f(a_{t,j},a_{t,\ell})$ is also an arc of $H_1$.
\end{observation}


In order to obtain a min ordering for $H_1$, using $f$, we further modify $f$
so that $f(a_{t,i},a_{t,i'})=a_{t,i}$ iff $f(a_{s,i},a_{s,i'})=a_{s,i}$, and keeping $f$ being a semilattice with respect to the arcs of $H_1$ (inside each copy of $U$), as well as with arcs of $H_2$. Notice that $f$ defines a min ordering on each copy of $U$; that is for every $1 \le t \le r$,
we obtain ordering $\prec_t$, by setting $a_{t,i} \prec_t a_{t,j}$ if and only if $f(a_{t,i},a_{t,j})=a_{t,i}$. 

Let $G$ be a graph constructed as follows. 
$V(G)=\{ (x,y) \mid x,y \in V(H_1)\}$.
The edge set of $G$ consists of the union of the following,
\begin{align*}
    E(G)
    &=
    \begin{multlined}[t]
    \{(a_{t,i},a_{t,j})(a_{s,i},a_{s,j}) \mid \\ f(a_{t,i},a_{t,j})=a_{t,i} \text{ , } f(a_{s,i},a_{s,j})=a_{s,i}  \text{ and }
    a_{t,i}a_{t,j}, a_{s,i}a_{s,j} \in A(H_1)\} 
    \end{multlined}
    \\
    & \cup 
    \begin{multlined}[t]
    \{(a_{t,i},a_{t,j})(a_{s,i},a_{s,j}) \mid \\ f(a_{t,i},a_{t,j})=a_{t,j} \text{ , } f(a_{s,i},a_{s,j})=a_{s,j}  \text{ and }
    a_{t,i}a_{t,j}, a_{s,i}a_{s,j} \in A(H_1)\} 
    \end{multlined}
    \\
   & \cup 
   \begin{multlined}[t]
   \{ (a_{t,i},a_{t,j})(a_{t,j},a_{t,\ell}), (a_{t,i},a_{t,j})(a_{t,i},a_{t,\ell}),(a_{t,j},a_{t,\ell})(a_{t,i},a_{t,\ell})  \mid \\ [f(a_{t,i},a_{t,j})=a_{t,i},f(a_{t,j}a_{t,\ell})=a_{t,j}] \textbf{ or } [f(a_{t,i},a_{t,j})=a_{t,j}, f(a_{t,j}a_{t,\ell})=a_{t,\ell}] \}
    \end{multlined}
\end{align*}


Suppose for some arc $a_{t,i}a_{t,j}$ of $H_1$, $f(a_{t,i},a_{t,j})=a_{t,i}$ while for some arc $a_{s,i}a_{s,j}$ of $H_1$ with $t<s$, we have $f(a_{s,i},a_{s,j})=a_{s,j}$. We may assume $t$ is the smallest subscript, and secondly, $s$ is the smallest subscript. Let $G_1$ be the set of vertices in $G$ that are reachable from $(a_{s,i},a_{s,j})$ in $G$, i.e., a connected component of $G$ containing $(a_{s,i},a_{s,j})$. 

Now for every 
$(x,y) \in G_1$, set $f(x,y)=f(y,x)=x$. Notice that by the construction of $G$, and since $f$ is a also CC polymorphism, there is no path from $(x,y) \in G_1$ to $(y,x) \in G_1$. Therefore, the changes to $f$ would be consistent.


Notice that after this modification for a fixed pair of indices $i,j$, $(a_{t,i},a_{t,j})$, $1 \le t \le r$, $f(a_{t,i},a_{t,j})=a_{t,i}$ when $a_{t,i}a_{t,j} \in A(H_1)$ (note that direction of the arcs is not according to $f$).  
Next we consider another arc $a_{t',i'}a_{t',j'} \in A(H_1)$, with $f(a_{t',i'},a_{t',j'})=a_{t',i'}$ while for some arc $a_{s',i'},a_{s',j'}$ with $t'<s'$, we have $f(a_{s',i'},a_{s',j'})=a_{s',j'}$. Let $G_2$ be the vertices that are reachable from $(a_{s',i'},a_{s',j'})$. Again for every pair $(x,y) \in G_2$, we set $f(x,y)=x$. 
Since $G$ is a graph, there is no vertex in $G_1$ that is reachable from a vertex in $G_2$.
Thus, the $f$ value for the vertices in $G_1$ is not going to change anymore. In other words, the changes of $f$ on $G_2$ would be consistent with the changes of $f$ on $G_1$. Notice that again since $f$ is a semilattice, there is no path from $(x,y) \in G_2$ to $(y,x) \in G_2 \cup G_1$.  
We repeat the above procedure until no such pairs $(a_{t',i'},a_{t',j'}),(a_{s',i'},a_{s',j'})$ where $a_{t',i'}a_{t',j'}, a_{s',i'}a_{s',j'}$ are arcs of $H_1$, and $f(a_{t',i'},a_{t',j'}) =a_{t',j'}$, $f(a_{s',i'},a_{s',j'})=a_{s',j'}$, can be found. 

In the next step we look for some $(a_{t,i},a_{t,j})$, where $t$ is the smallest index so that there exists a pair $(a_{s,i},a_{s,j})$, $t <s$ where  $(a_{s,i},a_{s,j})$ is not an isolated vertex in $G$, and  $f(a_{t,i},a_{t,j})=a_{t,i}$ while $f(a_{s,i},a_{s,j})=a_{s,j}$. Let $G_3$ be connected component of $G$, containing $(a_{s,i},a_{s,j})$. We further modify $f$ on the vertices of $G_3$, by setting $f(x,y)=x$ for every $(x,y) \in G_3$. Notice that as we argued above the changes are consistent with the previous changes on $f$. 

At the final stage, we consider pairs $(a_{t',i'},a_{t',j'})$, and $(a_{s',i'},a_{s',j'})$ so that both are isolated vertices in $G$, $t'<s'$, and $f(a_{t',i'},a_{t',j'})=a_{t',i'}$ while $f(a_{s',i'},a_{s',j'})=a_{s',j'}$ (assuming $t'$ is the smallest index, and then $s'$ is then smallest index). In this case we set $f(a_{s',i'},a_{s',j'})=a_{s',i'}$.

Finally, we define an ordering $<$ on the vertices $H_1$ by setting $x<y$ iff $f(x,y)=x$. 
This means that we would have 
$a_{t,1} < a_{t,2} < \dots <a_{t,n} < a_{t+1,1} < a_{t+1,2} <\dots < a_{t+1,n}$, $1 \le t \le r$

Notice that this ordering is a min ordering for $H_1$, and it is easy to see that is also a min ordering for $H_2$. 
\end{proof}

\subsection{Higher arities and a dichotomy}

\begin{theorem}\label{np-more-than-3}
Let $R$ be a relation of arity $r>3$. Then deciding whether $R$ admits a CSL is NP-complete. 
\end{theorem}
\begin{proof}
We use reduction from deciding whether a ternary relation has a CLS. Let $R_1$ be an arbitrary ternary relation on set $A$. 
Let $R$ be a relation of arity $r$, with the tuples $(\overbrace{a,a,\dots,a}^{r-3},a_1,a_2,a_3)$ for every $a \in A$, and every $(a_1,a_2,a_3) \in R_1$. Suppose $f$ is a CSL on $R_1$. Then we show that $f$ is also a CSL for $R$. We need to show that for every two tuples $t_1=(a,a,\dots,a,a_1,a_2,a_3),t_2=(b,b,\dots,b,b_1,b_2,b_3)$, \[(f(a,b),f(a,b),\dots,f(a,b),f(a_1,b_1),f(a_2,b_2),f(a_3,b_3)) \in R.\] Since $R_1$ is closed under $f$, we have $(f(a_1,b_1),f(a_2,b_2),f(a_3,b_3)) \in R_1$, and hence, $f(t_1,t_2) \in R$. Therefore,  $f$ is a CLS for $R$.
Conversely, suppose $f$ is a CSL on $R$. Then the projection of $f$ on the last three coordinates of $R$ is a CSL on $R_1$. 
\end{proof}
The above theorem together with Theorem~\ref{inv} provide a full complexity classification of {\sc Problem~\ref{problem1}} and yield us the following dichotomy theorem.
\begin{theorem}[Dichotomy Theorem]
Deciding if a relational structure $\mathbb{H}=\langle V, R_1,\dots, R_k \rangle$ admits a CSL polymorphism is polynomial-time solvable if all relations $R_i$ are unary, except possibly one binary relation. In all other cases, the problem is NP-complete. 
\end{theorem}

\section{Conclusions}

We have provided polynomial time algorithm, obstruction characterizations, for digraphs admitting a min ordering,
i.e., a CSL polymorphism. We believe they are a useful generalization of interval graphs, encompassing adjusted
interval digraphs, monotone proper interval digraphs, complements of circular arcs of clique covering number two,
two-directional orthogonal ray graphs, and other well-known classes. We have also similarly characterized digraphs
admitting a CC polymorphism. We showed that the class of digraphs admitting a set polymorphism, i.e., CTS polymorphisms of all arities, coincides with the the class of digraphs with a min ordering, and so is equal to the class of
bi-arc digraphs. Our algorithm can be adapted to recognize the digraphs that admit extension of min ordering, so
called $k$-min ordering ($k \geq 2$).

\begin{open-problem} 
What is the complexity of deciding whether a digraph admits a (not necessarily conservative)
semilattice polymorphism?
\end{open-problem}

\bibliographystyle{alpha}
\bibliography{main.bib}

\newcommand{\etalchar}[1]{$^{#1}$}
\begin{thebibliography}{HMNR12}

\bibitem[BDFG10]{bagan}
Guillaume Bagan, Arnaud Durand, Emmanuel Filiot, and Olivier Gauwin.
\newblock Efficient enumeration for conjunctive queries over x-underbar
  structures.
\newblock In {\em International Workshop on Computer Science Logic}, pages
  80--94. Springer, 2010.

\bibitem[BFH{\etalchar{+}}08]{nuf1}
Richard~C Brewster, Tomas Feder, Pavol Hell, Jing Huang, and Gary MacGillivray.
\newblock Near-unanimity functions and varieties of reflexive graphs.
\newblock {\em SIAM Journal on Discrete Mathematics}, 22(3):938--960, 2008.

\bibitem[BK12]{BK}
Libor Barto and Marcin Kozik.
\newblock Robust satisfiability of constraint satisfaction problems.
\newblock In {\em STOC}, volume~12, pages 931--940, 2012.

\bibitem[BKW17]{BKW}
Libor Barto, Andrei Krokhin, and Ross Willard.
\newblock Polymorphisms, and how to use them.
\newblock In {\em Dagstuhl Follow-Ups}, volume~7. Schloss
  Dagstuhl-Leibniz-Zentrum fuer Informatik, 2017.

\bibitem[BL76]{booth}
Kellogg~S Booth and George~S Lueker.
\newblock Testing for the consecutive ones property, interval graphs, and graph
  planarity using pq-tree algorithms.
\newblock {\em Journal of Computer and System Sciences}, 13(3):335--379, 1976.

\bibitem[Bul17]{bulatov-dichotomy}
Andrei~A Bulatov.
\newblock A dichotomy theorem for nonuniform csps.
\newblock In {\em 2017 IEEE 58th Annual Symposium on Foundations of Computer
  Science (FOCS)}, pages 319--330. IEEE, 2017.

\bibitem[CEJN15]{catarina}
Catarina Carvalho, Laszlo Egri, Marcel Jackson, and Todd Niven.
\newblock On maltsev digraphs.
\newblock {\em Electronic Journal of Combinatorics}, 2015.

\bibitem[CL17]{benoit}
Hubie Chen and Benoit Larose.
\newblock Asking the metaquestions in constraint tractability.
\newblock {\em ACM Transactions on Computation Theory (TOCT)}, 9(3):11, 2017.

\bibitem[COS09]{corneil}
Derek~G Corneil, Stephan Olariu, and Lorna Stewart.
\newblock The lbfs structure and recognition of interval graphs.
\newblock {\em SIAM Journal on Discrete Mathematics}, 23(4):1905--1953, 2009.

\bibitem[CVY16]{ChenVY16}
Hubie Chen, Matthew Valeriote, and Yuichi Yoshida.
\newblock Testing assignments to constraint satisfaction problems.
\newblock In {\em {IEEE} 57th Annual Symposium on Foundations of Computer
  Science, {FOCS} 2016, 9-11 October 2016, Hyatt Regency, New Brunswick, New
  Jersey, {USA}}, pages 525--534, 2016.

\bibitem[DKK{\etalchar{+}}17]{DalmauKKMMO17}
V{\'{\i}}ctor Dalmau, Marcin Kozik, Andrei~A. Krokhin, Konstantin Makarychev,
  Yury Makarychev, and Jakub Oprsal.
\newblock Robust algorithms with polynomial loss for near-unanimity csps.
\newblock In {\em Proceedings of the Twenty-Eighth Annual {ACM-SIAM} Symposium
  on Discrete Algorithms, {SODA} 2017, Barcelona, Spain, Hotel Porta Fira,
  January 16-19}, pages 340--357, 2017.

\bibitem[DSRW89]{sen-west}
Sandip Das, M~Sen, AB~Roy, and Douglas~B West.
\newblock Interval digraphs: An analogue of interval graphs.
\newblock {\em Journal of Graph Theory}, 13(2):189--202, 1989.

\bibitem[FG65]{fg}
Delbert Fulkerson and Oliver Gross.
\newblock Incidence matrices and interval graphs.
\newblock {\em Pacific journal of mathematics}, 15(3):835--855, 1965.

\bibitem[FHH03]{bi-arc}
Tomas Feder, Pavol Hell, and Jing Huang.
\newblock Bi-arc graphs and the complexity of list homomorphisms.
\newblock {\em Journal of Graph Theory}, 42(1):61--80, 2003.

\bibitem[FHHR12]{adjusted}
Tom{\'a}s Feder, Pavol Hell, Jing Huang, and Arash Rafiey.
\newblock Interval graphs, adjusted interval digraphs, and reflexive list
  homomorphisms.
\newblock {\em Discrete Applied Mathematics}, 160(6):697--707, 2012.

\bibitem[FHL{\etalchar{+}}13]{nuf2}
Tom{\'a}s Feder, Pavol Hell, Beno{\^\i}t Larose, Cynthia Loten, Mark Siggers,
  and Claude Tardif.
\newblock Graphs admitting k-nu operations. part 1: The reflexive case.
\newblock {\em SIAM Journal on Discrete Mathematics}, 27(4):1940--1963, 2013.

\bibitem[FV93]{fv}
Tom{\'a}s Feder and Moshe~Y Vardi.
\newblock Monotone monadic snp and constraint satisfaction.
\newblock In {\em Proceedings of the twenty-fifth annual ACM symposium on
  Theory of computing (STOC)}, pages 612--622. ACM, 1993.

\bibitem[FV98]{fv98}
Tom{\'a}s Feder and Moshe~Y Vardi.
\newblock The computational structure of monotone monadic snp and constraint
  satisfaction: A study through datalog and group theory.
\newblock {\em SIAM Journal on Computing}, 28(1):57--104, 1998.

\bibitem[Gol04]{gol}
Martin~Charles Golumbic.
\newblock {\em Algorithmic graph theory and perfect graphs}, volume~57.
\newblock Elsevier, 2004.

\bibitem[GZ11]{GuruswamiZ11}
Venkatesan Guruswami and Yuan Zhou.
\newblock Tight bounds on the approximability of almost-satisfiable horn {SAT}
  and exact hitting set.
\newblock In {\em Proceedings of the Twenty-Second Annual {ACM-SIAM} Symposium
  on Discrete Algorithms, {SODA} 2011, San Francisco, California, USA, January
  23-25, 2011}, pages 1574--1589, 2011.

\bibitem[HH04]{jing}
Pavol Hell and Jing Huang.
\newblock Interval bigraphs and circular arc graphs.
\newblock {\em Journal of Graph Theory}, 46(4):313--327, 2004.

\bibitem[HHML88]{don}
Roland H{\"a}ggkvist, Pavol Hell, Donald~J. Miller, and V~Neumann Lara.
\newblock On multiplicative graphs and the product conjecture.
\newblock {\em Combinatorica}, 8(1):63--74, 1988.

\bibitem[HHMR]{minorder-digraphs}
Pavol Hell, Jing Huang, Ross~M. McConnell, and Arash Rafiey.
\newblock Min-orderable digraphs.
\newblock {\em Accepted in SIAM Journal of Discrete Math}.

\bibitem[HHMR18]{mfcs2018}
Pavol Hell, Jing Huang, Ross~M McConnell, and Arash Rafiey.
\newblock Interval-like graphs and digraphs.
\newblock In {\em 43rd International Symposium on Mathematical Foundations of
  Computer Science (MFCS 2018)}. Schloss Dagstuhl-Leibniz-Zentrum fuer
  Informatik, 2018.

\bibitem[HMNR12]{esa2012}
Pavol Hell, Monaldo Mastrolilli, Mayssam~Mohammadi Nevisi, and Arash Rafiey.
\newblock Approximation of minimum cost homomorphisms.
\newblock In {\em European Symposium on Algorithms}, pages 587--598. Springer,
  2012.

\bibitem[HMPV00]{habib}
Michel Habib, Ross McConnell, Christophe Paul, and Laurent Viennot.
\newblock Lex-bfs and partition refinement, with applications to transitive
  orientation, interval graph recognition and consecutive ones testing.
\newblock {\em Theoretical Computer Science}, 234(1-2):59--84, 2000.

\bibitem[HN04]{homobook}
Pavol Hell and Jaroslav Nesetril.
\newblock {\em Graphs and homomorphisms}.
\newblock Oxford University Press, 2004.

\bibitem[HR11]{soda}
Pavol Hell and Arash Rafiey.
\newblock The dichotomy of list homomorphisms for digraphs.
\newblock In {\em Proceedings of the twenty-second annual ACM-SIAM symposium on
  Discrete Algorithms}, pages 1703--1713. Society for Industrial and Applied
  Mathematics, 2011.

\bibitem[HR12]{monoton-proper}
Pavol Hell and Arash Rafiey.
\newblock Monotone proper interval digraphs and min-max orderings.
\newblock {\em SIAM Journal on Discrete Mathematics}, 26(4):1576--1596, 2012.

\bibitem[JCG97]{JeavonsCG97}
Peter Jeavons, David~A. Cohen, and Marc Gyssens.
\newblock Closure properties of constraints.
\newblock {\em J. {ACM}}, 44(4):527--548, 1997.

\bibitem[Kaz11]{kazda}
Alexandr Kazda.
\newblock Maltsev digraphs have a majority polymorphism.
\newblock {\em European Journal of Combinatorics}, 32(3):390--397, 2011.

\bibitem[KOT{\etalchar{+}}12]{kotyz}
Gabor Kun, Ryan O'Donnell, Suguru Tamaki, Yuichi Yoshida, and Yuan Zhou.
\newblock Linear programming, width-1 csps, and robust satisfaction.
\newblock In {\em Proceedings of the 3rd Innovations in Theoretical Computer
  Science Conference (ITCS 2012)}, pages 484--495. ACM, 2012.

\bibitem[Lar17]{Benoit2017}
Beno{\^\i}t Larose.
\newblock Algebra and the complexity of digraph csps: a survey.
\newblock In {\em Dagstuhl Follow-Ups}, volume~7. Schloss
  Dagstuhl-Leibniz-Zentrum fuer Informatik, 2017.

\bibitem[LB62]{lekk}
C~Lekkeikerker and J~Boland.
\newblock Representation of a finite graph by a set of intervals on the real
  line.
\newblock {\em Fundamenta Mathematicae}, 51(1):45--64, 1962.

\bibitem[MM08]{maroti}
Mikl{\'o}s Mar{\'o}ti and Ralph McKenzie.
\newblock Existence theorems for weakly symmetric operations.
\newblock {\em Algebra universalis}, 59(3):463--489, 2008.

\bibitem[Opa79]{betweenness}
Jaroslav Opatrny.
\newblock Total ordering problem.
\newblock {\em SIAM Journal on Discrete Mathematics}, 8(1):111--114, 1979.

\bibitem[PS82]{papa}
Christos~H Papadimitrou and Kenneth Steiglitz.
\newblock Combinatorial optimization: algorithms and complexity.
\newblock 1982.

\bibitem[Rag08]{Raghavendra08}
Prasad Raghavendra.
\newblock Optimal algorithms and inapproximability results for every csp?
\newblock In {\em Proceedings of the 40th Annual {ACM} Symposium on Theory of
  Computing, Victoria, British Columbia, Canada, May 17-20, 2008}, pages
  245--254, 2008.

\bibitem[RRS19a]{arxiv-approximation}
Akbar Rafiey, Arash Rafiey, and Thiago Santos.
\newblock Toward a dichotomy for approximation of h-coloring.
\newblock {\em CoRR}, abs/1902.02201, 2019.

\bibitem[RRS19b]{RRS}
Akbar Rafiey, Arash Rafiey, and Tiago Santos.
\newblock Toward a dichotomy for approximation of h-coloring.
\newblock In {\em 46th International Colloquium on Automata, Languages and
  Programming (ICALP 2019)}. Schloss Dagstuhl-Leibniz-Zentrum fuer Informatik,
  2019.

\bibitem[STU10]{Ueno}
Anish Man~Singh Shrestha, Satoshi Tayu, and Shuichi Ueno.
\newblock On orthogonal ray graphs.
\newblock {\em Discrete Applied Mathematics}, 158(15):1650--1659, 2010.

\bibitem[Zhu92]{zhu}
Xuding Zhu.
\newblock A simple proof of the multiplicativity of directed cycles of prime
  power length.
\newblock {\em Discrete applied mathematics}, 36(3):313--316, 1992.

\bibitem[Zhu17]{zhuk}
Dmitriy Zhuk.
\newblock A proof of csp dichotomy conjecture.
\newblock In {\em 2017 IEEE 58th Annual Symposium on Foundations of Computer
  Science (FOCS)}, pages 331--342. IEEE, 2017.

\bibitem[Zwi98]{Zwick98}
Uri Zwick.
\newblock Finding almost-satisfying assignments.
\newblock In {\em Proceedings of the Thirtieth Annual {ACM} Symposium on the
  Theory of Computing, Dallas, Texas, USA, May 23-26, 1998}, pages 551--560,
  1998.

\end{thebibliography}

\end{document}